\documentclass[a4paper,11pt]{article}
\usepackage{iftex}
\ifPDFTeX
  \pdfoutput=1
\else
  \ifXeTeX\PassOptionsToPackage{xetex}{hyperref}\fi
\fi
\usepackage{amsmath,amssymb,graphicx}
\usepackage[numbers,sort&compress]{natbib}
\usepackage[a4paper,textwidth=151.2mm,top=25mm,bottom=25mm]{geometry}
\usepackage[colorlinks=true,linkcolor=blue,citecolor=blue,urlcolor=blue,linktocpage=true]{hyperref}
\usepackage[font=small,labelfont=bf]{caption}
\numberwithin{equation}{section}
\makeatletter
\newcommand{\affiliation}[1]{\gdef\hnl@affiliation{#1}}
\renewcommand{\abstract}[1]{\gdef\hnl@abstract{#1}}
\newcommand{\keywords}[1]{\gdef\hnl@keywords{#1}}
\newcommand{\acknowledgments}{\section*{Acknowledgments}}
\renewcommand{\maketitle}{%
 \begin{titlepage}
 \vspace*{12mm}
 \begin{center}
 {\LARGE\bfseries \@title\par}
 \vspace{12mm}
 {\large \@author\par}
 \vspace{4mm}
 {\small\itshape \hnl@affiliation\par}
 \end{center}
 \vspace{12mm}
 \noindent\textbf{Abstract}\par\medskip
 \noindent\hnl@abstract\par
 \vspace{7mm}
 \noindent{\small\textbf{Keywords:} \hnl@keywords\par}
 \vfill
 \end{titlepage}
 \pagenumbering{roman}
 \tableofcontents
 \clearpage
 \pagenumbering{arabic}
}
\makeatother
\date{}

\usepackage{hyperref}
\usepackage{slashed}

\usepackage[T1]{fontenc}
\usepackage{bm}
\usepackage{mathtools}
\usepackage{booktabs,tabularx,array}
\usepackage{tikz}
\let\fulltableofcontents\tableofcontents
\renewcommand{\tableofcontents}{\begingroup\footnotesize\fulltableofcontents\endgroup}

\usetikzlibrary{decorations.pathmorphing}
\tikzset{ferm/.style={thick},
 boson/.style={thick,decorate,decoration={snake,amplitude=1.0pt,segment length=5pt}},
 scalar/.style={thick,dashed},
 dip/.style={rectangle,fill=black,inner sep=2.2pt}}

\newcommand{\MSbar}{\overline{\mathrm{MS}}}
\newcommand{\cW}{c_W}
\newcommand{\sW}{s_W}
\newcommand{\dB}{d_B}
\newcommand{\dthree}{d_3}
\newcommand{\dgamma}{d_\gamma}
\newcommand{\dZ}{d_Z}
\newcommand{\dW}{d_W}
\newcommand{\MN}{M_N}
\newcommand{\Shat}{\widehat S}
\newcommand{\That}{\widehat T}
\newcommand{\Uhat}{\widehat U}

\title{Indirect probes of heavy neutral leptons through the dipole portal at FCC-ee}

\author{Natascia Vignaroli}

\affiliation{Dipartimento di Matematica e Fisica ``E. De Giorgi'', Universit\`a del Salento, Lecce, Italy}

\abstract{
We study indirect probes of the gauge-invariant dipole portal of Majorana
heavy neutral leptons (HNLs) at FCC-ee. Virtual HNLs modify electroweak
gauge-boson propagation and lepton couplings through loops with two
dipole interactions. We calculate the resulting finite matching
corrections and their effects on $Z$-pole observables. For real,
flavour-aligned couplings, the local response is specified by four
combinations of coupling shifts. Independent local ultraviolet
interactions can reproduce all four, making the dipole interpretation
dependent on assumptions about those interactions. With improved
Standard Model predictions, FCC-ee can probe equal, same-sign
hypercharge and weak dipole coefficients as small as
$0.14\,\mathrm{TeV}^{-2}$ at an HNL mass of 1 TeV. This projection
assumes dipole dominance, with independent ultraviolet coefficients
set to zero at twice the HNL mass. We quantify how additional
ultraviolet contributions can weaken or strengthen the sensitivity.
At 1 TeV, the projected reach is competitive with HL-LHC same-sign
dilepton searches for Majorana HNLs, while larger masses allow smaller
dipole coefficients to be probed within the effective theory's domain
of validity. Radiative Higgs decays offer a modest further improvement.
The FeynRules, FeynArts and UFO models used in this analysis are
publicly available at
\href{https://doi.org/10.5281/zenodo.23667622}{doi:10.5281/zenodo.23667622}.
}


\usepackage{etoolbox}
\makeatletter
\patchcmd{\l@section}{1.0em}{0.55em}{}{}
\makeatother
\begin{document}
\maketitle
\raggedbottom

\section{Introduction}
\label{sec:introduction}

Heavy neutral leptons (HNLs) are a natural extension of the neutrino
sector and an important target of searches for physics beyond the Standard
Model (SM). Most searches rely on their mixing with active neutrinos.
If further particles exist at a higher scale, however, HNLs can also
interact through higher-dimensional operators. Transition dipoles are
particularly interesting because they couple a neutral fermion directly
to electroweak field strengths, opening production and decay channels
even when mixing-induced interactions are small. Their collider
phenomenology has recently been explored in a gauge-invariant framework
in ref.~\cite{Vignaroli:2025pwn}.

The experimental programme spans very different physical regimes.
For light sterile states, neutrino scattering, stellar environments and
cosmological evolution probe the photon dipole
\cite{Magill:2018jla,Brdar:2020quo}. Meson decays and fixed-target
experiments extend the laboratory coverage
\cite{Barducci:2023hzo,Barducci:2024nvd}.
At higher masses, collider searches provide a direct probe of the dipole portal, through HNL production in Higgs decays~\cite{Butterworth:2019iff} and in lepton~\cite{Ovchynnikov:2023wgg,Zhang:2023nxy,Barducci:2024kig,Brdar:2025iua} and hadron collisions~\cite{Vignaroli:2025pwn,Beltran:2024twr}. The relevant observables range from displaced or radiative decays to
charged-lepton and electroweak-boson final states. These channels probe
different combinations of couplings and depend on the HNL lifetime,
flavour structure and kinematic accessibility.

Precision electroweak measurements address a complementary question.

Even when HNLs are too heavy to be produced directly at a given collider, their virtual effects can modify the decay rates and angular asymmetries of Standard Model particles.

The baseline FCC-ee programme envisages approximately $6\times10^{12}$
$Z$ bosons~\cite{FCC:2025lpp}, enabling precise measurements
of a broad set of electroweak observables sensitive to small virtual corrections.
Unlike a direct search, this test does not
require reconstructing an HNL decay. Its interpretation does require
understanding which corrections can be attributed to the dipole portal
and which could instead arise from other interactions of the particles
that generate it. Comparing direct and indirect searches is therefore
useful both for their mass coverage and for interpreting a possible
signal.

A consistent calculation must use the electroweak-invariant portal.
The hypercharge and weak dipoles involve a Higgs doublet before symmetry
breaking; their photon, $Z$ and $W$ interactions are consequently related.
The weak field strength also generates vertices with two gauge bosons.
These contacts are fixed by the same operator as the three-point dipole,
and enter the virtual corrections together with the familiar triangle
diagrams. This makes precision tests sensitive to the gauge completion
of the portal, rather than only to a photon transition moment.

The effective-theory literature provides several ingredients for this
analysis. Reference~\cite{Chala:2020pbn} studied the running
of Higgs--neutrino operators and the consequences of large dipoles.
Gauge and Yukawa anomalous dimensions have been developed further
\cite{Datta:2020ocb,Ardu:2024tzb}, including numerical evolution in
\texttt{wilson}~\cite{Aebischer:2024csk}.

The FCC-ee study of ref.~\cite{Bolton:2025tqw} considers the Standard
Model effective field theory extended with gauge-singlet right-handed
neutrinos ($\nu$SMEFT). It explores a broad set of production and decay
interactions, with dipoles outside the operator set considered.
Our work complements that study by determining the finite electroweak
corrections generated by two dipole insertions.

This is a different perturbative order from single-insertion running:
it involves both dimension-six thresholds proportional to the heavy mass
squared and dimension-eight terms. Work on dimension-eight bases and
universal electroweak corrections
\cite{Murphy:2020rsh,Chala:2021cgt,Corbett:2024yoy}
and on double dipole insertions in the low-energy effective field theory
below the electroweak scale (LEFT)
\cite{Jenkins:2017dyc} supplies the relevant methodological context.

The ultraviolet interpretation is equally important. Explicit loop completions
\cite{Butterworth:2019iff,Brdar:2020quo,Pages:2025odc} generally generate
interactions besides the dipoles. Reference~\cite{Chala:2020vqp}
gives a renormalizable-model example and its
one-loop matching onto the sterile-neutrino EFT. A symmetry that suppresses
neutrino masses or active--sterile mixing~\cite{Huang:2022pce} can still allow
operators that modify electroweak gauge propagation and lepton currents.
The size of these additional operators determines whether precision data
primarily probe the dipoles or a combination of interactions in the completion. Studies combining mixing
and dipoles~\cite{Bertuzzo:2024eds} illustrate the importance of this
distinction. We therefore examine how the interpretation changes when
independent ultraviolet contributions are allowed, rather than assuming
from the outset that every precision limit applies to every completion.

We take the HNLs to be Majorana fermions. To isolate the
flavour-conserving electroweak response, the numerical analysis uses three
mass-degenerate states with real, universal, flavour-aligned dipoles and
negligible active--sterile mixing. It permits a study of the diagonal response without a simultaneous
fit to flavour-changing transitions or a specified model of neutrino masses. Nonaligned flavour structures and their charged-lepton
flavour-violation constraints are left to dedicated studies. The
radiative stability of the mass and mixing assumptions is discussed
explicitly, and the relation of the retained amplitudes to a Dirac HNL
is given in section~\ref{sec:majorana-dirac}.

Our calculation relates the gauge-invariant portal to a correlated
pattern of widths and asymmetries. At momenta below the HNL mass, the
virtual amplitudes admit a local expansion in interactions among SM
fields. Matching expresses these interactions in a common operator
basis, connecting the corrections to gauge propagation and lepton
couplings with the observables and with possible contributions from
heavier particles. The resulting local response spans four physical combinations. Independent ultraviolet coefficients
can reproduce all four, so the pole data constrain the dipoles only once
these additional contributions are restricted. We quantify the FCC-ee
reach under a common dipole-dominance boundary, its sensitivity to
correlated ultraviolet effects, and the additional information from
radiative Higgs decays. The comparison with the HL-LHC projections of
ref.~\cite{Vignaroli:2025pwn} illustrates the complementarity of
virtual and direct probes within matched flavour and operator conventions.

Section~\ref{sec:portal} defines the model and conventions.
Sections~\ref{sec:selfenergies}--\ref{sec:oblique} develop the loop
calculation, matching and oblique description. Sections~\ref{sec:ewinputs}
and~\ref{sec:higgs} connect these results to pole observables and radiative
Higgs decays, respectively. They are followed by the FCC-ee results and their
ultraviolet interpretation in section~\ref{sec:fcc}; the direct-search
comparison is presented at its end. The appendices give the technical
checks and further comparisons with related operator sectors.
Table~\ref{tab:claims-scope} summarizes the scope of the results.

\section{Gauge-invariant HNL dipole portal}
\label{sec:portal}

We consider a sterile Majorana fermion $N$ of mass $\MN$ and neglect
active--sterile mixing. Its interactions with SM particles are mediated
by the dipole portal. Above electroweak symmetry breaking, the Higgs
doublet is needed to combine the active lepton and sterile state into
a gauge-invariant operator. For one SM lepton flavour, the two
independent dimension-six interactions are
\begin{align}
 {\cal O}_{B} &=
 g'\,\overline L\,\sigma^{\mu\nu}N\,\widetilde H\,B_{\mu\nu},
 &
 {\cal O}_{W} &=
 g\,\overline L\,\sigma^{\mu\nu}N\,\tau^I\widetilde H\,W^I_{\mu\nu},
 \label{eq:gauge-invariant-dipoles}
\end{align}
with
\begin{equation}
 {\cal L}_{\rm dip}=C_B{\cal O}_B+C_W{\cal O}_W+\mathrm{h.c.},
 \qquad
 \tau^I=\frac{\sigma^I}{2}.
 \label{eq:Ldip}
\end{equation}
Here $L$ is the SM left-handed lepton doublet, $H$ is the Higgs doublet
and $\widetilde H=i\sigma^2H^*$ its conjugate. The couplings $g'$ and $g$
belong to hypercharge and weak isospin, respectively; $B_{\mu\nu}$ and
$W^I_{\mu\nu}$ are their field strengths. The matrices $\sigma^I$ act on
weak isospin, whereas $\sigma^{\mu\nu}$ acts on spinor indices.
The coefficients $C_B$ and $C_W$ have mass dimension $-2$.

After electroweak symmetry breaking, $\dgamma$ and $\dZ$ multiply
the dimension-five interactions $\bar\nu_L\sigma^{\mu\nu}N_R A_{\mu\nu}$
and $\bar\nu_L\sigma^{\mu\nu}N_R Z_{\mu\nu}$, respectively,
with their Hermitian conjugates understood. Here $A_{\mu\nu}=\partial_\mu A_\nu-\partial_\nu A_\mu$ and
$Z_{\mu\nu}=\partial_\mu Z_\nu-\partial_\nu Z_\mu$; the two-gauge
interactions are displayed below. With
\begin{equation}
 A_\mu=\sW W^3_\mu+\cW B_\mu,\qquad
 Z_\mu=\cW W^3_\mu-\sW B_\mu,
 \label{eq:AZrotation}
\end{equation}
we define the corresponding gauge-basis combinations
\begin{align}
 \dB &\equiv \cW\dgamma-\sW\dZ
      =\frac{g'v}{\sqrt2}\,C_B,
 \label{eq:dB}\\
 \dthree &\equiv \sW\dgamma+\cW\dZ
      =\frac{gv}{2\sqrt2}\,C_W.
 \label{eq:d3}
\end{align}
The charged-current dipole satisfies
\begin{equation}
 \dW=\sqrt2\,\dthree
 =\frac{gv}{2}C_W
 =\sqrt2\,(\cW\dZ+\sW\dgamma).
 \label{eq:dWrelation}
\end{equation}
Equations~\eqref{eq:dB}--\eqref{eq:dWrelation} will be useful for displaying the
constraints either in the gauge-invariant $(C_B,C_W)$ basis or in the broken-phase
$(\dgamma,\dZ)$ basis.

\subsection{Majorana benchmark and the Dirac limit}
\label{sec:majorana-dirac}
The sterile field in eq.~\eqref{eq:Ldip} is Majorana, $N=N^c$, with
$\mathcal L_N=\tfrac12\bar N(i\slashed\partial-M_N)N$.
Only its right-handed component enters the portal. The factors associated
with Majorana contractions are included in the amplitudes and must not
be appended to the final Wilson coefficients.

The relation to a Dirac HNL follows from the chiral structure of the
propagator connecting two dipole vertices. Consider a Dirac state $N_D$ with the
same mass and dipole coefficients, coupled only through $N_{D,R}$, with
its other chiral component carrying no additional interactions. Assigning
it lepton number one makes the dipoles lepton-number conserving, as in
ref.~\cite{Magill:2018jla}. In both theories the heavy contraction
between a portal and its Hermitian conjugate is
\begin{equation}
 P_R\frac{i(\slashed k+M_N)}{k^2-M_N^2+i0}P_L
 =\frac{i\slashed kP_L}{k^2-M_N^2+i0}.
\end{equation}
The additional Majorana contraction carries lepton number violation.
It does not contribute to the external lepton-number-conserving currents
or the bosonic bubbles considered here in the limit of vanishing
light-neutrino masses and mixing.

Equivalently, write $N_D=(N_1+iN_2)/\sqrt2$ in terms of two degenerate
Majorana fields. Their dipole coefficients are
$C_{Xa}=C_X(1,i)_a/\sqrt2$, and hence
\begin{equation}
 \sum_{a=1}^2 C_{Xa}C_{Ya}^*=C_XC_Y^*,\qquad
 \sum_{a=1}^2 C_{Xa}C_{Ya}=0,
 \qquad X,Y\in\{B,W\}.
 \label{eq:Dirac-Majorana-bilinears}
\end{equation}
The first identity preserves the lepton-number-conserving response;
the second cancels the lepton-number-violating one. The normalization
of the two Majorana couplings compensates their multiplicity, preserving
the strength of the original Dirac interaction.
At equal physical dipole normalization and flavour bilinears, the pole
amplitudes and the two-dipole radiative Higgs bubbles retained in this
work therefore have the same Dirac and Majorana values. For example,
three aligned Dirac states give the same conditional pole contours as
three aligned Majorana states with the same $C_B,C_W$. This statement
applies to these lepton-number-conserving amplitudes at quadratic dipole
order.

Lepton-number-violating observables do distinguish the two cases.
In particular, the Majorana mass naturalness estimate relies on lepton-number
violation, and the HL-LHC projections used here exploit same-sign dileptons.
Their interpretation is specific to the Majorana scenario, even though
the precision amplitudes discussed above also apply to a lepton-number-conserving
Dirac HNL with the matched couplings. Conservation of lepton number
protects against the Majorana mass term, but does not by itself forbid
Dirac Yukawa mixing with the active neutrinos.

\subsection{Gauge completion and contact interactions}
With $Y_L=-1/2$, $Y_{\widetilde H}=-1/2$ and $Y_N=0$, both operators in
 eq.~\eqref{eq:gauge-invariant-dipoles} are hypercharge neutral.
The contraction $\bar L\widetilde H$ is an $SU(2)_L$ singlet, whereas
$\bar L\tau^I\widetilde H$ transforms as an adjoint and contracts with
$W^I_{\mu\nu}$. Thus the unbroken action is gauge invariant before a gauge
or a Higgs background is chosen~\cite{Vignaroli:2025pwn}.

For $D_\mu=\partial_\mu+ig\tau^I W^I_\mu+ig'YB_\mu$, define the field strength
by $[D_\mu,D_\nu]=ig\tau^I W^I_{\mu\nu}+ig'YB_{\mu\nu}$.
Its non-Abelian part is $-g\epsilon^{IJK}W^J_\mu W^K_\nu$.
Consequently ${\cal O}_W$ necessarily contains $N\ell W\gamma$,
$N\ell WZ$ and $N\nu W^+W^-$ contact interactions; their coefficients
are fixed by the same $C_W$ as the three-point dipoles.
To display them explicitly, let $N_R=P_RN$,
$\sigma^{\mu\nu}=i[\gamma^\mu,\gamma^\nu]/2$,
$W^\pm=(W^1\mp iW^2)/\sqrt2$, and use
the derivative field strengths defined above.
Before choosing a scalar background, the complete two-gauge part is
\begin{equation}
 {\mathcal L_{2V}=-g^2C_W\epsilon^{IJK}
 (\bar L_a\sigma^{\mu\nu}N_R)(\tau^I)_{ab}
 \widetilde H_b W^J_\mu W^K_\nu+\mathrm{h.c.}}
 \label{eq:two-gauge-unbroken}
\end{equation}
In the background $\widetilde H=(v/\sqrt2,0)^T$, the interactions split as
\begin{align}
 \mathcal L_{1V}={}&\bar\nu_L\sigma^{\mu\nu}N_R
 (d_\gamma A_{\mu\nu}+d_Z Z_{\mu\nu})
 +d_W\bar\ell_L\sigma^{\mu\nu}N_R (\partial_\mu W^-_\nu-\partial_\nu W^-_\mu)
 +\mathrm{h.c.},\\
 \mathcal L_{2V}^{(v)}={}&2ig d_W\bar\ell_L\sigma^{\mu\nu}N_R
 W^-_\mu(s_W A_\nu+c_W Z_\nu)\nonumber\\
 &+\sqrt2 ig d_W\bar\nu_L\sigma^{\mu\nu}N_R
 W^+_\mu W^-_\nu+\mathrm{h.c.}
 \label{eq:two-gauge-broken}
\end{align}
The signs in eq.~\eqref{eq:two-gauge-broken} follow from the declared
$D_\mu=\partial_\mu+ig\tau\cdot W_\mu+ig'YB_\mu$ convention.
For example,
\begin{align}
 W^3_{\mu\nu}&=(\partial_\mu W^3_\nu-\partial_\nu W^3_\mu)
 +ig(W^+_\mu W^-_\nu-W^-_\mu W^+_\nu),\\
 W^-_{\mu\nu}&=(\partial_\mu W^-_\nu-\partial_\nu W^-_\mu)
 +ig(W^-_\mu W^3_\nu-W^3_\mu W^-_\nu).
\end{align}
Contracting with the antisymmetric $\sigma^{\mu\nu}$ gives the factor
of two. Changing the covariant-derivative sign requires changing the
field-strength and SM gauge-vertex conventions consistently.
There are no independent $C_B$ two-gauge contacts: the hypercharge field
strength is Abelian. In particular the $W\gamma$ and $WZ$ contact
coefficients are fixed to $2ie d_W$ and $2igc_Wd_W$ in this convention.

These vertices are part of the dimension-six operator, despite their
vev-dependent dimension-five broken-phase form. They are not optional
higher-order corrections to the single-boson dipole. The necessity of
their inclusion in direct multi-boson production was demonstrated in
ref.~\cite{Vignaroli:2025pwn}; the same completion supplies the charged
contact diagrams in the present lepton-current calculation. Omitting
them changes the theory and spoils the gauge relations between amplitudes.
An electromagnetic dipole-only description can be appropriate below the
weak scale, but cannot replace this completion for electroweak loops.
The contacts are also included in the collider study of
ref.~\cite{Brdar:2025iua}, as noted in ref.~\cite{Vignaroli:2025pwn}.

The scalar interactions are obtained by restoring the Higgs and Goldstone
fields around the vacuum,
\begin{equation}
 H=\begin{pmatrix}\phi^+\\(v+h+i\phi^0)/\sqrt2\end{pmatrix},\qquad
 \widetilde H=\begin{pmatrix}(v+h-i\phi^0)/\sqrt2\\-\phi^-\end{pmatrix}.
\end{equation}
Substitution into the two operators fixes all Higgs and Goldstone vertices,
including the Higgs/Goldstone partners of the contact interactions.
The neutral-scalar partners of eq.~\eqref{eq:two-gauge-broken} follow by
$v\to v+h-i\phi^0$ in its non-Hermitian part, with the conjugate
replacement in its Hermitian conjugate. The charged-Goldstone partners
are fixed by the second component $-\phi^-$ in
eq.~\eqref{eq:two-gauge-unbroken}; retaining that compact expression
specifies them without any additional Wilson coefficient.
The $v$-dependent vertices therefore cannot be varied independently:
\begin{align}
 d_\gamma&=\frac{v}{\sqrt2}(g'c_W C_B+\tfrac12gs_W C_W),&
 d_Z&=\frac{v}{\sqrt2}(-g's_W C_B+\tfrac12gc_W C_W),\\
 C_B&=\frac{\sqrt2}{g'v}(c_Wd_\gamma-s_Wd_Z),&
 C_W&=\frac{2\sqrt2}{gv}(s_Wd_\gamma+c_Wd_Z).
\end{align}
These relations reduce the broken-phase couplings to two independent portal
coefficients. The same operator structure also simplifies the loop amplitudes.
In particular, the antisymmetry of the dipole gives $q_\mu\sigma^{\mu\nu}q_\nu=0$ at a three-point vertex.
Because the longitudinal part of a gauge propagator is proportional to its
momentum at each endpoint, this identity removes that part whenever one
endpoint is a single-gauge dipole. It therefore simplifies the direct-current
diagrams considered here, as demonstrated in
appendix~\ref{app:gauge-current-check}. Higher-point Slavnov--Taylor
identities also involve contact and Goldstone interactions; they cannot
be inferred from the transverse three-point vertex alone.

For several lepton flavours the quadratic neutral self-energies generalize as
\begin{equation}
 |\dB|^2\to\sum_\alpha|\dB^\alpha|^2,\qquad
 |\dthree|^2\to\sum_\alpha|\dthree^\alpha|^2,\qquad
 \dB\dthree\to
 \sum_\alpha {\rm Re}\!\left(\dB^\alpha\dthree^{\alpha *}\right).
 \label{eq:flavour-generalization}
\end{equation}
Unless stated otherwise, below we display the one-flavour, real-coupling result.
Flavour multiplicities are inserted explicitly when constructing the
universal benchmark.

\paragraph{One-loop topology.}
The implemented gauge-invariant action is Hermitian, and its
three-point broken-phase couplings reproduce eqs.~\eqref{eq:dB}--\eqref{eq:dWrelation}.
At the one-loop order considered here, adding a propagating scalar between
two dipole vertices that already contain two internal fermion lines would
increase the loop number to two. Higgs and Goldstone vertices therefore
add no such one-loop vacuum-polarization graph. Their role depends on the
external states: the Higgs partners are essential for the radiative decays
studied in section~\ref{sec:higgs}.

\subsection{Calculation strategy and model implementation}
\label{sec:computational-methods}
The gauge-invariant dipole Lagrangian is implemented in
\textsc{FeynRules}~\cite{Alloul:2013bka}, with explicit electroweak
index expansion before exporting the interaction vertices. The model,
called \texttt{dipole\_GI\_FA}, supplies generic and classes files with the vertices
for diagram generation with \textsc{FeynArts}~\cite{Hahn:2000kx}.
The amplitudes are converted to \textsc{FeynCalc}~\cite{Shtabovenko:2023idz}
for Dirac and Lorentz algebra, external-state projection and
Passarino--Veltman reduction.
Scalar-integral expansions are evaluated analytically, with
\textsc{Package-X}~\cite{Patel:2016fam} used for the scalar checks.
Dimensional dependence is retained until pole subtraction and finite-part
extraction; the renormalization prescription is specified in
appendix~\ref{app:local-pole-closure}.

Several complementary checks test the normalization and symmetry relations
that enter the observable predictions. These include Born-level normalization of the physical projectors,
neutral and charged two-point transversality, gauge-basis rotations,
Majorana orientations, and reconstruction of the direct currents with
external residues. The contact diagrams are included throughout, so these
checks apply to the electroweak-invariant portal. We also test selected
tree-level matrix elements using the UFO
format~\cite{Degrande:2011ua} and
\textsc{MadGraph5\_aMC@NLO}~\cite{Alwall:2014hca}; these tests check
the vertex normalizations and coupling dependence. The likelihood and figures are evaluated
from the resulting analytic responses using Python numerical routines.
The direct-search comparison uses the projections of
ref.~\cite{Vignaroli:2025pwn}.

We use \textsc{Matchete}~\cite{Fuentes-Martin:2022jrf} to check an example of
the additional lepton currents that can accompany a UV completion.
The example contains a heavy charged vector-like singlet $E$ coupled to
the SM lepton and Higgs doublets. Integrating it out at tree level yields
correlated singlet and triplet currents. Field redefinitions simplify the effective action through dimension
six, and the Pauli-matrix identity expresses its current operators in
the Warsaw basis. This reproduces the $E$ direction in
eq.~\eqref{eq:VLL-response-patterns}, providing an independent check
of the tree-level current benchmark, distinct from the dipole-loop
thresholds.

\paragraph{Model files and reuse for direct searches.}
The FeynRules implementation and its \texttt{dipole\_GI\_FA} and
\texttt{dipole\_GI\_UFO} exports describe the same gauge-invariant portal
Lagrangian used to relate the indirect effects to the direct-search
processes of ref.~\cite{Vignaroli:2025pwn}. The Higgs, Goldstone and
non-Abelian interactions follow from the gauge-invariant operators,
rather than being added as independent dipoles after symmetry breaking.
The UFO can therefore also be used for direct-production studies with
matched flavour assignments, parameter cards and normalization
conventions. The FeynRules source, FeynArts models and UFO implementation
used in the associated studies are publicly available, together with
usage instructions and reproducible validation scripts, in the
\href{https://github.com/vignaron/hnl-dipole-portal-models}{public model repository}.
Version~1.0.0 is archived on Zenodo at
\href{https://doi.org/10.5281/zenodo.23667622}{doi:10.5281/zenodo.23667622}.

\section{One-loop electroweak vacuum polarizations}
\label{sec:selfenergies}

\begin{figure}[tbp]
\centering
\begin{tikzpicture}[every node/.style={font=\small}]
\foreach \x/\toplabel/\botlabel/\leftlabel/\rightlabel in
{0/{$N$}/{$\nu$}/{$V$}/{$V'$},6/{$N$}/{$\ell$}/{$W^+$}/{$W^-$}} {
 \begin{scope}[xshift=\x cm]
 \coordinate (a) at (0,0);\coordinate (b) at (2.3,0);
 \draw[boson] (-1,0)--(a);\draw[boson] (b)--(3.3,0);
 \draw[ferm] (a) to[bend left=50] node[above] {\toplabel}(b);
 \draw[ferm] (a) to[bend right=50] node[below] {\botlabel}(b);
 \node[dip] at (a) {};\node[dip] at (b) {};
 \node[left] at (-1,0) {\leftlabel};\node[right] at (3.3,0) {\rightlabel};
 \end{scope}
}
\end{tikzpicture}
\caption{Two-dipole vacuum polarizations. The neutral graph has
$V,V'\in\{\gamma,Z\}$ and includes the two Majorana orientations; the charged
graph has the normalization in eq.~\eqref{eq:FWW}. Squares indicate dipole
insertions. The relative group and orientation factors give
$F_{WW}=d_W^2F_0/2=d_3^2F_0$ for massless charged leptons.}
\label{fig:vacuum-polarizations}
\end{figure}
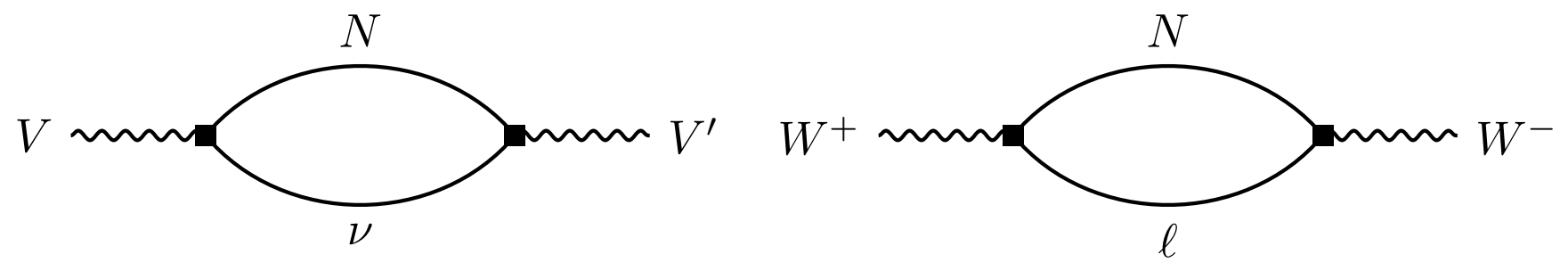

A gauge boson can fluctuate into an HNL and a light lepton and then
return to a gauge boson, as shown in figure~\ref{fig:vacuum-polarizations}.
The resulting self-energy, or vacuum polarization, changes its
propagation and can mix the neutral gauge fields. We derive the common
loop function governing these fluctuations and separate its constant
and momentum-dependent terms. This organization exposes the
correlations among neutral and charged channels and provides the bosonic
input for the effective-theory matching in the next section.

\subsection{Conventions and transverse decomposition}

We define the transverse gauge-boson self-energy by
\begin{equation}
 i\Pi^{\mu\nu}_{VV'}(q)
 =
 i\left(q^\mu q^\nu-q^2g^{\mu\nu}\right)F_{VV'}(q^2),
 \label{eq:selfenergy-convention}
\end{equation}
For local bosonic terms the coefficient is that of the quadratic action
insertion; see appendix~\ref{app:signaudit}. We define the corresponding
dimension-two function by
\begin{equation}
 \Pi_{VV'}(q^2)=q^2F_{VV'}(q^2).
 \label{eq:Pi-dimension-two}
\end{equation}
The explicit one-loop calculation is performed in dimensional regularization,
$D=4-2\epsilon$.  Details of the tensor projection, Passarino--Veltman reduction
and Majorana orientations are given in appendix~\ref{app:loop}.

The three-point dipole vertex contains $\sigma^{\mu\nu}q_\nu$.
Consequently $q_\mu\sigma^{\mu\nu}q_\nu=0$ by antisymmetry, so the two-dipole
bubble is transverse at each external gauge leg. This fixes its tensor
structure but neither its finite constant nor the numerical UV residue.

\subsection{Neutral sector and universal loop function}

The dipole--dipole contributions to the neutral self-energies are controlled by
a single scalar function.  In the $(A,Z)$ basis we find
\begin{equation}
 F_{AA}=\dgamma^2F_0,\qquad
 F_{AZ}=\dgamma\dZ F_0,\qquad
 F_{ZZ}=\dZ^2F_0,
 \label{eq:neutral-universality-AZ}
\end{equation}
where each external gauge field couples through a dipole vertex.
The ordinary SM $Z\nu\nu$ vertex instead enters the direct-current
triangles discussed below. The exact dipole--dipole loop function is
\begin{align}
 F_0(q^2)=\frac{1}{2\pi^2(D-1)q^2}
 \Big\{&
 A_0(\MN^2)\big[(D-2)\MN^2-q^2\big]
 \nonumber\\
 &-(\MN^2-q^2)\big[(D-2)\MN^2+q^2\big]
 B_0(q^2,0,\MN^2)\Big\}.
 \label{eq:F0-exact}
\end{align}
A nontrivial feature of the $q^2\to0$ expansion is that the finite constant term
must be obtained before setting $D=4$: a contribution of the form
$(D-4)\times1/\epsilon$ survives.  The result is
\begin{equation}
 F_0(0)=-\frac{\MN^2}{2\pi^2},
 \label{eq:F0zero}
\end{equation}
and
\begin{equation}
 F_0'(0)=\frac{1}{6\pi^2}
 \left[
 \frac{1}{\epsilon}
 +\ln\frac{\mu_{\rm PX}^2}{\MN^2}
 +\frac{11}{6}-\gamma_E-\ln\pi
 \right],
 \label{eq:F0prime}
\end{equation}
in the dimensional-regularization normalization of the displayed
Package-X scalar integrals, whose scale is denoted by $\mu_{\rm PX}$.
The common $\MSbar$ scale used below is $\mu=\mu_{\rm PX}/(2\pi)$.
The measure-dependent constant accompanying the pole is absorbed when
converting to the common $\MSbar$ convention; the numerator contribution
to the finite $11/6$ is preserved. Appendix~\ref{app:renormalization}
specifies this conversion. The renormalized slope and all Wilson
coefficients below use the same subtraction convention.

Rotating eq.~\eqref{eq:neutral-universality-AZ} to the $(B,W^3)$ basis gives
\begin{equation}
 F_{BB}=\dB^2F_0,\qquad
 F_{33}=\dthree^2F_0,\qquad
 F_{3B}=\dthree\dB F_0.
 \label{eq:gauge-basis-universality}
\end{equation}
Thus the neutral self-energy matrix factorizes:
\begin{equation}
 {\bm F}(q^2)
 =
 F_0(q^2)
 \begin{pmatrix}
 \dB^2 & \dB\dthree\\
 \dB\dthree & \dthree^2
 \end{pmatrix}
 =
 F_0(q^2)
 \begin{pmatrix}\dB\\ \dthree\end{pmatrix}
 \begin{pmatrix}\dB&\dthree\end{pmatrix}.
 \label{eq:rank-one-F}
\end{equation}
It therefore has rank at most one (rank one for a nonzero matrix) at any
$q^2$ within this contribution. Factorization uses a common loop kernel and
one real coupling vector, not gauge invariance alone. A sum over nonaligned
flavour vectors, complex relative phases in the CP-even matrix, or independent
local matching terms can spoil the rank-one equality.

\subsection{Charged sector}

For one charged-lepton flavour and neglecting the charged-lepton mass, the
charged-current result is
\begin{equation}
 F_{WW}(q^2)=\dW^2 F_{\rm single}(q^2)
            =\frac{\dW^2}{2}F_0(q^2)
            =\dthree^2F_0(q^2).
 \label{eq:FWW}
\end{equation}
Consequently,
\begin{equation}
 F_{WW}(q^2)=F_{33}(q^2)
 \qquad (m_\ell=0).
 \label{eq:SU2-check}
\end{equation}
This equality provides a useful check of the $SU(2)_L$ normalization, in
particular of the relation $\dW=\sqrt2\,\dthree$.

The gauge-invariant dipole Lagrangian also contains non-Abelian four-point
interactions such as $N\ell W\gamma$, $N\ell WZ$ and $N\nu WW$.  An explicit
diagram-by-diagram inspection shows that no additional one-loop two-point
topology involving these contact interactions contributes to the $WW$
self-energy in the absence of active--sterile mixing.  Such interactions will,
however, become relevant for higher-point amplitudes, including the $Zee$ contact diagrams of appendix~\ref{sec:zpole} and
$e^+e^-\to W^+W^-$.

\section{EFT interpretation and matching}
\label{sec:matching}

At momenta well below $M_N$, virtual HNL propagation can be represented
by local interactions among SM fields. Matching determines their
coefficients by requiring this low-energy description to reproduce the
expanded amplitudes of the theory containing $N$. The difference between
the coefficients above and below the HNL threshold is the threshold
correction calculated here.

There are therefore two effective descriptions in this study. The
starting dipole EFT contains the HNL and represents still heavier
particles through $C_B,C_W$ and other possible operators. After removing
the HNL as a propagating field, the low-energy EFT contains SM fields
only. The dipole calculation fixes an additive contribution to its
coefficients; the heavier particles can supply independent contributions
to the same operators. Keeping these two sources distinct is essential
for interpreting the precision reach. We begin with gauge propagation
and then consider the direct lepton couplings.

The low-momentum expansion of eq.~\eqref{eq:gauge-basis-universality} separates
two qualitatively different EFT structures:
\begin{equation}
 F_{ij}(q^2)=d_i d_j
 \left[
 -\frac{\MN^2}{2\pi^2}
 +q^2F_0'(0)+{\cal O}(q^4)
 \right],
 \qquad i,j=B,3.
 \label{eq:low-q-expansion}
\end{equation}
Since $\Pi_{ij}=q^2F_{ij}$, the constant in $F_{ij}$ corrects the
ordinary gauge kinetic terms, while its slope adds two powers of
momentum. These contributions correspond to dimension-six and
dimension-eight operators, respectively. Both arise from two insertions
of dimension-six dipoles: the factor $M_N^2$ supplies the two mass powers
needed by the dimension-six coefficient. Thus the dimension of the
induced operator and the number of dipole insertions describe different
aspects of the expansion. The first term is finite; the slope contains
a UV divergence that requires a local counterterm.

\subsection{Finite dimension-six threshold}

The constant term in the transverse form factor modifies the gauge kinetic
terms in a Higgs background. It is therefore represented by operators
containing two Higgs fields and two field strengths. We use $\Delta{\cal L}=+\sum_i C_iQ_i$ and the standard Warsaw normalization
\begin{align}
 Q_{HB}&=(H^\dagger H)B_{\mu\nu}B^{\mu\nu},\nonumber\\
 Q_{HW}&=(H^\dagger H)W^I_{\mu\nu}W^{I\mu\nu},\nonumber\\
 Q_{HWB}&=(H^\dagger\sigma^I H)W^I_{\mu\nu}B^{\mu\nu}.
 \label{eq:dim6-bosonic}
\end{align}
The Warsaw normalization uses a Pauli matrix in $Q_{HWB}$, giving twice
the isospin contraction of the generator in the dipole operator. With $H=(0,v/\sqrt2)^T$,
$H^\dagger\sigma^3H=-v^2/2$. We identify the displayed transverse coefficient with the quadratic
effective-action insertion, as detailed in
appendix~\ref{app:signaudit}. In this convention,
\begin{equation}
 F_{BB}^{HB}(0)=-2v^2C_{HB},\qquad
 F_{33}^{HW}(0)=-2v^2C_{HW},\qquad
 F_{3B}^{HWB}(0)=+v^2C_{HWB}.
 \label{eq:dim6-F}
\end{equation}
Equating these local insertions to the loop form factors therefore gives,
for one HNL coupled to one lepton flavour,
\begin{align}
 \Delta C_{HB}&=+\frac{g'^2\MN^2}{8\pi^2}C_B^2,
 \label{eq:match-HB}\\
 \Delta C_{HW}&=+\frac{g^2\MN^2}{32\pi^2}C_W^2,
 \label{eq:match-HW}\\
 \Delta C_{HWB}&=-\frac{gg'\MN^2}{8\pi^2}C_BC_W.
 \label{eq:match-HWB}
\end{align}
The mixed form factor gives a useful physical consequence. Denoting the
HNL-loop contribution by $F_{3B}^{\rm loop}$, one obtains
\begin{equation}
 F_{3B}^{\rm loop}(0)=-\frac{gg'v^2\MN^2}{8\pi^2}C_BC_W,
 \qquad \Shat=\frac{\cW}{\sW}F_{3B}(0)
 =+\frac{\cW}{\sW}v^2\Delta C_{HWB}.
 \label{eq:Warsaw-S-check}
\end{equation}
The sign of the induced oblique correction is consequently fixed: positive
$C_BC_W$ gives negative $\Shat$. To translate the loop form factor into a
Warsaw coefficient, we use the quadratic action in momentum space, whose
second derivative with respect to the gauge fields is the two-point
insertion. This accounts for the relative factors of diagonal and mixed
field strengths. Equations~\eqref{eq:match-HB}--\eqref{eq:match-HWB}
then specify the dipole contribution to the bosonic coefficients; the UV
boundary terms introduced above are added separately.

\subsection{Finite lepton-current thresholds}
\label{sec:finitecurrentmatching}
The portal couples to the left-handed SM doublet. With negligible
charged-lepton Yukawas, there is no chirality-changing interaction that
would turn its direct correction into a right-handed charged-lepton
current. Electroweak invariance then organizes the leading direct
response into a singlet and a triplet. We use the Warsaw operators
\cite{Grzadkowski:2010es},
\begin{align}
 Q_{Hl}^{(1)}&=(H^\dagger i\overleftrightarrow D_\mu H)
 (\bar L\gamma^\mu L),\\
 Q_{Hl}^{(3)}&=(H^\dagger i\overleftrightarrow D_\mu^I H)
 (\bar L\sigma^I\gamma^\mu L),
\end{align}
where $H^\dagger i\overleftrightarrow D_\mu^I H
=iH^\dagger\sigma^I D_\mu H-i(D_\mu H)^\dagger\sigma^I H$.
Both triplet currents use the Pauli-matrix normalization. We extract their
coefficients from amplitudes with external lepton and Higgs fields and
compare the resulting currents with the broken-phase vertex and external-leg
contributions shown in figure~\ref{fig:zee-topologies}; details are given
in appendix~\ref{app:currentaudit}. For one real flavour, negligible
charged-lepton Yukawa coupling and the local heavy-mass expansion, this gives
\begin{align}
 {\Delta C_{Hl}^{(1)}(\mu)=\frac{\MN^2}{128\pi^2}
 (4g'^2C_B^2+3g^2C_W^2)K_{\MSbar}}\,,\label{eq:finite-Hl1}\\
 {\Delta C_{Hl}^{(3)}(\mu)=-\frac{\MN^2}{128\pi^2}
 (4g'^2C_B^2-g^2C_W^2)K_{\MSbar}}\,,\label{eq:finite-Hl3}\\
 K_{\MSbar}&=1-6\ln\frac{\mu^2}{\MN^2}.
\end{align}
At $\mu=\MN$ the logarithm vanishes. Together with
 eqs.~\eqref{eq:match-HB}--\eqref{eq:match-HWB}, these are the five finite
thresholds used in the dimension-six matching. All have mass dimension $-2$,
but arise at second order in the dimension-six portal, $M_N^2 C^2$.
The full coefficients are $C_i^{\rm EFT}=C_i^{\rm independent}+\Delta C_i$,
where the independent terms encode ultraviolet boundary conditions.
We use naive dimensional regularization (NDR) for the open fermion
chains. The constant $1$ in $K_{\MSbar}$ is specific to this prescription
together with $\MSbar$ subtraction: a finite scheme change shifts it together with an opposite
shift of $C_i^{\rm independent}$, leaving the full amplitude unchanged.
The logarithmic slope is fixed by the one-loop anomalous dimension in
this operator convention; the logarithm alone is not an observable.
Our dipole-dominance forecasts refer specifically to vanishing independent
boundary coefficients in this prescription.

The relative weights of the singlet and triplet follow from their
electroweak group contractions. Their common finite normalization contains
additional dynamical information: the constant in $K_{\MSbar}$ includes
products of $D$-dependent numerator terms and UV poles. There is no mixed $C_BC_W$ term in this
unbroken two-insertion bubble because there is no tree-level $B$--$W^I$
propagator. Mixed coefficients may appear in other topologies or at
higher perturbative orders.

For ${\cal L}_Z=-Z_\mu\bar f\gamma^\mu(g_{fL}P_L+g_{fR}P_R)f$,
$g_Z=e/(s_Wc_W)$, the local direct current shifts are
\begin{align}
 \Delta g_{\nu L}&=\frac{g_Zv^2}{2}(-\Delta C_{Hl}^{(1)}+\Delta C_{Hl}^{(3)}),\label{eq:current-map-neutrino}\\
 \Delta g_{eL}&=-\frac{g_Zv^2}{2}(\Delta C_{Hl}^{(1)}+\Delta C_{Hl}^{(3)}).
 \label{eq:current-map}
\end{align}
These direct shifts already include the external-lepton residues
associated with the current matching. The observable response also
requires gauge propagation, gauge-field normalization and the conversion
to measured electroweak inputs, assembled in section~\ref{sec:ewinputs}.
For the charged current
written as $-{g}/{\sqrt2}\,W^+_\mu\bar\nu_L\gamma^\mu e_L+\mathrm{h.c.}$,
the direct relative shift is $v^2\Delta C_{Hl}^{(3)}$.
This relation follows from the triplet operator and does not require
a separate charged-current matching coefficient at dimension six.

The two contact diagrams in figure~\ref{fig:zee-topologies} contain the
four-point $ZeNW$ interaction fixed by the non-Abelian weak field strength.
They contribute at the same order as the triangles and are needed for the
normalization and symmetry relations of the complete vertex. External-leg
residues normalize the asymptotic lepton states, while input corrections
express the amplitude in terms of measured electroweak parameters. We
combine these contributions using Born-normalized physical couplings and
common $\MSbar$ subtraction, as detailed in appendix~\ref{app:local-pole-closure}.

\begin{figure}[tbp]
\centering
\begin{tikzpicture}[every node/.style={font=\small}]
\begin{scope}
 \coordinate (a) at (0,0);\coordinate (b) at (2,0);\coordinate (t) at (1,1.6);
 \draw[ferm] (-.6,-.6)--(a);\draw[ferm] (b)--(2.6,-.6);
 \draw[ferm] (a)--node[above left] {$N$}(t);
 \draw[ferm] (t)--node[above right] {$\nu$}(b);
 \draw[boson] (a)--node[below] {$W$}(b);
 \draw[boson] (1,2.3)--(t);
 \node[above] at (1,2.3) {$Z$};\node[left] at (-.6,-.6) {$e$};
 \node[right] at (2.6,-.6) {$\bar e$};
 \node[dip] at (a) {};\node[dip] at (t) {};\fill (b) circle(1.5pt);
 \node at (1,-1.15) {(a) $d_Wd_Z$; two orientations};
\end{scope}
\begin{scope}[xshift=4.4cm]
 \coordinate (a) at (0,0);\coordinate (b) at (2,0);\coordinate (t) at (1,1.6);
 \draw[ferm] (-.6,-.6)--(a);\draw[ferm] (b)--(2.6,-.6);
 \draw[boson] (a)--node[above left] {$W$}(t);
 \draw[boson] (t)--node[above right] {$W$}(b);
 \draw[ferm] (a)--node[below] {$N$}(b);
 \draw[boson] (1,2.3)--(t);\node[above] at (1,2.3) {$Z$};
 \node[dip] at (a) {};\node[dip] at (b) {};\fill (t) circle(1.5pt);
 \node at (1,-1.15) {(b) $d_W^2$; $ZWW$};
\end{scope}
\begin{scope}[yshift=-4.1cm,xshift=2.2cm]
 \coordinate (a) at (0,0);\coordinate (b) at (2,0);
 \draw[ferm] (-1,0)--(a);\draw[ferm] (b)--(3,0);
 \draw[ferm] (a) to[bend left=50] node[above] {$N$}(b);
 \draw[boson] (a) to[bend right=50] node[below] {$W$}(b);
 \draw[boson] (-.8,1.3)--(a);\node[above left] at (-.8,1.3) {$Z$};
 \node[dip] at (a) {};\node[dip] at (b) {};
 \node[left] at (-1,0) {$e$};\node[right] at (3,0) {$\bar e$};
 \node at (1,-1.3) {(c) $d_W^2$; contact and its conjugate};
\end{scope}
\end{tikzpicture}
\caption{The five one-particle-irreducible (1PI) HNL contributions to $Zee$, organized into three
classes. Squares indicate dipole vertices, including the four-point
$ZeNW$ contact in (c); dots denote SM vertices. Fermion lines are schematic
and Majorana arrows are suppressed. The sum of (b) and the two contacts,
not (b) alone, gives the $d_W^2$ term of eq.~\eqref{eq:Zee-1PI}.
External-leg and input-parameter terms are added separately.}
\label{fig:zee-topologies}
\end{figure}
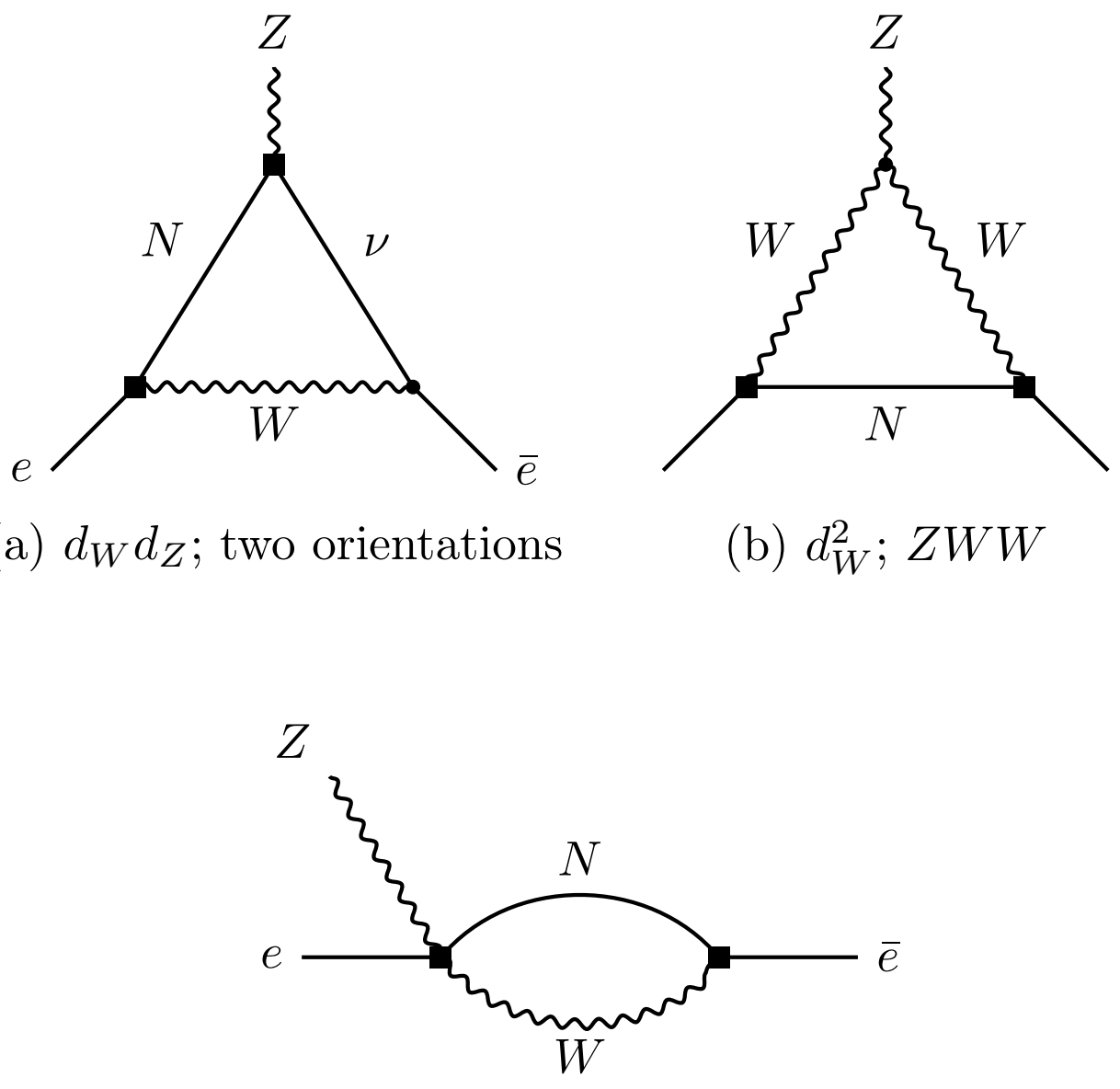

\subsection{Dimension-eight counterterms and leading logarithms}
Beyond the leading terms discussed above, the momentum expansion probes
the curvature of the gauge-boson self-energies. These curvature terms
are conventionally described by $W$ for weak isospin,
$Y$ for hypercharge and $X$ for their mixed component. The parameter
$\Shat$ introduced above instead measures the leading neutral gauge
mixing. Their precise derivative definitions are collected in
section~\ref{sec:oblique}.

The divergent $q^4$ terms arise at order $C_{B,W}^2\sim\Lambda^{-4}$ and require
dimension-eight counterterms in a consistent EFT treatment.  For this purpose we keep a set of operators that reproduces the
amplitudes before imposing the external equations of motion, often called
a Green basis. It makes the propagator corrections explicit through
higher-derivative structures of the form
\begin{align}
 Q_{2B,H}&=(H^\dagger H)
 (\partial_\mu B^{\mu\nu})(\partial^\rho B_{\rho\nu}),\nonumber\\
 Q_{2W,H}&=(H^\dagger H)
 (D_\mu W^{I\mu\nu})(D^\rho W^I_{\rho\nu}),\nonumber\\
 Q_{2WB,H}&=(H^\dagger\sigma^I H)
 (D_\mu W^{I\mu\nu})(\partial^\rho B_{\rho\nu}),
 \label{eq:dim8-green}
\end{align}
up to basis choices, integration by parts and equations of motion
~\cite{Chala:2021cgt,Corbett:2024yoy}.  With precisely the definitions
in eq.~\eqref{eq:dim8-green}, cancellation of the poles in our convention
requires
\begin{align}
 \delta C_{2B,H}&=
 +\frac{g'^2 C_B^2}{12\pi^2\epsilon},
 \nonumber\\
 \delta C_{2W,H}&=
 +\frac{g^2 C_W^2}{48\pi^2\epsilon},
 \nonumber\\
 \delta C_{2WB,H}&=
 -\frac{gg'C_BC_W}{12\pi^2\epsilon}.
 \label{eq:dim8-cts}
\end{align}
The signs and numerical factors in eq.~\eqref{eq:dim8-cts} are tied to the
operator definitions in eq.~\eqref{eq:dim8-green}; translating to a different
dimension-eight basis requires the corresponding field redefinitions and EOM
relations.

A physical prediction for $W$, $Y$ and $X$ therefore requires a boundary
condition for the independent dimension-eight Wilson coefficients.  Under the
minimal assumption that the relevant bosonic dimension-eight coefficients
vanish at a UV scale $\Lambda$, the calculable leading logarithm is obtained
from
\begin{equation}
 F_0'(0)\big|_{\rm LL}
 =\frac{1}{6\pi^2}\ln\frac{\Lambda^2}{\MN^2}.
 \label{eq:F0prime-LL}
\end{equation}
The logarithm measures the evolution between the chosen UV boundary
and the HNL mass. Equation~\eqref{eq:F0prime-LL} follows by combining
that evolution with the stated vanishing boundary coefficient. A
different boundary adds its own local contribution to the same
observable response. The logarithm isolates the scale-dependent part of
the curvature; finite matching contributes at the same order. The
forecast retains both, including the $11/6$ constant in the bosonic
slope and the finite lepton-current terms given in
appendix~\ref{app:local-pole-closure}.

\subsection{Dimension-six lepton-current poles}
\label{sec:currentmatching}

The current thresholds of section~\ref{sec:finitecurrentmatching}
also carry scale dependence. Their UV poles determine the local terms
needed to subtract the divergence and the running that connects the
boundary scale to $M_N$. To make the group structure and normalization
explicit, we write the Warsaw operators~\cite{Grzadkowski:2010es}
with Pauli matrices in both triplet currents:
\begin{align}
 Q_{Hl}^{(1)}&=(H^\dagger i\overleftrightarrow D_\mu H)
                  (\bar L\gamma^\mu L),\nonumber\\
 Q_{Hl}^{(3)}&=(H^\dagger i\overleftrightarrow D_\mu^I H)
                  (\bar L\sigma^I\gamma^\mu L),\\
 H^\dagger i\overleftrightarrow D_\mu^I H
 &=iH^\dagger\sigma^I D_\mu H-i(D_\mu H)^\dagger\sigma^I H.
 \label{eq:Hl-definitions}
\end{align}
Thus the $\sigma^I$ in $Q_{Hl}^{(3)}$ must not be replaced by the gauge generator
$\tau^I=\sigma^I/2$ used in ${\cal O}_W$. We use
$D_\mu=\partial_\mu+ig\tau^I W_\mu^I+ig'YB_\mu$ and define physical neutral
couplings by ${\cal L}_Z=-Z_\mu\bar f\gamma^\mu(g_{fL}P_L+g_{fR}P_R)f$.
This fixes the sign of the coupling shifts below.

Let $C_i^{\rm pole}$ denote the local Wilson coefficient that reproduces the
loop UV pole, rather than its cancelling counterterm. Direct four-point
matching, with external $\bar L,L,H,H^\dagger$, gives
\begin{align}
 C_{Hl}^{(1),\rm pole}
 &=-\frac{3\MN^2}{64\pi^2\epsilon_{\rm UV}}
       (4g'^2C_B^2+3g^2C_W^2),\label{eq:CHl1-pole}\\
 C_{Hl}^{(3),\rm pole}
 &=+\frac{3\MN^2}{64\pi^2\epsilon_{\rm UV}}
       (4g'^2C_B^2-g^2C_W^2),\label{eq:CHl3-pole}\\
 \delta C_{Hl}^{(a)}&=-C_{Hl}^{(a),\rm pole},\qquad a=1,3.
 \label{eq:Hl-counterterms}
\end{align}
Both expressions have dimension $-2$: the two coefficients of dimension $-2$
are accompanied by $\MN^2$. These are divergent dimension-six local terms at
order $\Lambda^{-4}$, unlike the finite bosonic thresholds in
eqs.~\eqref{eq:match-HB}--\eqref{eq:match-HWB}. The poles fix the scale dependence of the current operators. Their finite
thresholds are given in section~\ref{sec:finitecurrentmatching}, while
the independent UV terms are specified by a boundary prescription when
constructing the phenomenological benchmark.

\paragraph{Electroweak group structure.}
The relative singlet and triplet weights follow from weak-isospin
algebra, independently of the scalar loop integral. The two portal vertices carry weak-isospin
indices, and projecting their product onto the current operators separates
the singlet and triplet contributions. For $\widetilde H_i=\varepsilon_{ik}H_k^*$, $\varepsilon=i\sigma^2$, the two
relevant $SU(2)$ tensors reduce to
\begin{align}
 \varepsilon_{ik}\varepsilon^*_{jl}
 &=\tfrac12\delta_{ij}\delta_{kl}
   -\tfrac12\sum_A\sigma^A_{ij}\sigma^A_{kl},\nonumber\\
 \sum_A(\tau^A\varepsilon)_{ik}(\tau^A\varepsilon)^*_{jl}
 &=\tfrac38\delta_{ij}\delta_{kl}
   +\tfrac18\sum_A\sigma^A_{ij}\sigma^A_{kl}.
 \label{eq:su2-current-tensors}
\end{align}
Using the same index ordering in the two current projections, these
identities explain the singlet/triplet ratios $-1$ for $B^2$ and $3$ for $W^2$.
They do not determine the common loop normalization or its sign; those are
fixed by the loop integral and a tree-level current-operator benchmark.

In the unbroken two-dipole bubble, the internal gauge propagator is either
$B$ or $W^I$. There is no tree-level $B$--$W^I$ propagator mixing in the
unbroken theory, so this topology produces no $C_BC_W$ current pole at
dimension six. The cancellation of the corresponding $\MN^2$ terms between
$\gamma$ and $Z$ exchange in the broken calculation is an independent check.
This statement is specific to this order and topology; gauge symmetry alone
does not forbid mixed coefficients in every amplitude. In particular the
mixed bosonic threshold and the dimension-eight vertex residues are nonzero.

\begin{figure}[tbp]
\centering
\begin{tikzpicture}[scale=1, every node/.style={font=\small}]
 \coordinate (a) at (0,0); \coordinate (b) at (3,0);
 \draw[ferm] (-1.2,-.55)--(a); \draw[ferm] (b)--(4.2,-.55);
 \draw[scalar] (-1.2,1.1)--(a); \draw[scalar] (b)--(4.2,1.1);
 \draw[ferm] (a) to[bend left=48] node[above] {$N$} (b);
 \draw[boson] (a) to[bend right=48] node[below] {$B\ \mathrm{or}\ W^I$} (b);
 \node[dip] at (a) {}; \node[dip] at (b) {};
 \node[left] at (-1.2,-.55) {$L$}; \node[right] at (4.2,-.55) {$\bar L$};
 \node[left] at (-1.2,1.1) {$H^*$}; \node[right] at (4.2,1.1) {$H$};
 \node at (5.2,0) {$\Longrightarrow$};
 \node[draw,align=center,rounded corners,inner sep=8pt] at (8,0)
 {$\MN^2 C^2:\ Q_{Hl}^{(1,3)}$\\[4pt]$P^2 C^2:\ \text{dimension eight}$};
\end{tikzpicture}
\caption{Schematic unbroken four-point matching. Each square is a dipole
insertion with one external Higgs field. The graph displays field content;
conjugate insertions and the appropriate external crossings are understood.
The local momentum expansion separates the $\MN^2$ current poles from
higher-derivative terms.}
\label{fig:unbroken-current}
\end{figure}
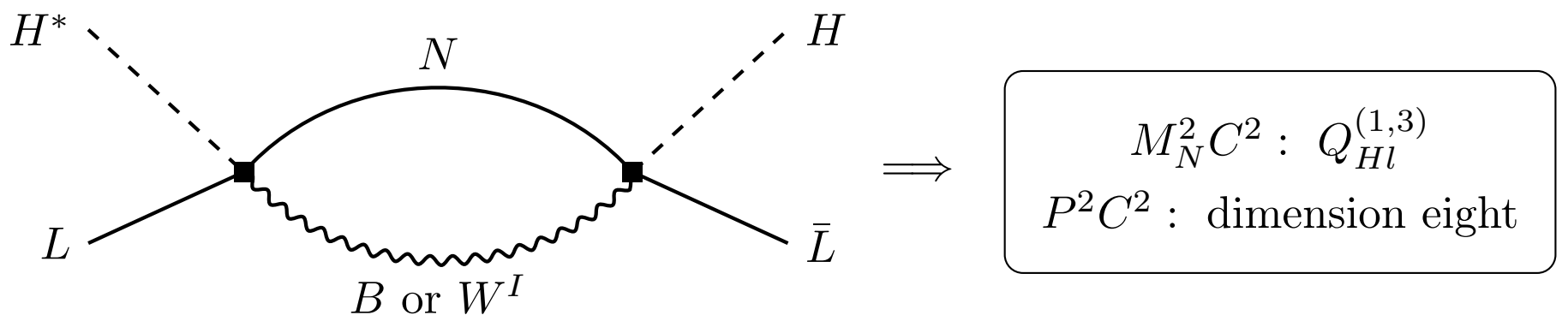

\paragraph{Separating the Higgs current from an EOM term.}
The four-point amplitude in figure~\ref{fig:unbroken-current} carries
independent external Higgs momenta. Its momentum dependence distinguishes a
physical current correction from a change of field variables. With all external momenta incoming, $p_1+p_2+p_3+p_4=0$, the bubble momentum
can be decomposed as
\begin{equation}
 \slashed p_1+\slashed p_3
 =\tfrac12(\slashed p_3-\slashed p_4)
  +\tfrac12(\slashed p_1-\slashed p_2).
 \label{eq:current-momentum-split}
\end{equation}
The first structure projects onto the Higgs current. The second is associated
with a redundant fermion-derivative operator, for example
\begin{equation}
 R_{Hl}^{(1)}=(H^\dagger H)\bar L i\overleftrightarrow{\slashed D}L
 \ \xrightarrow{\rm EOM}\
 (H^\dagger H)(\bar L Y_e H e_R+\mathrm{h.c.}).
 \label{eq:Hl-EOM}
\end{equation}
It vanishes in the charged-lepton Yukawa limit used here, or maps to a
Yukawa-type operator when $Y_e$ is retained. The operator is therefore eliminated in the current matching used here. Consequently a lepton self-energy with a constant Higgs background
alone is insufficient to determine the two current coefficients; the
four-point momentum dependence and the group tensors are required.

After symmetry breaking, the current operators predict
\begin{align}
 \Delta g_{\nu L}^{(6)}&=\frac{g_Zv^2}{2}
                  (-C_{Hl}^{(1)}+C_{Hl}^{(3)}),\nonumber\\
 \Delta g_{eL}^{(6)}&=-\frac{g_Zv^2}{2}
                  (C_{Hl}^{(1)}+C_{Hl}^{(3)}),\qquad
 g_Z=\frac{e}{s_Wc_W},\label{eq:Hl-coupling-map}\\
 \Delta g_{\nu L}^{(6),\rm pole}
 &=\frac{3e^3v^2\MN^2}{64\pi^2c_W^3s_W^3\epsilon_{\rm UV}}
       (4s_W^2C_B^2+c_W^2C_W^2),\nonumber\\
 \Delta g_{eL}^{(6),\rm pole}
 &=\frac{3e^3v^2\MN^2}{32\pi^2c_Ws_W^3\epsilon_{\rm UV}}C_W^2.
 \label{eq:current-broken-check}
\end{align}
These coincide with the independently calculated broken-phase residues in
appendix~\ref{sec:zpole}. The absence of $C_B^2$ in the electron result follows
from $C_{Hl}^{(1),B^2}+C_{Hl}^{(3),B^2}=0$.

\section{Extended oblique parameters}
\label{sec:oblique}

The same propagator corrections affect several fermion species.
Oblique parameters describe this shared part of the electroweak response
through a small set of self-energy coefficients: gauge mixing, shifts
of mass relations and curvature with momentum. They provide a useful
physical summary of the bosonic matching. We use the extended description of
ref.~\cite{Corbett:2025oqk}, including the mixed higher-derivative
parameter $X$. This gives a compact physical interpretation of the
self-energies before the direct lepton currents are included. Our
normalizations are
\begin{align}
 \Shat &=\frac{g}{g'}\Pi'_{3B}(0)
       =\frac{\cW}{\sW}F_{3B}(0),
 \label{eq:Sdef}\\
 \That &=\frac{\Pi_{33}(0)-\Pi_{11}(0)}{m_W^2},
 \label{eq:Tdef}\\
 W &=\frac{m_W^2}{2}\Pi''_{33}(0)
    =m_W^2F'_{33}(0),
 \label{eq:Wdef}\\
 Y &=\frac{m_W^2}{2}\Pi''_{BB}(0)
    =m_W^2F'_{BB}(0),
 \label{eq:Ydef}\\
 X &=\frac{m_W^2}{2}\Pi''_{3B}(0)
    =m_W^2F'_{3B}(0).
 \label{eq:Xdef}
\end{align}
The sign of mixed quantities is convention dependent and should always be read
together with eqs.~\eqref{eq:AZrotation} and \eqref{eq:selfenergy-convention}.

The finite HNL threshold gives
\begin{equation}
 {
 \Shat=
 -\frac{\cW}{\sW}
 \frac{\MN^2}{2\pi^2}\dB\dthree
 }.
 \label{eq:Sresult}
\end{equation}
The dipole vertices force the two-point insertion to be transverse.
Since its scalar form factor is regular at zero momentum, the insertion
vanishes there and cannot generate a gauge-boson mass term. Thus, for the
dipole--dipole two-point functions considered here,
\begin{equation}
 \That=0,
 \label{eq:Tzero}
\end{equation}
and eq.~\eqref{eq:SU2-check} implies $\Uhat=V=0$ in the massless
charged-lepton limit.

Separately from the dimension-six thresholds, and imposing $C_{d=8}(\Lambda)=0$ for the relevant curvature coefficients in the specified scheme, the leading-logarithmic
curvature parameters are
\begin{align}
 Y_{\rm LL}&=
 \frac{m_W^2\dB^2}{6\pi^2}
 \ln\frac{\Lambda^2}{\MN^2},
 \label{eq:YLL}\\
 W_{\rm LL}&=
 \frac{m_W^2\dthree^2}{6\pi^2}
 \ln\frac{\Lambda^2}{\MN^2}
 =
 \frac{m_W^2\dW^2}{12\pi^2}
 \ln\frac{\Lambda^2}{\MN^2},
 \label{eq:WLL}\\
 X_{\rm LL}&=
 \frac{m_W^2\dB\dthree}{6\pi^2}
 \ln\frac{\Lambda^2}{\MN^2}.
 \label{eq:XLL}
\end{align}
Equivalently,
\begin{align}
 Y_{\rm LL}&=
 \frac{m_W^2g'^2v^2}{12\pi^2}C_B^2 L,
 &
 W_{\rm LL}&=
 \frac{m_W^2g^2v^2}{48\pi^2}C_W^2 L,
 \nonumber\\
 X_{\rm LL}&=
 \frac{m_W^2gg'v^2}{24\pi^2}C_BC_W L,
 &
 L&\equiv\ln\frac{\Lambda^2}{\MN^2}.
 \label{eq:oblique-Cbasis}
\end{align}
The finite threshold can similarly be written as
\begin{equation}
 \Shat=-\frac{g^2v^2\MN^2}{8\pi^2}C_BC_W.
 \label{eq:S-Cbasis}
\end{equation}

\subsection{Rank-one correlations}

Equations~\eqref{eq:YLL}--\eqref{eq:XLL} imply the exact correlation
\begin{equation}
 {X_{\rm LL}^2=W_{\rm LL}Y_{\rm LL}}.
 \label{eq:XYW}
\end{equation}
This follows directly from the rank-one form of the self-energy matrix and is
therefore independent of the choice between the gauge-invariant and broken
parameterizations.  At leading-logarithmic order in the curvature terms, fixed $\MN$ and
$\Lambda$ also relate $\Shat$ to $X_{\rm LL}$:
\begin{equation}
 {
 \Shat=
 -3\,\frac{\cW}{\sW}
 \frac{\MN^2}{m_W^2\ln(\Lambda^2/\MN^2)}\,X_{\rm LL}
 }.
 \label{eq:S-X-relation}
\end{equation}
Thus this induced two-point contribution occupies a restricted subspace.
The relation assumes the stated UV boundary conditions and $\ln(\Lambda^2/\MN^2)\ne0$;
independent bosonic Wilson coefficients can violate it. It is not a
constraint on the total EFT or on the nonuniversal lepton-current sector.

In the broken basis,
\begin{align}
 Y_{\rm LL}&=
 \frac{m_W^2L}{6\pi^2}
 (\cW\dgamma-\sW\dZ)^2,
 \nonumber\\
 W_{\rm LL}&=
 \frac{m_W^2L}{6\pi^2}
 (\sW\dgamma+\cW\dZ)^2,
 \nonumber\\
 X_{\rm LL}&=
 \frac{m_W^2L}{6\pi^2}
 (\cW\dgamma-\sW\dZ)(\sW\dgamma+\cW\dZ),
 \label{eq:broken-obliques}\\
 \Shat&=
 -\frac{\cW}{\sW}\frac{\MN^2}{2\pi^2}
 (\cW\dgamma-\sW\dZ)(\sW\dgamma+\cW\dZ).
 \nonumber
\end{align}
Within the same leading-logarithmic contribution, the two curvature
blind directions are therefore
\begin{equation}
 Y=0:\quad \dZ=\frac{\cW}{\sW}\dgamma,
 \qquad
 W=0:\quad \dZ=-\frac{\sW}{\cW}\dgamma.
 \label{eq:blind-lines}
\end{equation}

For the illustrative direction $\dZ=0$,
\begin{equation}
 \frac{|\Shat|}{Y}
 =
 \frac{3\MN^2}
 {m_W^2\ln(\Lambda^2/\MN^2)}.
 \label{eq:SoverY}
\end{equation}
Using $m_W=80.379~{\rm GeV}$ as an illustrative input and
$\Lambda=1~{\rm TeV}$, the equality $|\Shat|=Y$ occurs at
$\MN\simeq99.66~{\rm GeV}$.  This comparison identifies where the two theory contributions have
comparable size. Their relative importance in a precision measurement
also depends on the uncertainties and correlations of the observables
that probe each oblique direction.

\subsection{Relation to epsilon combinations}

The traditional epsilon variables offer an alternative description of
the bosonic contribution to pole observables. In the convention used
here, the extended oblique parameters give, for $\That=\Uhat=V=0$,
\begin{align}
 \delta\epsilon_1 &=
 -W+2\frac{\sW}{\cW}X-\frac{\sW^2}{\cW^2}Y,
 \nonumber\\
 \delta\epsilon_2 &=
 -W+2\frac{\sW}{\cW}X,
 \nonumber\\
 \delta\epsilon_3 &=
 \Shat-W+\frac{X}{\sW\cW}-Y.
 \label{eq:eps-combinations}
\end{align}
For $\dZ=0$ these reduce to
\begin{align}
 \delta\epsilon_1&=0,\nonumber\\
 \delta\epsilon_2&=
 \frac{m_W^2\sW^2\dgamma^2}{6\pi^2}
 \ln\frac{\Lambda^2}{\MN^2},
 \nonumber\\
 \delta\epsilon_3&=
 -\frac{\cW^2\MN^2\dgamma^2}{2\pi^2}.
 \label{eq:eps-dZzero}
\end{align}
These relations summarize the bosonic part of the pole response. The
forecast below uses the pole pseudo-observables directly and also includes
the lepton-current corrections, which the oblique parameters alone do
not describe.

\section{From matching to electroweak observables}
\label{sec:ewinputs}

The matching coefficients affect measured quantities in two ways: they change
the gauge-boson propagators and they modify the lepton vertices directly.
We now combine these effects at fixed electroweak inputs and translate
them into partial widths, ratios and asymmetries. This also specifies which
measurements carry independent information for the FCC-ee forecast.

\subsection{Input scheme and aligned-flavour benchmark}
We use the measured quantities $\{\alpha,M_W,M_Z\}$ to define the
electroweak parameters~\cite{Biekotter:2023xle,Denner:1991kt}. Holding
these inputs fixed is important: a loop correction changes the relation
between a measured mass or charge and the parameters in the Lagrangian.
The associated parameter shifts contribute to every predicted coupling.
The reference parameters obey
\begin{equation}
 s_W^2=1-\frac{M_W^2}{M_Z^2},\qquad e^2=4\pi\alpha,
 \qquad v=\frac{2M_Ws_W}{e}.
\end{equation}
$G_F$ is predicted, not imposed as an additional input. We take three degenerate
HNLs with real, flavour-aligned dipoles $C_X^{ia}=C_X\delta^{ia}$, $X=B,W$.
The bosonic trace has multiplicity $n_F=3$, while the current coefficient of
each external lepton has one contribution. In particular,
\begin{equation}
 C_{HWB}^{\rm univ}=-3\frac{e^2\MN^2}{8\pi^2s_Wc_W}C_BC_W.
 \label{eq:CHWB-universal}
\end{equation}
This flavour choice isolates the flavour-conserving electroweak response
and avoids introducing additional flavour parameters into the forecast.
At this order the aligned matrices obey
$C_XC_Y^\dagger=C_XC_Y I_3$: the response is diagonal and universal.
A general flavour texture, and its charged-lepton flavour-violating
constraints, require a separate analysis and are not included here.

For comparison, a single HNL with universal real couplings has
$C_X^{i1}=C_X$. It obeys
\begin{equation}
 (C_XC_Y^\dagger)_{ii}=C_XC_Y,\qquad
 \operatorname{Tr}(C_XC_Y^\dagger)=3C_XC_Y,
 \qquad X,Y\in\{B,W\},
 \label{eq:diagonal-flavour-equivalence}
\end{equation}
just as the three-HNL aligned benchmark. With the same heavy mass and
independent local boundary prescription, the two realizations therefore
give identical diagonal pole shifts at quadratic order in the dipoles.
The one-HNL realization nevertheless has nonzero off-diagonal flavour
bilinears. Its flavour-changing observables and allowed parameter space
therefore differ from those of the aligned three-HNL benchmark.
The common boundary prescription used for the final forecast is specified
in section~\ref{sec:final-conditional-fit}; it applies consistently to all
local terms through the stated order. The matched direct currents and universal input
corrections are assembled as in appendix~\ref{app:local-pole-closure}.
Both chiralities of the universal response are included; in particular,
$\delta g_{eR}$ is not set to zero. No $G_F$ or $W$-width constraint is
added to the pole likelihood.

\subsection{Linear response of widths and asymmetries}
The pole observables measure effective couplings, not Wilson coefficients
directly. A partial width measures the sum of squared left- and
right-handed couplings, whereas an asymmetry measures their relative
weight. Their combination separates changes in normalization from
changes in chirality. To compare with precision data, the loop-induced
shifts must first be expressed at fixed measured inputs: a correction
to a propagator also changes the relation between Lagrangian parameters
and those inputs. This is why propagator, vertex and external-residue
contributions must be assembled before forming the observable response.

For dimensionless chiral couplings $a_{fL,R}=g_{fL,R}/g_Z$, define
$D_f=a_{fL}^2+a_{fR}^2$ and $r_f=\delta\Gamma_f/\Gamma_f$. At linear order in the matched response,
\begin{align}
 r_f&=\frac{2(a_{fL}\delta a_{fL}+a_{fR}\delta a_{fR})}{D_f},\\
 \delta A_f&=\frac{4a_{fL}a_{fR}
 (a_{fR}\delta a_{fL}-a_{fL}\delta a_{fR})}{D_f^2}.
 \label{eq:input-current-map}
\end{align}
Universal normalization changes are included in $\delta a_f$. For
$A_{\rm FB}^{0,f}=3A_e A_f/4$, the derivative of the forward--backward asymmetry includes both factors:
$\delta A_{\rm FB}^{0,f}=3(A_f\delta A_e+A_e\delta A_f)/4$.
We evaluate the BSM shifts at the declared reference inputs and add them
to the precision-SM prediction. The latter supplies the reference central
values and SM uncertainties used in the forecast.

Let $f_u$ be the SM up-type fraction of the hadronic width, and let
$B_h,B_\nu,B_L$ denote hadronic, invisible and total charged-lepton branching
fractions. The ratio $R_\ell$ compares the hadronic and leptonic widths, while
$R_b$ and $R_c$ are heavy-quark fractions of the hadronic width.
The quantity $\sigma_h$ is the hadronic peak cross section; their definitions are given below. At linear order in the induced shifts,
\begin{align}
 r_h&=f_u r_u+(1-f_u)r_d,\qquad
 r_Z=B_h r_h+B_\nu r_{\rm inv}+B_Lr_\ell,\\
 r_{R_\ell}&=r_h-r_\ell,\qquad
 r_{\sigma_h}=r_\ell+r_h-2r_Z,\\
 r_{R_b}&=r_d-r_h,\qquad r_{R_c}=r_u-r_h.
 \label{eq:hadronic-combinations}
\end{align}
These relations allow the same coupling response to be expressed in
different observable bases. The likelihood uses one nonredundant basis,
with its correlations, so that each measurement contributes once.

\subsection{Observable content of the likelihood and the role of epsilon parameters}
\label{sec:fit-content}
Near the $Z$ resonance, cross sections and angular distributions are
summarized by pole pseudo-observables: widths, ratios and asymmetries
extracted after accounting for radiation and nonresonant contributions.
Several such quantities share the same measured information, so their
correlations are essential when choosing an independent set for a fit. We use the twelve pseudo-observables in
section~\ref{sec:fcc-channel-extension}, including electron, muon and tau
information with the adopted covariance. The defining relations include
\begin{equation}
 \sigma_h^0=\frac{12\pi}{M_Z^2}
 \frac{\Gamma_e\Gamma_{\rm had}}{\Gamma_Z^2},\qquad
 R_\ell^0=\frac{\Gamma_{\rm had}}{\Gamma_\ell},\qquad
 A_{\rm FB}^{0,f}=\frac34 A_e A_f.
\end{equation}
For heavy quarks the ratios instead use the hadronic width in the
denominator: $R_b=\Gamma_b/\Gamma_{\rm had}$ and
$R_c=\Gamma_c/\Gamma_{\rm had}$, as in eq.~\eqref{eq:hadronic-combinations}.
LEP measurements~\cite{ALEPH:2005ab} established this pole-observable
framework. They also provide a reference for the improvement attainable
at FCC-ee, which we quantify in section~\ref{sec:fcc-sensitivity} using
the same dipole response and UV assumptions for both experiments.
The FCC-ee scenario with current SM uncertainties retains the projected
FCC-ee experimental precision; it is distinct from the LEP reference.
Lepton universality of the signal does not eliminate the experimental
correlations between flavours.

$\Gamma_Z$ includes the hadronic, charged-lepton and invisible widths. The
neutrino corrections therefore enter the fit through $\Gamma_Z$ and the
peak cross section. We do not add $\Gamma_{\rm inv}$ or $\Gamma_\ell$ as extra independent
data derived from the same pole dataset. Likewise, a flavour-averaged
leptonic asymmetry would duplicate information already carried by
$A_e,A_\mu,A_\tau$ in the adopted observable vector.
The electromagnetic input is the Thomson coupling evolved with SM
contributions only, $\bar\alpha_Z=\alpha(0)/(1-\Delta\alpha_{\rm SM})$;
we use $\{\bar\alpha_Z,M_W,M_Z\}$ throughout. The dipole contribution to
photon propagation remains explicit in the amplitudes. The adopted
precision on this input is conditional on controlling additional BSM
effects in its experimental extraction.\footnote{Determinations from
low-energy hadronic data and direct scattering measurements need not
have the same BSM dependence~\cite{Mildner:2024wbl,Riembau:2025ppc}.
Our forecast does not simulate the extraction of the electromagnetic
input from scattering distributions.}
No additional $G_F$ measurement is included in the pole forecast.
The measured $M_W$ constrains the prediction through the input covariance.
Charged-current observables such as $W$ widths and branching fractions
would provide complementary information, but are outside the pole dataset
considered here.
The joint covariance in eq.~\eqref{eq:theory-input-conversion}
propagates these input uncertainties without duplicating them.

The $\epsilon_i$ relations in eq.~\eqref{eq:eps-combinations} provide a
compact interpretation of the bosonic contribution. The complete dipole
response also contains the direct lepton currents $C_{H\ell}^{(1,3)}$,
so an equivalent epsilon-based analysis would require vertex parameters
and a consistently transformed covariance. We instead fit the pole
pseudo-observables directly, including both local orders. An epsilon
fit derived from the same measurements would re-express this information
rather than add an independent dataset.

\subsection{EFT validity and the heavy-mass expansion}
\label{sec:lowmass-validity}
Two scale hierarchies and one perturbative expansion control the
interpretation. The hierarchy between electroweak scales and $M_N$
controls the local approximation; the hierarchy between the HNL and the
heavier mediators controls the dipole EFT. Within those descriptions,
the number of portal insertions specifies the perturbative order.
The expansion used for the pole response requires electroweak momenta
and masses to be small relative to the HNL mass. The dipole EFT, in turn,
requires the states that generate the portal to lie above the energies
and masses being described. A naive estimate of the latter condition,
following ref.~\cite{Vignaroli:2025pwn}, is obtained by writing
\begin{equation}
 C_i=c_i/\Lambda_i^2,\qquad M_N<\Lambda_i
 \ \Longrightarrow\ |C_i|<|c_i|/M_N^2
 \label{eq:naive-validity}
\end{equation}
where $c_i$ is a dimensionless coupling and $\Lambda_i$ is the scale
associated with the operator. Setting $|c_i|=1$ gives a reference line
against which to compare the inferred coefficients. Its location depends
on this coupling assumption; it is a scale diagnostic rather than a
unitarity bound. The accuracy of the expansion in electroweak scales over
$M_N$ can be investigated separately using the exact loop functions.
A third issue is the truncation in portal insertions: the prediction
includes the leading $C^2$ amplitude but not amplitudes with additional
dipole insertions. Their size depends on the gauge factors, masses,
loop order and independent higher-dimensional coefficients, and cannot
be inferred from $|C|M_N^2$ alone. In particular, the observable-map
linearization test in appendix~\ref{app:forecast-methods} does not
estimate this portal truncation error. An extra loop with two dipole interactions is expected to be suppressed
by a dimensionless factor of order $g_X^2|C_X|^2E^4/(16\pi^2)$,
where $E$ is the characteristic energy in the loop. This estimate gives
the possible size of a higher-order correction, but its actual size also
depends on the particles and vertices in the diagram. In particular,
replacing Higgs fields by their vacuum values replaces some powers of
$E$ by $v$. A quantitative error estimate therefore requires the omitted
amplitudes. This question is separate from whether the couplings of the
heavy particles that generate the dipole can be treated perturbatively. Likewise, $|c_i|>1$ by itself is
not a breakdown criterion; interpreting such values requires a UV
power counting consistent with the assumed hierarchy.

\subsection{Finite-mass diagnostic}
\label{sec:finite-mass-summary}
The twice-subtracted two-point remainder
$R(t)=F_0(t)-F_0(0)-tF_0'(0)$ isolates the nonlocal part beyond the
constant and slope. Its size at $t=M_Z^2$ is
\begin{center}
\begin{tabular}{rrrrrrr}
\toprule
$\MN$ [GeV] & 100 & 125 & 150 & 200 & 300 & 1000\\
$100|R/F_0(0)|$ & 16.4906 & 5.7279 & 2.5807 & 0.7698 & 0.1464 & 0.00116\\
\bottomrule
\end{tabular}
\end{center}
The decrease with $M_N$ shows how rapidly the two-point function approaches
its local form. These ratios isolate one contribution to the full
prediction; the corresponding vertex, external-field and input corrections
are needed to quantify the error of a complete pole observable. The exact
remainder is discussed in appendix~\ref{app:finite-mass-remarks}, and the
large-mass convergence of the assembled response is tested in
appendix~\ref{app:assembled-heavy-check}.

The same gauge-invariant operators also couple the HNL to the physical
Higgs field. In the next section we examine whether radiative Higgs decays
supply additional information to the pole observables. Production
processes such as $WW$ and $ZH$ offer further possibilities, discussed
later as directions for future studies.

\section{Conditional radiative Higgs decays}
\label{sec:higgs}

We now examine the dipole-induced contributions to $h\to\gamma\gamma$
and $h\to Z\gamma$ in the limit of vanishing light-neutrino masses and
mixing. These decays test whether the same gauge-invariant interactions that
shift the pole couplings leave an additional signal involving the
physical Higgs. Their origin is already contained in the portal Lagrangian:
replacing the Higgs vacuum value by the physical field produces the
single-Higgs partner of each dipole vertex, with coefficient $d_V/v$.
At zero Higgs momentum this is a soft-Higgs insertion: differentiating
the vacuum-dependent dipole couplings generates the Higgs vertex, in
the spirit of the low-energy theorems of ref.~\cite{Kniehl:1995tn}.
Here the HNL mass is held fixed, so the derivative acts on the dipoles
rather than on the heavy mass. The explicit finite-momentum calculation
below is needed for the physical decay. Thus the same coefficients
probed by pole observables also determine a radiative Higgs signal. We first establish the relation to the local
bosonic operators, then include the finite-momentum bubble and its
interference with the SM amplitude. A ratio of decay rates isolates the
additional information while reducing dependence on Higgs production.
The auxiliary two-Higgs projections are described in
appendix~\ref{sec:higgs-matching-update}; the analysis here concerns the
neutral two-dipole decay contribution.

\subsection{Local matching and correlated Higgs couplings}
The neutral Higgs expansion $H=(0,(v+h)/\sqrt2)^T$ replaces each
three-point dipole by a Higgs-dependent vertex with coefficient $d_V/v$.
The local interaction follows by expanding the matched Warsaw operators
around the vacuum. We write its coefficients as
\begin{equation}
 \Delta\mathcal L_h=h\left[
 c_{\gamma\gamma}F_{\mu\nu}F^{\mu\nu}
 +c_{Z\gamma}Z_{\mu\nu}F^{\mu\nu}
 +c_{ZZ}Z_{\mu\nu}Z^{\mu\nu}
 +c_{WW}W^+_{\mu\nu}W^{-\mu\nu}\right].
 \label{eq:higgs-local-definition}
\end{equation}
The coefficients $c_{VV'}$ have dimension $-1$ and include the full
normalization of the displayed field-strength products, with no implicit
factors $1/4$ or $1/2$. They describe the induced Higgs interactions,
whereas the dimension-$-2$ coefficients $C_B,C_W$ define the portal. At first order in the induced local operators, the neutral
rotation gives
\begin{align}
 c_{\gamma\gamma}&=v(c_W^2 C_{HB}+s_W^2 C_{HW}-s_Wc_W C_{HWB}),\\
 c_{Z\gamma}&=v[2s_Wc_W(C_{HW}-C_{HB})
                         -(c_W^2-s_W^2)C_{HWB}],\\
 c_{ZZ}&=v(s_W^2 C_{HB}+c_W^2 C_{HW}+s_Wc_W C_{HWB}),\\
 c_{WW}&=2vC_{HW}.
 \label{eq:higgs-Warsaw-rotation}
\end{align}
The Wilson coefficients in these equations include the chosen HNL
multiplicity. Inserting only their dipole-induced thresholds, for $n_F$
degenerate aligned HNLs with real couplings, yields
\begin{align}
 {c_{\gamma\gamma}=+\frac{n_F M_N^2}{4\pi^2v}d_\gamma^2},\qquad
 &{c_{Z\gamma}=+\frac{n_F M_N^2}{2\pi^2v}d_\gamma d_Z},\\
 c_{ZZ}=+\frac{n_F M_N^2}{4\pi^2v}d_Z^2,\qquad
 &c_{WW}=+\frac{n_F M_N^2}{4\pi^2v}d_W^2.
 \label{eq:higgs-local-dipoles}
\end{align}
The neutral local coefficients obey $c_{Z\gamma}^2=4c_{\gamma\gamma}c_{ZZ}$,
reflecting the rank-one loop threshold; arbitrary independent UV operators
can violate this relation. The coefficients multiplying the displayed
field-strength operators are not, by themselves, full $hZZ$ or $hWW$
amplitudes: input shifts, field normalization and other tensor structures
must also be included where relevant.

This result motivates studying $h\to\gamma\gamma$ together with
$h\to Z\gamma$. The former local contribution vanishes on $C_W=-2C_B$,
where $d_\gamma=0$. The latter also vanishes on
$C_W=2(s_W^2/c_W^2)C_B$, where $d_Z=0$ and $h\gamma\gamma$ is generally
nonzero. These are zeros of the retained local dipole threshold, not a
claim that all Higgs observables become insensitive there. In particular,
no sign-definite $Z\gamma$ interference follows from this matching alone,
since $d_\gamma d_Z$ changes sign between coefficient directions.

For on-shell photons we use $e(0)$ in $d_\gamma$ and the declared
weak-input coupling in $d_Z$. This convention is also used in the SM
radiative amplitudes; a local threshold with a single common coupling
must be translated before quoting numerical decay interference.

\subsection{Interference and perturbative order}
With the normalization of eq.~\eqref{eq:higgs-local-definition}, an on-shell
CP-even amplitude described by the corresponding total coefficient has
\begin{align}
 \Gamma(h\to\gamma\gamma)&=\frac{m_h^3}{4\pi}
                      |c_{\gamma\gamma}^{\rm tot}|^2,\\
 \Gamma(h\to Z\gamma)&=\frac{m_h^3}{8\pi}
       \left(1-\frac{M_Z^2}{m_h^2}\right)^3|c_{Z\gamma}^{\rm tot}|^2.
 \label{eq:higgs-width-normalization}
\end{align}
Here the SM amplitude must be expressed in exactly the same tensor and
Lagrangian convention before adding the HNL correction. At the retained
order $C_{B,W}^2\sim\Lambda^{-4}$, its interference gives
\begin{equation}
 \frac{\delta\Gamma_i}{\Gamma_i^{\rm SM}}
 =2\,\mathrm{Re}\left(\frac{\delta c_i}{c_i^{\rm SM}}\right),
 \qquad i=\gamma\gamma,Z\gamma.
\end{equation}
For reference, in the convention above the one-loop SM coefficients are
\begin{align}
 c_{\gamma\gamma}^{\rm SM}&=(-7.80909+0.05273\,i)\,10^{-6}\;\mathrm{GeV}^{-1},\nonumber\\
 c_{Z\gamma}^{\rm SM}&=(-2.87048+0.00112\,i)\,10^{-5}\;\mathrm{GeV}^{-1}.
\end{align}
They include $W,t,b,c,\tau$ loops.\footnote{We use the electroweak inputs
of the forecast, $m_t=172.57$ GeV, $m_h=125.2$ GeV, and illustrative
one-loop masses $m_b=3.0$ GeV, $m_c=0.65$ GeV, $m_\tau=1.777$ GeV.
The same mass enters each light-fermion propagator and Yukawa coupling;
no higher-order mass conversion is implied. External photons use
$e_0^2=4\pi\alpha(0)$ and the weak coupling uses $\bar\alpha_Z$.}
At $M_N=1$ TeV, $r_B=2$ and $C_B=C_W=0.1\,\mathrm{TeV}^{-2}$,
the interference gives $(r_{\gamma\gamma},r_{Z\gamma})
=(-4.88391,-0.72134)\,10^{-3}$, where $r_i=\delta\Gamma_i/\Gamma_i^{\rm SM}$.
The squared HNL amplitude is of order $\Lambda^{-8}$ and is not included
in this truncation. The same-order independent local matching contributions
remain separate assumptions, as in the electroweak analysis.

\subsection{Finite-momentum neutral two-dipole bubble}
\label{sec:higgs-finite-bubble}
The finite-momentum decay amplitude is particularly simple because the
Higgs attaches to a dipole vertex rather than to a separate neutrino
Yukawa interaction. At the other vertex only a gauge boson enters, so
its momentum fixes the momentum flowing through the two-point fermion
loop. Attaching the Higgs at either end therefore probes the same scalar
kernel at the two gauge-boson virtualities. The tensor algebra leading
to this result is given in appendix~\ref{app:higgs-tensor-proof}.
For on-shell photons the virtuality is zero, whereas an on-shell $Z$
probes $M_Z^2$. This explains why the two radiative decays have different
finite-momentum corrections even though they originate from the same
loop function.

Summing the two possible Higgs attachments, with normalization fixed by
 eq.~\eqref{eq:higgs-local-dipoles}, gives the on-shell bubble coefficients
\begin{align}
 \delta c_{\gamma\gamma}^{\rm bub}
 &=-\frac{n_Fd_\gamma^2}{2v}F_0^{\rm ren}(0),\\
 \delta c_{Z\gamma}^{\rm bub}
 &=-\frac{n_Fd_\gamma d_Z}{2v}
       [F_0^{\rm ren}(0)+F_0^{\rm ren}(M_Z^2)].
 \label{eq:higgs-finite-bubble}
\end{align}
Both reproduce the local threshold. For two on-shell photons the loop
coefficient is therefore equal to its local value within this contribution.
The Higgs mass enters its tensor kinematics and decay phase space but does
not supply a third denominator or an additional scalar-loop invariant in
this contact-vertex topology. This property must not be generalized to
Higgs triangles with a separate Yukawa vertex or to independent derivative
Higgs operators.

Writing $z=M_Z^2/M_N^2$, the remainder is
$R(M_Z^2)=M_N^2r(z)/(6\pi^2)$. For $0<z<1$, the function entering the
decay amplitude is
\begin{align*}
 r(z)&=\frac{2-z-(1-z)(2+z)b(z)}{z}+1-\frac76z,\\
 b(z)&=2+\frac{1-z}{z}\ln(1-z).
\end{align*}
Its derivation and small-$z$ expansion are given in
appendix~\ref{sec:nonlocal-check}. The $Z\gamma$ bubble relative to the
local threshold is
\begin{equation}
 \frac{\delta c_{Z\gamma}^{\rm bub}}{c_{Z\gamma}^{\rm local}}
 =1-\frac{z}{6}\left(\ln\frac{\mu^2}{M_N^2}+\frac{11}{6}\right)
    -\frac{r(z)}{6}.
 \label{eq:higgs-bubble-ratio}
\end{equation}
This ratio is understood away from a zero of $d_\gamma d_Z$; the coefficient
itself remains regular and vanishes at that zero. At $\mu=M_N$ it equals
$0.93263$, $0.98974$ and $0.99745$ for $M_N=200$, 500 and 1000 GeV,
respectively. The corresponding width shifts follow by interference with the SM amplitudes.

The momentum-dependent bubble also produces a UV-divergent local
contribution to $c_{Z\gamma}$, with residue
$-n_Fd_\gamma d_ZM_Z^2/(12\pi^2v)$. Renormalization absorbs this pole
into a local Higgs operator. Its finite coefficient describes an
additional short-distance contribution, so it must be specified along
with the dipole threshold. If its independent
renormalized projected contribution is set to zero at $\Lambda$, the logarithm
in eq.~\eqref{eq:higgs-bubble-ratio} is replaced by
$\ln(\Lambda^2/M_N^2)$ after combining loop and local running. This is an
additional Higgs boundary assumption, not a consequence of setting the
$Z$-pole local combinations to zero. At 200 GeV the resulting ratios are
$0.88460$ and $0.82111$ for $\Lambda/M_N=2$ and 5; at 1 TeV they are
$0.99553$ and $0.99299$. Arbitrary independent finite Higgs matching terms
can modify these numbers. The photon bubble has no corresponding
momentum-dependent UV pole on shell, but independent local $h\gamma\gamma$
contributions remain possible.

For HNL masses above the Higgs mass the channel $h\to N\nu$ is kinematically closed.

\subsection{Production-independent decay contrast}
\label{sec:decay-contrast}
The SM and direct dipolar radiative amplitudes both start at one loop.
We include $2\operatorname{Re}(\mathcal A_{\rm SM}^{(1)*}
\mathcal A_{\rm dip}^{(1)})$. A dipole-induced one-loop input or
external-residue correction multiplying $\mathcal A_{\rm SM}^{(1)}$
is a two-loop amplitude and is not selectively added. This counting does
not apply automatically to independent UV operators generated at tree level.

For a fixed production category $p$, the linear signal-strength response is
$\delta\mu_{pi}=n_p+r_i$, where $n_p$ absorbs production and the total
width. Profiling one free $n_p$ per category leaves the contrast
\begin{equation}
 R_H=\frac{\Gamma(h\to Z\gamma)}{\Gamma(h\to\gamma\gamma)},\qquad
 \frac{\delta R_H}{R_H}=r_{Z\gamma}-r_{\gamma\gamma}.
 \label{eq:decay-contrast}
\end{equation}
This construction uses consistently defined, unfolded production
categories; it does not identify experimental acceptances between decays.
The projected fractional errors on $(\sigma_p\mathcal B_{\gamma\gamma},
\sigma_p\mathcal B_{Z\gamma})$ are $(3.6,11.8)\%$ for $ZH$ at 240 GeV,
$(13,22)\%$ for $ZH$ at 365 GeV and $(15,23)\%$ for fusion at
365 GeV~\cite{Selvaggi:2025kmd}. These are rate errors, not errors on absolute
partial widths. With independent experimental categories and channels,
the combined contrast error is $10.30\%$. Within-category correlations
$\rho=\pm0.5$, considered only as sensitivity tests, change it to
$8.37$--$11.86\%$; these correlations are not provided by the source.

The combined forecast retains the twelve-observable $Z$ covariance
and adds the decay contrast with its shared SM-input correlations.
The contrast derivatives use the six input directions of the source
covariance, with the $G_F\to\alpha$ conversion applied jointly to
the pole and decay observables. This includes the common input
uncertainties once. Intrinsic decay uncertainties are common
across production categories and independent between the two decay
channels. We adopt the upper-end conservative estimates of $1\%$ for
$\gamma\gamma$ and $1.5\%$ for $Z\gamma$ from
ref.~\cite{deBlas:2025gyz}, table 3.3. The aggressive scenario uses
$0.4\%$ for $\gamma\gamma$ and treats the intrinsic $Z\gamma$ theory
error as negligible, since the source gives no aggressive projection
for that channel. This is an assumption about the future precision,
not a statement that the theory error vanishes. The resulting contrast
errors are $10.46\%$ (conservative) and $10.31\%$ (aggressive).
Retaining instead a $1.5\%$ intrinsic $Z\gamma$ error in the aggressive
scenario gives $10.42\%$. The experimental uncertainty dominates in
all three cases, so this choice has little impact on the additional
information supplied by the Higgs decays.

The SM derivatives are evaluated at one loop with $W,t,b,c,\tau$,
holding light masses fixed. The resulting comparison uses a leading
response together with a projected SM uncertainty budget and the
additional boundary specified above for independent radiative coefficients.
A precision-Higgs fit would additionally require higher-order
$\alpha_s$ derivatives and a correspondingly improved amplitude calculation.

\begin{table}[tbp]
\centering
\begin{tabular}{lrr}
\toprule
SM scenario & $\Lambda/M_N=2$ & $\Lambda/M_N=5$\\
\midrule
Conservative & $6.78\%$ & $1.24\%$\\
Aggressive & $0.97\%$ & $0.16\%$\\
Aggressive, $1.5\%$ $Z\gamma$ error & $0.95\%$ & $0.15\%$\\
\bottomrule
\end{tabular}
\caption{Largest relative reduction of the conditional coefficient-region
radius in an angular scan at $M_N=1$ TeV after adding the decay contrast.
The pole and combined results use the same joint 95\% C.L. construction. The entries give the largest improvement over the scanned coupling directions.}
\label{tab:decay-gain}
\end{table}

Table~\ref{tab:decay-gain} shows that this complement is modest when
future SM theory precision at the pole is achieved. The shared-input
correlations with the contrast have magnitude below $4\times10^{-4}$
in the stated covariance model. Both direct radiative contributions
vanish for $C_W=-2C_B$; the single contrast has a further cancellation
direction. Independent finite $h\gamma\gamma$ and $hZ\gamma$ terms can
absorb its signal. It consequently does not remove the unrestricted UV
degeneracy. We therefore present the pole reach separately in the
following section, with table~\ref{tab:decay-gain} quantifying the
additional information supplied by the decay contrast.

\section{FCC-ee reach and its physical interpretation}
\label{sec:fcc}
\label{sec:final-conditional-fit}

The precision programme of FCC-ee can test dipole interactions through
their virtual effects, including masses beyond its direct-production
reach. We now quantify this reach and identify the
assumptions needed to interpret it. The central issue is that the dipoles
and additional short-distance interactions can change the same measured
couplings. We therefore present the sensitivity under dipole dominance
first, then examine how it changes when additional contributions are
allowed, and finally compare it with direct HNL searches.

\subsection{Measurements and benchmark assumptions}
\label{sec:fcc-channel-extension}
Our forecast combines the total $Z$ width, the hadronic peak cross section,
leptonic ratios and asymmetries, and heavy-quark ratios and asymmetries:
\begin{equation}
 (\Gamma_Z,\sigma_h^0,R_e,R_\mu,R_\tau,
 A_e,A_\mu,A_\tau,R_b,R_c,A_b,A_c).
\end{equation}
Together, these measurements distinguish changes in the overall decay
rate from changes in the relative strengths of left- and right-handed
couplings. All three charged-lepton flavours contribute. We use the
experimental projections and correlated SM uncertainties supplied with
ref.~\cite{Armadillo:2026mvp}, in the input scheme introduced in
section~\ref{sec:ewinputs}.

The theoretical uncertainty of the SM prediction is important because the
expected experimental errors are very small. We compare the current SM
uncertainties with conservative and aggressive projections for future
calculations. In ref.~\cite{Armadillo:2026mvp}, the conservative
scenario assumes moderate improvements achievable by extending existing
higher-order techniques; the aggressive scenario assumes larger gains
requiring advances in those methods. Both include uncertainties in SM
predictions and in the theory input needed to extract pole observables.
The experimental-only result
shows the sensitivity that would be obtained if these SM uncertainties
were negligible. Correlations between observables and shared inputs are
included throughout; the covariance conversion is detailed in
appendix~\ref{app:forecast-methods}. Changing the normalization of the electromagnetic input coordinate
shifts the quoted coefficient reaches at 1 TeV by less than $0.36\%$.

To specify dipole dominance quantitatively, we set the independent local
UV coefficients to zero at a common scale $\Lambda_B$ and evolve them
consistently to the HNL threshold. We consider $r_B=\Lambda_B/M_N=2,5$.
The choice of $r_B$ specifies the interval over which the independent
coefficients evolve before the HNL is integrated out. The two benchmarks
illustrate this UV dependence, with a wider range discussed in
appendix~\ref{app:forecast-methods}; neither $r_B=2$ nor the bands between
the two curves define an EFT validity boundary or a statistical error. The HNL flavour assignment and the local expansion
are those established in section~\ref{sec:ewinputs}.

We quote joint 95\% C.L. regions for an expected dataset centred on the SM.
The leading shifts depend on three combinations, $C_B^2$, $C_BC_W$
and $C_W^2$. We first construct a joint confidence region for their
linear response, then select the points obtainable from two real dipole
coefficients. This avoids assuming a nonzero linear dependence on
$C_B,C_W$ at the SM point, where that dependence vanishes. For the
fixed-UV benchmark this gives the threshold $\chi^2_3(0.95)\simeq7.81$.
Appendix~\ref{app:forecast-methods} explains the construction and its
coverage. Axis and diagonal values quoted below are sections through
these joint regions.

\subsection{Sensitivity under dipole dominance}
\label{sec:fcc-sensitivity}
\begin{figure}[htbp]\centering
\includegraphics[width=\textwidth]{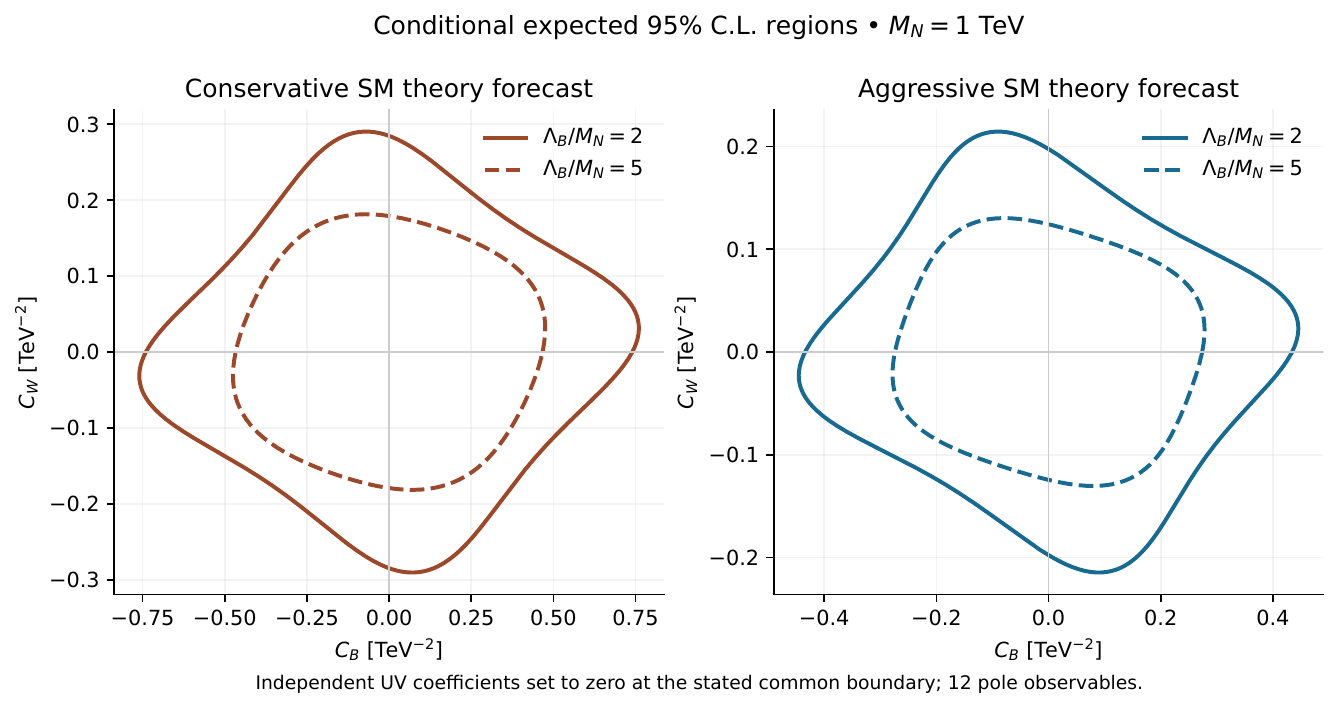}
\caption{Conditional expected 95\% C.L. regions with conservative and
aggressive SM uncertainty forecasts. Solid and dashed lines denote the
two common UV boundary hypotheses.}
\label{fig:final-FCC-contours}\end{figure}

The projected reach is illustrated in figure~\ref{fig:final-FCC-contours}
and quantified at $M_N=1$ TeV in table~\ref{tab:final-FCC}. The contours
show that FCC-ee is sensitive to both portal coefficients, with different
strengths along the axes and the two diagonals. Their shape reflects the
relative weights of the hypercharge and weak interactions and the
mixed contribution proportional to $C_BC_W$.

For $C_B=C_W$ and $r_B=2$, the reach with current SM uncertainties is
$0.5044\,\mathrm{TeV}^{-2}$. It improves to $0.2201\,\mathrm{TeV}^{-2}$
for the conservative scenario or $0.1431\,\mathrm{TeV}^{-2}$ for the
aggressive scenario. The corresponding
experimental-only value is $0.0650\,\mathrm{TeV}^{-2}$. The gap between
these results makes the physical role of improved SM calculations clear:
the experimental sample alone does not determine the attainable dipole
sensitivity. The conservative forecast probes
coefficients well below $1\,\mathrm{TeV}^{-2}$ in several directions.
For a uniform rescaling of standard deviations, $\sigma\to a\sigma$
with correlations fixed, the quadratic response gives
$C_{\rm reach}\to\sqrt{a}C_{\rm reach}$. A tenfold reduction of all
errors would thus improve the coefficient reach by $\sqrt{10}\simeq3.16$.
The fourth-root scaling applies to the variance, or to inverse luminosity
when statistical errors dominate. The actual SM scenarios change different
error components by different amounts.

\begin{table}[htbp]\centering
\begin{tabular}{llrrrr}\toprule
Dataset / SM errors & $r_B$ & $C_B$ axis & $C_W$ axis & $C_B=C_W$ & $C_B=-C_W$\\\midrule
LEP reference & 2 & 2.7985 & 1.1409 & 0.9170 & 1.0512\\
LEP reference & 5 & 1.7677 & 0.7181 & 0.6392 & 0.6875\\\midrule
Experimental only & 2 & 0.3481 & 0.0677 & 0.0650 & 0.0619\\
Experimental only & 5 & 0.2189 & 0.0426 & 0.0426 & 0.0416\\
Current & 2 & 1.8722 & 0.8381 & 0.5044 & 0.4830\\
Current & 5 & 1.1771 & 0.5274 & 0.4449 & 0.4172\\
Conservative & 2 & 0.7401 & 0.2845 & 0.2201 & 0.2472\\
Conservative & 5 & 0.4669 & 0.1791 & 0.1614 & 0.1747\\
Aggressive & 2 & 0.4340 & 0.1976 & 0.1431 & 0.1836\\
Aggressive & 5 & 0.2738 & 0.1244 & 0.1069 & 0.1268\\
\bottomrule\end{tabular}
\caption{Sections of conditional joint 95\% C.L. regions at $M_N=1$ TeV,
using $\chi^2_3(0.95)\simeq7.81$,
with the common finite boundary and the covariance prescriptions of
appendix~\ref{app:forecast-methods}. Units are $\mathrm{TeV}^{-2}$;
on each diagonal the entry is $|C_B|=|C_W|$.
All entries are expected sensitivities for SM-centred data. The LEP
reference uses four flavour-universal pole observables with their
experimental, input and SM theory covariance; the remaining rows use
the twelve-observable FCC-ee projection. If
$\Lambda_{\rm UV}=\Lambda_B$, the aggressive $C_B$-axis entry at
$r_B=5$ corresponds to $|c_B|=|C_B|\Lambda_B^2\simeq6.85$;
its realization with perturbative loop-generated dipoles requires
additional UV assumptions (appendix~\ref{sec:dipole-dominance}).}
\label{tab:final-FCC}\end{table}

LEP provides a useful measure of the gain in sensitivity. We use the
flavour-universal combination $(\Gamma_Z,\sigma_h^0,R_\ell,A_{\rm FB}^{0,\ell})$,
including the luminosity update and the correlated input and SM theory
uncertainties supplied with refs.~\cite{Janot:2019oyi,Mildner:2024wbl}.
To isolate the improvement in precision, we centre both datasets on the
SM and apply the same finite dipole response, UV boundary and confidence
construction. The LEP entries in table~\ref{tab:final-FCC} are therefore
expected sensitivities at LEP precision, rather than observed exclusion
limits or a global LEP/SLC fit. The covariance prescription is given in
appendix~\ref{app:forecast-methods}.

At $M_N=1$ TeV and $r_B=2$, LEP precision probes equal, same-sign
coefficients at about $0.92\,\mathrm{TeV}^{-2}$. FCC-ee lowers this to
$0.22$ or $0.14\,\mathrm{TeV}^{-2}$, an improvement by factors of
$4.2$ and $6.4$ in the conservative and aggressive SM scenarios.
Across the two axes and two diagonals, and for both boundary choices,
the gains range from $3.8$--$4.3$ and $5.4$--$6.5$, respectively.
These factors compare the four-observable LEP reference with the
FCC-ee observable set adopted here, including its separate
lepton-flavour and heavy-quark measurements. They quantify the smaller dipole coefficients
that can be tested under the same UV hypothesis; the larger LEP values
do not, by themselves, establish perturbative realizations at that reach.

With current SM uncertainties, the improvement is only a factor of
$1.4$--$2.2$. The substantially greater reach of the future-theory
scenarios therefore requires progress in SM predictions as well as the
larger experimental sample. In this sense, FCC-ee extends the precision
test established at LEP into a region of appreciably weaker dipole
interactions, while retaining its dependence on the independent UV
contributions discussed below.

Changing the boundary from $r_B=2$ to $r_B=5$ strengthens the same-sign
reach to $0.1614\,\mathrm{TeV}^{-2}$ and
$0.1069\,\mathrm{TeV}^{-2}$ in the conservative and aggressive scenarios,
respectively. The boundary affects the finite local coefficients through
their running, and hence changes the observable response. We display
both choices to expose this theoretical assumption instead of treating
one as a unique prediction of a UV completion.

\begin{figure}[htbp]\centering
\includegraphics[width=\textwidth]{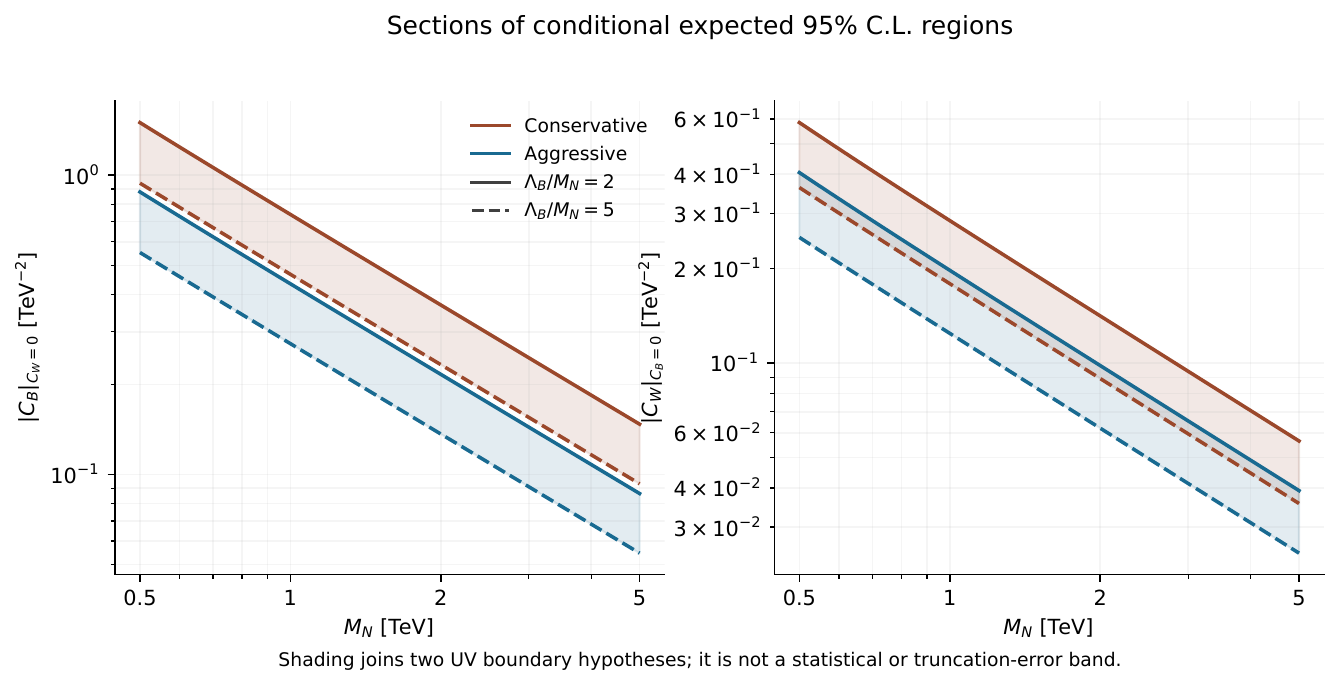}
\caption{Expected 95\% C.L. sensitivity bands on the Wilson coefficients
$C_B$ and $C_W$ as a function of the HNL mass, along the two coefficient axes.
Shading connects the two UV boundary hypotheses $r_B=2,5$. The lower
mass endpoint is a presentation choice; the accuracy of the local
expansion is discussed in section~\ref{sec:lowmass-validity}.}
\label{fig:final-FCC-mass}\end{figure}

Figure~\ref{fig:final-FCC-mass} shows how the axis reach evolves with the
HNL mass. At fixed dimensionful dipole coefficients the dominant
threshold grows as $M_N^2 C_XC_Y$, so the coefficient reach improves
approximately as $1/M_N$. This is the origin of the favourable mass
dependence of the indirect probe. A family of UV models in which the
mediator scale grows with $M_N$ and dimensionless couplings stay fixed
instead has a decoupling response, as discussed in
appendix~\ref{app:UV-interpretation-details}.

The mass scan starts at 500 GeV, where the hierarchy between the HNL
mass and electroweak scales is larger than near threshold. Its interpretation
at smaller masses requires the finite-mass qualifications in
section~\ref{sec:lowmass-validity}. The radiative Higgs-decay observable
of section~\ref{sec:higgs} adds a modest further improvement: at
$M_N=1$ TeV and $r_B=2$, the largest reductions of the coefficient-region
radius are about $7\%$ and $1\%$ in the conservative and aggressive
scenarios. The contours displayed here therefore present the pole reach,
with the separate Higgs contribution quantified in table~\ref{tab:decay-gain}.

\subsection{What changes when other ultraviolet interactions are allowed?}
The dipole coefficients predict correlated shifts across the pole
observables. However, other particles responsible for generating the
portal can also modify the lepton currents or gauge-boson propagation.
A precision measurement constrains the sum of these effects. To determine
how much of the dipole sensitivity survives, we allow selected additional
coefficients to vary and choose their values to give the best agreement
with the expected SM data. This is the profiling procedure used below.

Table~\ref{tab:profiled-current-summary} shows the progression. At
$M_N=1$ TeV, $r_B=2$ and with aggressive SM uncertainties, allowing the
two independent lepton currents to vary moves the same-sign section from
$0.1431$ to $0.2208\,\mathrm{TeV}^{-2}$. Some sensitivity remains because
these currents alone cannot reproduce every bosonic effect of the
dipoles. Allowing the mixed bosonic coefficient $C_{HWB}$ as well removes
the bound on the $C_W$ axis in this local analysis.

\begin{table}[htbp]\centering\small
\begin{tabularx}{\textwidth}{@{}l c X@{}}\toprule
Independent UV coefficients & Response rank & Consequence\\\midrule
Fixed & 3 & Same-sign section: $0.1431\,\mathrm{TeV}^{-2}$\\
Free $C_{Hl}^{(1,3)}$ & 2 & Same-sign section: $0.2208\,\mathrm{TeV}^{-2}$\\
Free currents and $C_{HWB}$ & 1 & The $C_W$ axis is unconstrained\\
All four directions free & 0 & No finite local-pole dipole bound\\
\bottomrule\end{tabularx}
\caption{Effect of profiling independent local contributions at $M_N=1$ TeV,
$r_B=2$, with aggressive SM uncertainties. The same-sign entries denote
$|C_B|=|C_W|$; the confidence threshold is recomputed for the surviving
response rank. The rank-zero statement allows arbitrary independent UV
coefficients in the linearized pole response; it does not establish a
perturbative completion realizing an arbitrarily large cancellation.}
\label{tab:profiled-current-summary}\end{table}

The disappearance of sensitivity has a direct coupling interpretation.
At the fixed pole, all local shifts are described by two universal
changes, proportional to weak isospin and electric charge, and two
direct changes of the left-handed neutrino and charged-lepton currents.
These are the four combinations derived in
appendix~\ref{app:local-pole-closure}. This span can already be
realized with $C_{Hl}^{(1)},C_{Hl}^{(3)},C_{HWB},C_{HD}$ at dimension six.
Independent local UV operators
span all four. When all are free, they can compensate the entire dipole
response at the pole and no finite bound on either dipole remains.
Additional measurements of the same pole couplings improve their
precision but do not identify which interaction produced them. This
four-dimensional space of possible coupling shifts is distinct from
the three-dimensional dipole signal space used in the statistical
construction. The dipole model explores only the patterns generated by
its three quadratic monomials; arbitrary UV operators span the larger
space and can reproduce those patterns.

This result defines the scope of the indirect reach. The numbers in
table~\ref{tab:final-FCC} apply when the dipole contribution dominates,
and the intermediate cases quantify how specific relaxations weaken that
assumption. A dipole interpretation can become more restrictive in a
specified UV model, where symmetries and coupling relations correlate
the extra contributions rather than leaving them arbitrary.

\subsection{Ultraviolet models and dipole dominance}
The possibility of a dominant dipole signal in a UV completion depends
on its particle content, couplings and symmetry structure. A loop-generated dipole typically
comes with other effective operators; some can arise already at tree
level. The benchmark reach applies when these additional contributions
are sufficiently suppressed or constrained by the model. The power counting
and examples used for this comparison are given in
appendix~\ref{app:UV-interpretation-details}.

The UV model of ref.~\cite{Butterworth:2019iff} contains two colour-singlet
vector-like fermions, $X_E$ (a weak doublet with hypercharge $1/2$) and
$X_N$ (a weak singlet with hypercharge $1$), and a singly charged scalar
$\phi$. All three are odd under a $Z_2$ symmetry, while the SM fields
and $N$ are even.
Its photon dipole is proportional to a product of three couplings. A coefficient $|C_B+C_W/2|=0.1\,\mathrm{TeV}^{-2}$ with
mediator masses of 2 TeV requires a product of about $7.6\times10^2$,
corresponding to a geometric mean of $9.1$. This estimate is already
outside a comfortably weakly coupled regime. It illustrates why moving
the mediators to a much higher scale does not readily isolate a large
loop-generated dipole: preserving the dipole requires increasing the
underlying couplings. The example is a scale comparison, not a full
constraint on that completion.

The assumption of negligible active--sterile mixing has a related
naturalness question. Dipole running can generate a Higgs--lepton--HNL
operator, while two Majorana dipoles can induce light-neutrino masses.
Near the projected reach, a viable realization therefore needs a
mechanism that protects the neutrino mass or compensates the induced
contribution. The size of the hierarchy is illustrated by the cutoff
estimate in appendix~\ref{app:UV-interpretation-details}: for
$|d_\gamma|=10^{-5}\,\mathrm{GeV}^{-1}$, $M_N=1$ TeV and a 2-TeV
cutoff, the induced mass is of order $10$ MeV, far above the observed
neutrino mass scale. Although this estimate is not a renormalized mass
prediction, it makes the need for protection apparent. Such protection
does not, on its own, constrain the size of
the independent electroweak currents. The quantitative estimates and
relevant UV constructions are discussed in the same appendix.

Thus our forecast supplies a target for UV models with controlled
additional electroweak effects. We do not identify the zero-boundary
benchmark with a generic weakly coupled completion. Its usefulness is
that it specifies the dipole contribution and the accuracy with which
other effects must be understood to interpret a precision constraint.

\subsection{How robust is the reach against extra contributions?}
\label{sec:UV-robustness}
Complete freedom in the UV coefficients is the most general case.
A model with specified interactions usually predicts a correlated extra
response, which can strengthen the signal as well as compensate it.
The relevant question is whether that response resembles the dipole
pattern across the measured observables, with a reinforcing or opposing
sign. A sign comparison of individual Wilson coefficients is insufficient
because the measurements weight combinations of couplings.

We compare the full observable responses using their experimental and
SM covariance. An extra contribution aligned with the dipole signal
strengthens the expected constraint. A sizeable reduction requires both
an opposing pattern and a comparable magnitude. For example, if the
extra response is at most $20\%$ or $50\%$ of the dipole response at
every coefficient value, the largest possible increases of the
coefficient limit are approximately $12\%$ and $41\%$, respectively,
at the same statistical threshold. These are assumptions about the
relative observable shifts; they are not probabilities for a cancellation.
The distinction between such relative budgets and fixed UV contributions
is derived in appendix~\ref{app:UV-interpretation-details}.

Published vector-like-lepton matching gives concrete examples of both
behaviours. At $M_N=1$ TeV, with the aggressive covariance and
$r_B=2,5$, nonzero contributions with the charged-singlet $E$ and triplet
$\Sigma_1$ patterns reinforce the signal on the tested axes and diagonals.
Allowing their strengths to vary freely down to zero therefore cannot
weaken the dipole-only constraint in these cases. By contrast,
the $\Sigma_0$ pattern can weaken the same-sign reach by a factor of
about $1.5$--$2.2$ when the original threshold is held fixed. Some axis
directions are nearly degenerate in these restricted examples, so the
residual sensitivity there depends strongly on effects outside the test.
The representations, current patterns and statistical comparison are
specified in appendix~\ref{app:UV-interpretation-details}.

The examples demonstrate that the effect of additional UV interactions
is controlled by their representation and coupling correlations. Symmetry and mediator quantum numbers can
select an opposing direction, while other representations reinforce the
dipole signal. Establishing robustness for a particular completion
requires its correlated matching. Keeping only the dipole logarithm does
not provide an independent lower floor: at a fixed pole momentum,
independent local coefficients can compensate the sum of logarithmic and
finite terms.

\subsection{Complementarity with direct HL-LHC searches}
\label{sec:direct-comparison-plan}
\begin{figure}[tbp]
 \centering
 \includegraphics[width=\textwidth]{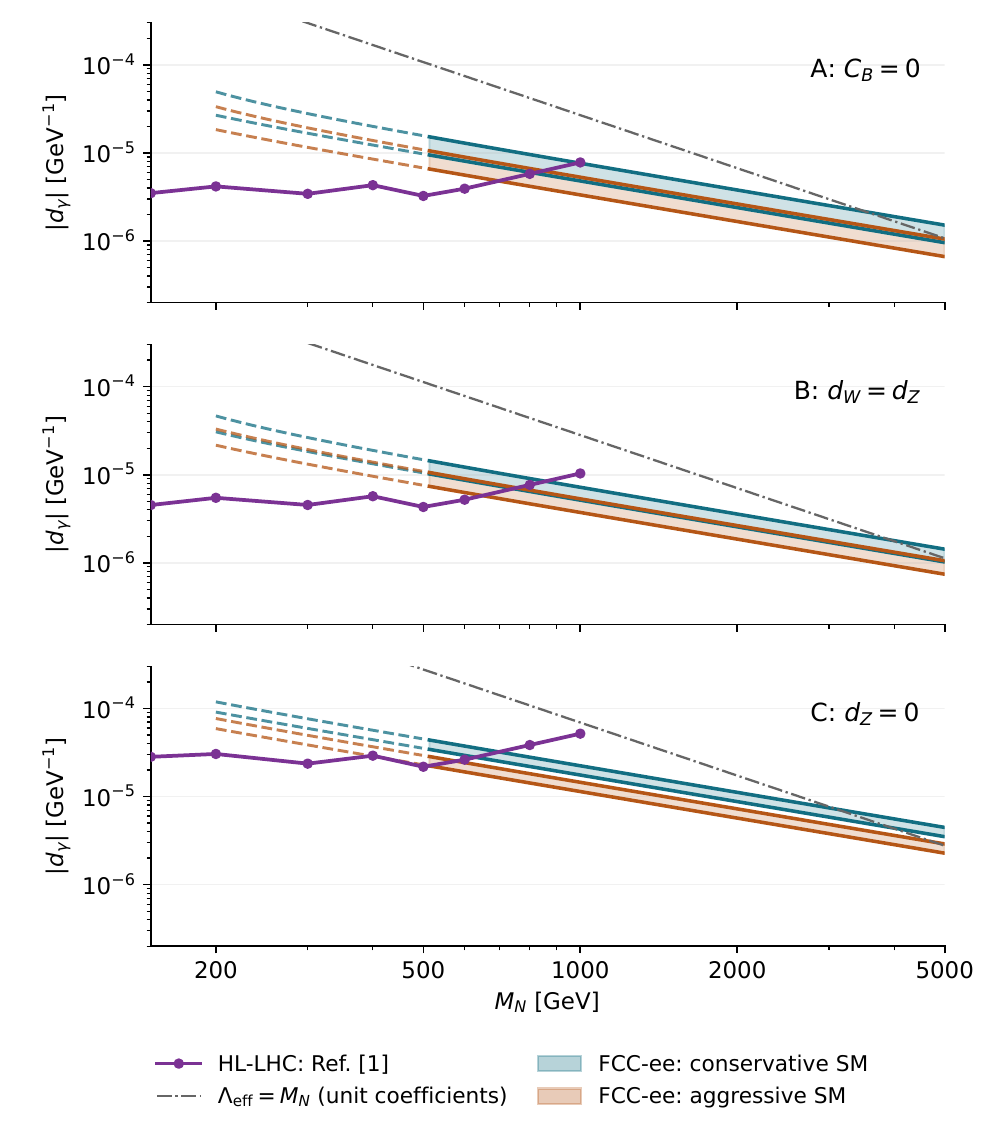}
 \caption{Direct and indirect reach in the three benchmark directions.
 Purple points are the HL-LHC projections shown in figure~6 of
 ref.~\cite{Vignaroli:2025pwn}. FCC-ee bands connect the conditional
 boundaries $\Lambda_B/M_N=2,5$ for each SM uncertainty scenario; they
 are not probability bands. Dashed coloured curves below 500 GeV flag
 local-expansion diagnostics, not a physical threshold. The grey
 dash-dotted curve marks $\Lambda_{\rm eff}=M_N$ for unit dimensionless
 dipole coefficients.
 }
 \label{fig:direct-indirect}
\end{figure}

Direct production and electroweak precision measurements test different
aspects of the portal. The former seeks an on-shell HNL and its decay
products; the latter measures the effect of virtual HNLs on ordinary
particle couplings. Figure~\ref{fig:direct-indirect} compares the FCC-ee
reach with the same-sign dilepton projections of
ref.~\cite{Vignaroli:2025pwn} for its 13-TeV,
$3\,\mathrm{ab}^{-1}$ HL-LHC benchmark. The three panels use that study's
coupling directions A--C, with a common operator normalization.
The conversion and flavour correspondence are detailed in
appendix~\ref{app:direct-comparison-details}.

Reference~\cite{Vignaroli:2025pwn} also presents direct-search
projections for FCC-hh and a 10-TeV muon collider. We focus here on
HL-LHC as the nearer-term comparison: its data are expected to precede
FCC-ee operation. FCC-hh and a high-energy muon collider represent
longer-term opportunities with less certain realization timescales,
especially for the latter. Their projected reach therefore addresses a
different experimental horizon from the comparison made here.

At $M_N=1$ TeV, the direct sensitivities in A, B and C are
$(7.8,10.3,51.7)\times10^{-6}\,\mathrm{GeV}^{-1}$ in $|d_\gamma|$.
For the conservative FCC-ee scenario and $r_B=2$, the corresponding
sections are $(7.6,7.2,22.3)\times10^{-6}\,\mathrm{GeV}^{-1}$.
Under the specified UV boundary, the conservative indirect reach is
therefore comparable in A and probes coefficients smaller by factors
of $1.4$ and $2.3$ in B and C, respectively. Changing to
$r_B=5$ gives $(4.8,5.1,17.5)\times10^{-6}\,\mathrm{GeV}^{-1}$.
\footnote{These values compare prospective sensitivities. The direct
study's two-standard-deviation criterion and our joint 95\% C.L.
construction are distinct; see appendix~\ref{app:direct-comparison-details}.}

The mass dependence gives a second source of complementarity. The
virtual threshold grows with the HNL mass at fixed dipole coefficients,
whereas direct production depends on kinematic accessibility and the
production and decay rates. The figure displays the published direct
mass interval alongside the indirect curves; its direct endpoints do
not define an upper mass reach. The dashed indirect curves at lower
masses mark the local-expansion qualification, while the grey guide
indicates the mediator-scale estimate for unit dimensionless dipoles.
These two features describe different limitations of the EFT and are
explained in appendix~\ref{app:direct-comparison-details}.

The complementarity also concerns discovery. A pole deviation can be
produced by several UV interactions, whereas on-shell HNL production
and measurements of its decay channels would help establish a dipole
origin. The comparison here is at matched diagonal flavour response,
not a combined likelihood or a flavour-violation analysis. The same-sign dilepton channel used here is particularly favourable
for Majorana HNLs because it suppresses the SM background
\cite{Vignaroli:2025pwn}. A lepton-number-conserving Dirac HNL does
not produce this signal, but can be sought through opposite-sign leptons
or radiative decays. For the latter, the HL-LHC sensitivity is estimated
at $|d_\gamma|\sim3\times10^{-5}\,\mathrm{GeV}^{-1}$ in the
$d_W=0$ benchmark, using $N\to\gamma\nu$ rather than same-sign
dileptons~\cite{Brdar:2025iua,Vignaroli:2025pwn}. This estimate
comes from extrapolating the LHC constraints of
ref.~\cite{Magill:2018jla}; it gives a quantitative reference for a
search that does not require lepton-number violation.
Although these estimates refer to different coupling directions, the
stronger same-sign reach in A and B illustrates the potential advantage
of a low-background signature. The radiative channel remains relevant
when that signature is absent.
By contrast, the lepton-number-conserving virtual response is unchanged
when the coupling bilinears are matched as in
section~\ref{sec:majorana-dirac}. The indirect probe thus retains its
sensitivity without relying on the low-background same-sign signature.

\subsection{Information beyond the pole observables}
\label{sec:offpole-identifiability}
\label{sec:WW-identifiability}
Measurements at different energies and with different external states
can distinguish effects that coincide at the $Z$ pole. Off-pole fermion
production probes the momentum dependence, while $W$-pair production
also probes charged currents and triple-gauge interactions. In the
present model, the latter has an additional tree-level HNL exchange
through the charged dipole, with characteristic angular dependence.
These channels provide well-motivated extensions of the analysis.

Their interpretation still depends on which independent UV operators
are allowed. At any fixed order in a local momentum expansion, free
coefficients with the same structures can reproduce that polynomial
dependence. Exact mass dependence, additional correlations from a UV
model, or new processes can provide further information. The necessary
amplitude content and the limits of the local argument are outlined in
appendix~\ref{app:beyond-pole-details}. We leave quantitative off-pole
and $W$-pair sensitivities to a dedicated study, keeping the present
results focused on the pole observables and the radiative Higgs decays.

\section{Conclusions}
\label{sec:conclusions}
Electroweak precision measurements can probe the Majorana HNL dipole
portal without requiring direct production of the heavy state. We have
determined how virtual HNLs modify gauge-boson propagation and lepton
couplings, and how these effects appear in the widths and asymmetries
measured at FCC-ee. For real, universal, flavour-aligned dipoles, the
analysis gives both a quantitative sensitivity target and the conditions
under which that target can be interpreted as a constraint on the portal.

The electroweak completion of the dipole interaction is essential to
this prediction. Gauge symmetry relates the neutral and charged
couplings and requires the two-gauge-boson contact vertices that enter
the lepton-current corrections. The same interaction therefore predicts correlated effects in several
precision observables and in radiative Higgs decays. Our calculation
includes both the leading mass-enhanced correction and the first
subleading terms projected onto the pole couplings, with a common
treatment of the gauge-boson and lepton contributions.

Under dipole dominance, FCC-ee can probe equal, same-sign coefficients
at approximately $0.22\,\mathrm{TeV}^{-2}$ with conservative SM
uncertainties and $0.14\,\mathrm{TeV}^{-2}$ with the more ambitious
theory projection, for $M_N=1$ TeV. These values assume that additional local contributions vanish at a
common scale twice the HNL mass ($\Lambda_B/M_N=2$). Choosing that boundary at five times
the HNL mass gives approximately $0.16$ and
$0.11\,\mathrm{TeV}^{-2}$, respectively. The two variations have different
meanings: improved SM predictions increase the experimental programme's
sensitivity to the portal, whereas moving the boundary scale changes the assumed contribution
of physics above the HNL mass. Keeping them distinct is necessary when comparing
these forecasts with a particular model. Relative to the four-observable
LEP reference evaluated with the same assumptions, the conservative
FCC-ee programme tests coefficients about four times smaller; the more
ambitious SM projection increases this gain to roughly five--six.
Improved SM predictions are essential to realizing this advance.

Larger HNL masses allow smaller dipole coefficients to be probed,
with an approximate $1/M_N$ reach driven by the leading correction
proportional to $M_N^2C_XC_Y$. This mass enhancement describes the
response at fixed dipole coefficients, rather than a variation with
all parameters of a UV completion held fixed. If the particles
that generate the dipoles become heavier together with the HNL, at fixed
mass ratios and dimensionless couplings, the dipole coefficients decrease
as the inverse mediator mass squared. Their suppression then outweighs
the factor $M_N^2$, and the precision effect decreases. Near the
electroweak scale, the full mass dependence of both gauge-boson and
lepton-current corrections becomes relevant. In the benchmark directions
tested here, it changes the coefficient reach by at most $0.15\%$ at
500 GeV and $4.3\%$ at 200 GeV.

The main qualification concerns what pole measurements can identify.
At the one-loop order retained here, the local dipole corrections
enter four independent combinations of the couplings measured at the
$Z$ pole. Other heavy
particles can induce exactly the same four changes through local
interactions, already at dimension six. Pole data therefore constrain
the sum of these effects and cannot, on their own, establish a dipole
origin. If the additional contributions are unrestricted, they can
completely cancel the dipole signal. In specific UV models, however,
their relative sizes and signs can be correlated by the mediator
representations and interactions. The vector-like-lepton examples
considered here illustrate the consequence: a charged-singlet contribution
reinforces the signal along the tested directions, whereas a
weak-triplet contribution can partly compensate it. A large
reduction requires an opposing pattern of comparable size across the
measured quantities, rather than simply an opposite sign in one Wilson
coefficient. The forecast is thus a target for models in which these
additional effects are controlled; its realization must also respect
perturbative power counting and the smallness of neutrino masses.

Radiative Higgs decays add a modest complementary constraint. At
$M_N=1$ TeV and $\Lambda_B/M_N=2$, the ratio of the $Z\gamma$ and
$\gamma\gamma$ widths improves the coefficient-region radius by at most
about $7\%$ and $1\%$ in the conservative and aggressive scenarios.
The pole measurements therefore provide most of the projected
sensitivity under the stated assumptions. Measurements away from the
pole and $W$-pair production could supply different information, but
their reach requires a dedicated treatment of the additional amplitudes
and UV coefficients.

Direct searches provide a particularly useful counterpart. At an HNL
mass of 1 TeV, the conservative FCC-ee sensitivities are comparable to
or stronger than the HL-LHC projections of
ref.~\cite{Vignaroli:2025pwn} in the three benchmark directions,
under the stated dipole-dominance assumption. Their different mass
dependence makes the comparison useful beyond this single benchmark.
More fundamentally, observing an HNL and measuring its decay pattern
would help distinguish a dipole interaction from other sources of a
precision deviation.
The HL-LHC comparison uses the favourable Majorana same-sign dilepton
channel, whose small SM background enhances the direct reach. A
lepton-number-conserving Dirac HNL must instead be sought in other final
states, including radiative decays. The available radiative projection
is weaker than the same-sign reach in A and B, but uses a different
coupling direction; it does not establish a universal loss of direct
sensitivity for Dirac HNLs. The indirect response remains unchanged
for matched coupling bilinears, preserving the role of precision
measurements when the same-sign signature is absent.

Combining direct production with precision measurements would test both
the heavy state and the correlated virtual effects of its interactions.
Our results identify the correlated electroweak signature of the
portal, quantify the precision needed to test it, and specify which
additional interactions can obscure its interpretation. Together with
direct searches, these predictions provide a framework for testing
whether a future HNL signal is accompanied by the virtual effects
expected from a gauge-invariant dipole interaction.

\acknowledgments
The author acknowledges the use of GPT-6 and Claude
for assistance with literature searches, the development of computational
cross-check scripts, and improvements to the clarity of the manuscript.
The tools are provided by \href{https://openai.com/}{OpenAI} and
\href{https://www.anthropic.com/}{Anthropic}, respectively. The author
takes full responsibility for the calculations, interpretation and final text.

\appendix

\section{Dipole normalization and bosonic loop functions}
\label{app:bosonic-details}
This appendix supports the model conventions of section~\ref{sec:portal}
and the vacuum-polarization matching of sections~\ref{sec:selfenergies}
and~\ref{sec:matching}. We specify the contraction and action
normalizations, the subtraction convention and the independent scalar
check of the finite loop function.

\subsection{Majorana contractions}
\label{app:majorana}

The neutral Majorana loop contains two equal orientations.  For a single
orientation,
\begin{equation}
 F_{\rm single}(q^2)=\frac12F_0(q^2),
\end{equation}
An explicit Dirac-trace calculation shows that the two opposite chiral
sequences give equal contributions. These are the two contractions needed for the
Hermitian transition interaction; the corresponding particle and
antiparticle contributions also occur for a Dirac HNL, with both
assignments of the transition vertices to the external gauge legs. Keeping just one of these contractions would therefore underestimate
the complete lepton-number-conserving amplitude. With the physical
coupling bilinears matched as in eq.~\eqref{eq:Dirac-Majorana-bilinears},
the Dirac and Majorana results for the processes considered here agree;
there is no additional factor of two in converting between them.

\subsection{Transverse projection}
\label{app:loop}

We use the transverse projector $P^{\mu\nu}=q^\mu q^\nu-q^2g^{\mu\nu}$,
whose contraction is $(D-1)q^4$. This standard tensor reduction isolates
the scalar function calculated in section~\ref{sec:selfenergies}.
The scalar transverse form factor can therefore be extracted as
\begin{equation}
 F(q^2)=
 \frac{P_{\mu\nu}\Pi^{\mu\nu}(q)}
 {(D-1)(q^2)^2}.
 \label{eq:projector}
\end{equation}
Keeping the full $D$ dependence through the Passarino--Veltman reduction gives
eq.~\eqref{eq:F0-exact}.  Setting $D=4$ before expanding the scalar integrals
would miss the finite contribution in eq.~\eqref{eq:F0zero}.

\subsection{Gauge rotation and the charged channel}
\label{app:gaugechecks}
The neutral amplitudes share the scalar kernel in
eq.~\eqref{eq:neutral-universality-AZ}. Applying the rotation in
eq.~\eqref{eq:AZrotation} gives eq.~\eqref{eq:gauge-basis-universality}.
The charged calculation independently gives $F_{WW}=d_W^2F_0/2=F_{33}$
in the massless-lepton limit, testing the weak-generator normalization
and the charged-dipole vertex together. These relations refer to the
two-dipole contributions specified in section~\ref{sec:selfenergies}.

\subsection{Zero-momentum threshold}
\label{app:threshold}

At $q^2=0$ the neutral threshold matrix is
\begin{equation}
 {\bm F}(0)
 =
 -\frac{\MN^2}{2\pi^2}
 \begin{pmatrix}
 \dB^2&\dB\dthree\\
 \dB\dthree&\dthree^2
 \end{pmatrix}.
 \label{eq:threshold-matrix}
\end{equation}
Its determinant vanishes.  The eigenvalues are
\begin{equation}
 \lambda_1=0,\qquad
 \lambda_2=-\frac{\MN^2}{2\pi^2}
 \left(\dB^2+\dthree^2\right),
\end{equation}
so the gauge-field combination orthogonal to $(\dB,\dthree)$ receives no
threshold from this loop contribution.

\subsection{Self-energy and Warsaw normalization}
\label{app:signaudit}
Let $P^{\mu\nu}=q^\mu q^\nu-q^2g^{\mu\nu}$, and let $K^{\mu\nu}$ be the
quadratic action kernel from $\Delta{\cal L}=+\sum_iC_iQ_i$.
Fourier transformation gives
$F_{\mu\nu}(q)F^{\mu\nu}(-q)=-2V_\mu(q)P^{\mu\nu}V_\nu(-q)$.
A diagonal $cFF$ interaction has kernel $-4cP$, whereas a mixed $cFG$
interaction has mixed kernel $-2cP$. In standard Warsaw normalization,
$H^\dagger\sigma^3H=-v^2/2$, hence
\begin{equation}
 K_{BB}=-2v^2C_{HB}P,\qquad K_{33}=-2v^2C_{HW}P,
 \qquad K_{3B}=+v^2C_{HWB}P.
\end{equation}
The action-normalized loop insertion has $\Pi=K$ in this dictionary.
It gives eq.~\eqref{eq:dim6-F} and the threshold signs in
 eqs.~\eqref{eq:match-HB}--\eqref{eq:match-HWB}. The positive coefficient
of $\mathcal X_{\mu\nu}^\dagger\mathcal X^{\mu\nu}$ in
 eq.~\eqref{eq:hh-alpha-update} provides a consistent Higgs-background check.
For the higher-derivative operators in eq.~\eqref{eq:dim8-green}, the same
choice gives local contributions
\begin{equation}
 F_{BB}=-v^2q^2C_{2B,H},\quad F_{33}=-v^2q^2C_{2W,H},\quad
 F_{3B}=+\frac{v^2q^2}{2}C_{2WB,H}.
\end{equation}
The counterterms in eq.~\eqref{eq:dim8-cts} cancel the positive loop poles
of eq.~\eqref{eq:F0prime} in this dictionary.

\subsection{Scalar-integral subtraction convention}
\label{app:renormalization}

The unrenormalized scalar integrals in eq.~\eqref{eq:F0prime} display
the combination $1/\epsilon-\gamma_E-\ln\pi$. At the same numerical
scale this differs from the conventional
$\Delta_{\MSbar}=1/\epsilon-\gamma_E+\ln(4\pi)$ by
$-\ln(4\pi^2)$. This is an integration-measure convention, distinct from
the finite contributions generated by $D$-dependent numerator factors.
Explicitly, with $\mu_{\rm PX}=2\pi\mu$,
\begin{equation}
 \frac1\epsilon-\gamma_E-\ln\pi+\ln\frac{\mu_{\rm PX}^2}{M_N^2}
 =\Delta_{\MSbar}+\ln\frac{\mu^2}{M_N^2}.
\end{equation}
Converting the scalar integrals and local subtraction terms together gives
the $\MSbar$-normalized finite slope
$[\ln(\mu^2/M_N^2)+11/6]/(6\pi^2)$ used in the phenomenological response.
The direct scalar quadrature in appendix~\ref{app:numerical-bubble}
uses this normalization from the outset. Equation~\eqref{eq:common-finite-boundary} then evolves the independent
coefficients from the chosen UV boundary. A consistent change of the
integration measure transforms the subtraction convention and those
coefficients together, preserving the physical prediction.

\subsection{Numerical scalar-integral check of the finite bosonic kernel}
\label{app:numerical-bubble}
We also evaluate the scalar integrals directly by Feynman-parameter
quadrature, independently of the Package-X closed forms. In units
$M_N^2=\mu^2=1$ and an $\MSbar$-normalized measure, define
\begin{equation}
 A(\epsilon)=\frac{e^{\gamma_E\epsilon}\Gamma(\epsilon)}{1-\epsilon},
 \quad
 B(z,\epsilon)=e^{\gamma_E\epsilon}\Gamma(\epsilon)
 \int_0^1 dx\,[x(1-(1-x)z)]^{-\epsilon}.
\end{equation}
Inserting these in eq.~\eqref{eq:F0-exact} with $D=4-2\epsilon$ and
subtracting the pole gives the dimensionless finite function
\begin{equation}
 \lim_{\epsilon\to0}\left\{
 \frac{3[(D-2-z)A-(1-z)(D-2+z)B]}{(D-1)z}
 -\frac{z}{\epsilon}\right\}
 =-3+\frac{11}{6}z+r(z).
\end{equation}
At $z=-1,-0.1,10^{-4},0.01,0.1,0.5,0.9$, quadrature at 65-digit
precision and linear extrapolation in $\epsilon$ agree with this expression
to better than $10^{-19}$ in absolute units. Independent small-$z$
extrapolation gives $-3$ and $11/6$ for the constant and slope. This tests
the finite scalar evaluation, including evanescent-times-pole terms,
and the nonlocal remainder; it shares the tensor reduction in
 eq.~\eqref{eq:F0-exact}. The fermion bubble contains no internal gauge
propagator, so a second gauge choice does not provide a distinct test of
this kernel. A separate generic-$R_\xi$ check of the direct current diagrams and
external residues is described below. The scalar quadrature and the
\textsc{Matchete} tree-level matching~\cite{Fuentes-Martin:2022jrf} have different scopes and are not independent
recalculations of the finite lepton-current coefficients.

\section{Lepton currents and the local pole response}
\label{app:consistency-checks}
The direct current contribution has two ingredients: the vertex
correction and the normalization of the external lepton states. The
latter is determined by the residue of the lepton propagator at its
pole. This appendix gives both ingredients of the matching in
section~\ref{sec:matching}, together with the identities that connect
them. Their combination with gauge propagation and measured inputs
produces the pole response used in section~\ref{sec:ewinputs}.
The Ward and gauge-parameter identities test this assembly. The
complementary checks, including those discussed in the other appendices,
are summarized in table~\ref{tab:consistency-summary}.

\begin{table}[tbp]\centering\small
\begin{tabularx}{\textwidth}{@{}lXX@{}}\toprule
Check & Method & Result\\\midrule
Electromagnetic Ward identity & Photon coupling replacement, UV algebra and finite scalar evaluation & Pole cancels exactly; finite zero-momentum limit agrees numerically\\
Subtraction-scale dependence & Differentiate vertices and residues with respect to the logarithm of the scale & Derivatives equal the UV pole coefficients\\
Heavy-mass expansion & Compare exact assembled amplitudes with the local response & Leading scaling and subleading coefficients agree\\
Bosonic finite kernel & Independent Feynman-parameter quadrature & Constant and slope reproduced\\
FCC-ee sections & Reconstruct information matrix from response and covariance & All FCC-ee entries of table~\ref{tab:final-FCC} reproduced\\\bottomrule
\end{tabularx}
\caption{Complementary consistency checks. Their assumptions and numerical
precision are specified in the corresponding subsections. Quadrature,
expansion tests and Ward identities probe different properties; numerical
agreement does not exclude an error common to the compared amplitudes.}
\label{tab:consistency-summary}\end{table}

\subsection{Lepton self-energies and neutral-current UV residues}
\label{sec:zpole}

\subsubsection{Conventions and external-leg factors}

We give the UV residues for one lepton doublet $(\nu_e,e)_L$, with
massless external leptons and real dipole coefficients. The calculation keeps
$\MN$, $m_W$, $m_Z$ and the external $Z$ momentum squared $q^2$ distinct.
The neutral-current residues combine the 1PI vertices, external-lepton
wave-function renormalization (WFR) and SM input-parameter counterterms. Local
EFT counterterms cancel the remaining poles. The finite observable
response also includes gauge-field normalization, neutral mixing and
input shifts, as assembled in appendix~\ref{app:local-pole-closure}.

The tree-level calibration of the FeynArts/FeynCalc (FC) projector gives
\begin{equation}
 \Gamma_{fL,\rm tree}^{\rm FC}=-g_{fL},\qquad
 g_{eL}=-\frac{e(c_W^2-s_W^2)}{2s_Wc_W},\qquad
 g_{\nu L}=\frac{e}{2s_Wc_W}.
 \label{eq:FC-sign}
\end{equation}
Accordingly the physical-coupling loop residue is
$\Delta g_{fL}^{\rm pole}=-\Gamma_{fL}^{\rm FC,UV}$.
The minus sign here converts conventions; it is \emph{not} UV subtraction.
A cancelling EFT counterterm has the further opposite sign,
$\delta g_{fL}^{\rm EFT,ct}=-\Delta g_{fL}^{\rm pole}$, in the same physical
coupling convention. We reserve $\Delta g$ for a loop-induced shift and
$\delta C$ for a Wilson counterterm.

Writing $\Sigma_f(p)=\slashed pP_L\Sigma_{fL}(p^2)$, the calibrated
self-energy coefficient extracted with FeynCalc is defined by
$C_f^{\rm FC}=-\Sigma_{fL}$. The poles are
\begin{align}
 \Sigma_{eL}^{\rm UV}(p^2)
 &=\frac{d_W^2}{8\pi^2\epsilon_{\rm UV}}
                  (6\MN^2-2p^2+3m_W^2),\label{eq:Sigma-e}\\
 \Sigma_{\nu L}^{\rm UV}(p^2)
 &=\frac{1}{8\pi^2\epsilon_{\rm UV}}
 \left[2d_\gamma^2(3\MN^2-p^2)
       +d_Z^2(6\MN^2-2p^2+3m_Z^2)\right].\label{eq:Sigma-nu}
\end{align}
The right-handed self-energy vanishes in this limit. The external-leg factors
used in the amplitude assembly are
\begin{align}
 \delta Z_{eL}&=\Sigma_{eL}^{\rm UV}(0)
       =\frac{3d_W^2(2\MN^2+m_W^2)}{8\pi^2\epsilon_{\rm UV}},\nonumber\\
 \delta Z_{\nu L}&=\Sigma_{\nu L}^{\rm UV}(0)
       =\frac{3[2\MN^2d_\gamma^2+(2\MN^2+m_Z^2)d_Z^2]}
                    {8\pi^2\epsilon_{\rm UV}}.
 \label{eq:lepton-WFR}
\end{align}
There are two external legs, each supplying $\delta Z_L/2$, so the net
coupling contribution is $g_{fL}\delta Z_{fL}$ and the FC coefficient is its
negative. These external-leg factors should not be identified with the
negative of the Wilson poles in eq.~\eqref{eq:Hl-counterterms}.

\subsubsection{Charged-lepton vertex and input corrections}

The five HNL 1PI diagrams consist of two $N\nu W$ triangles, one $WNW$
triangle containing the SM $ZWW$ vertex, and two $NW$ contact diagrams.
The contact vertices arise from the non-Abelian field strength in
${\cal O}_W$. Keeping only the triangle with $ZWW$ would not implement the
operator's gauge completion. The complete left-handed 1PI pole is
\begin{equation}
 \Gamma_{eL}^{\rm 1PI,FC,UV}
 =\frac{eq^2d_Wd_Z}{2\sqrt2\pi^2s_W\epsilon_{\rm UV}}
 -\frac{ec_Wd_W^2}{48\pi^2s_W\epsilon_{\rm UV}}
       (36\MN^2-5q^2+18m_W^2).
 \label{eq:Zee-1PI}
\end{equation}
The right-handed 1PI form factor is zero for $m_e=0$. Equation~\eqref{eq:Zee-1PI}
uses a projector with $\gamma^\mu P_L\mapsto1$ and
$\gamma^\mu P_R\mapsto0$. No additional Majorana multiplicity is to be applied
after the listed diagrams have been summed.

The WFR and parameter pieces in the same convention are
\begin{align}
 \Gamma_{eL}^{\rm WFR,FC,UV}
 &=+\frac{3e(c_W^2-s_W^2)d_W^2}{16\pi^2c_Ws_W\epsilon_{\rm UV}}
       (2\MN^2+m_W^2),\label{eq:Zee-WFR}\\
 \Gamma_{eL}^{\rm param,FC,UV}
 &=-\frac{e}{48\pi^2s_W^3c_W\epsilon_{\rm UV}}
       (2d_Z^2m_Z^2-d_W^2m_W^2).
 \label{eq:Zee-param}
\end{align}
Here the on-shell input relations used in the calculation are
\begin{equation}
 m_W^2=c_W^2m_Z^2,\qquad
 \left.\frac{\delta e}{e}\right|_{\rm UV}=0,\qquad
 \delta s_W^2\big|_{\rm UV}
 =\frac{c_W^2(2d_Z^2m_Z^2-d_W^2m_W^2)}{12\pi^2\epsilon_{\rm UV}}.
 \label{eq:input-poles}
\end{equation}
Differentiating $g_{eL}$ with respect to $s_W^2$ gives
$e/(4s_W^3c_W^3)$ and reproduces eq.~\eqref{eq:Zee-param}, including the FC
sign. The vanishing charge pole follows from the absence of a divergent
constant in $F_{AA}$ and from $\Pi_{AZ}(0)=0$ in this sector.

Substituting
\begin{equation}
 d_\gamma=\frac{ev}{\sqrt2}(C_B+C_W/2),\quad
 d_Z=\frac{ev}{\sqrt2}\left(-\frac{s_W}{c_W}C_B+
                              \frac{c_W}{2s_W}C_W\right),\quad
 d_W=\frac{ev}{2s_W}C_W,
 \label{eq:dipole-substitution}
\end{equation}
we obtain the final residue
\begin{align}
 \Gamma_{eL}^{\rm FC,UV}
 ={}&-\frac{3e^3v^2\MN^2}{32\pi^2c_Ws_W^3\epsilon_{\rm UV}}C_W^2
 \nonumber\\
 &+\frac{e^3v^2q^2}{192\pi^2c_Ws_W^3\epsilon_{\rm UV}}
       (17c_W^2C_W^2-24s_W^2C_BC_W)\nonumber\\
 &-\frac{e^3v^2m_W^2}{192\pi^2c_W^5s_W^3\epsilon_{\rm UV}}
       (9c_W^4C_W^2-4c_W^2C_BC_W+4s_W^2C_B^2).
 \label{eq:Zee-final}
\end{align}
The leading mass term agrees with
$-\Delta g_{eL}^{(6),\rm pole}$ in eq.~\eqref{eq:current-broken-check}.

\subsubsection{Neutrino vertex and independent cross-check}

The two HNL 1PI graphs are $N\nu Z$ triangles with conjugate dipole
orientations. The independently normalized projector confirms
\begin{equation}
 \Gamma_{\nu L}^{\rm 1PI,FC,UV}
 =\frac{e d_Z^2q^2}{4\pi^2s_Wc_W\epsilon_{\rm UV}},\qquad
 \Gamma_{\nu R}^{\rm 1PI,FC,UV}=0.
 \label{eq:Znu-1PI}
\end{equation}
There is no $\MN^2$ or $m_Z^2$ in this 1PI pole. This is an explicit result
of the loop reduction, not a consequence of gauge symmetry alone, and it
must not be inferred by rescaling the charged-lepton projector correction.
The remaining terms are
\begin{align}
 \Gamma_{\nu L}^{\rm WFR,FC,UV}
 &=-\frac{3e}{16\pi^2s_Wc_W\epsilon_{\rm UV}}
   [2\MN^2d_\gamma^2+(2\MN^2+m_Z^2)d_Z^2],\label{eq:Znu-WFR}\\
 \Gamma_{\nu L}^{\rm param,FC,UV}
 &=+\frac{e(c_W^2-s_W^2)}{48\pi^2s_W^3c_W\epsilon_{\rm UV}}
      (2d_Z^2m_Z^2-d_W^2m_W^2).
 \label{eq:Znu-param}
\end{align}
The parameter term follows from
$\partial g_{\nu L}/\partial s_W^2=-e(c_W^2-s_W^2)/(4s_W^3c_W^3)$.
After the substitutions in eq.~\eqref{eq:dipole-substitution}, the sum is
\begin{align}
 \Gamma_{\nu L}^{\rm FC,UV}
 ={}&-\frac{3e^3v^2\MN^2}{64\pi^2c_W^3s_W^3\epsilon_{\rm UV}}
                   (4s_W^2C_B^2+c_W^2C_W^2)\nonumber\\
 &+\frac{e^3v^2q^2}{32\pi^2c_W^3s_W^3\epsilon_{\rm UV}}
                   (c_W^2C_W-2s_W^2C_B)^2\nonumber\\
 &+\frac{e^3v^2m_Z^2}{384\pi^2c_W^3s_W^3\epsilon_{\rm UV}}
 \Big[-9c_W^4C_W^2+4c_W^2(11s_W^2-2c_W^2)C_BC_W\nonumber\\
 &\hspace{47mm}+4s_W^2(2c_W^2-11s_W^2)C_B^2\Big].
 \label{eq:Znu-final}
\end{align}
All of its $\MN^2$ term comes from WFR. Orthogonality of the neutral gauge
rotation gives
\begin{equation}
 d_\gamma^2+d_Z^2=d_B^2+d_3^2
 =\frac{g'^2v^2}{2}C_B^2+\frac{g^2v^2}{8}C_W^2.
 \label{eq:no-mixed-neutrino}
\end{equation}
This explains the cancellation of $C_BC_W$ in that term and connects the
broken-phase calculation to the unbroken current matching. Together the
$\MN^2$ parts of eqs.~\eqref{eq:Zee-final} and \eqref{eq:Znu-final} reproduce
both coefficients in eqs.~\eqref{eq:CHl1-pole}--\eqref{eq:CHl3-pole}.

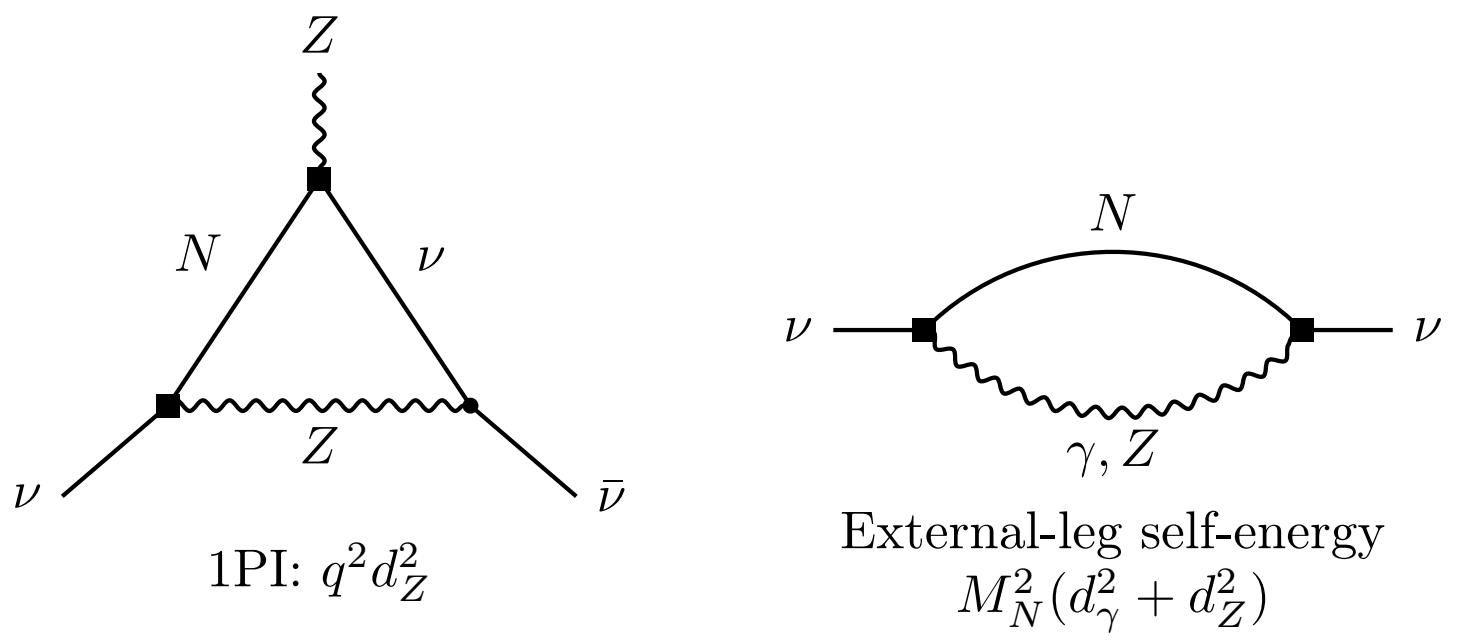
\begin{figure}[tbp]
\centering
\begin{tikzpicture}[every node/.style={font=\small}]
 \coordinate (a) at (0,0);\coordinate (b) at (2,0);\coordinate (t) at (1,1.5);
 \draw[ferm] (-.7,-.6)--(a);\draw[ferm] (b)--(2.7,-.6);
 \draw[ferm] (a)--node[above left] {$N$}(t);
 \draw[ferm] (t)--node[above right] {$\nu$}(b);
 \draw[boson] (a)--node[below] {$Z$}(b);
 \draw[boson] (1,2.2)--(t);\node[above] at (1,2.2) {$Z$};
 \node[dip] at (a) {};\node[dip] at (t) {};\fill (b) circle(1.5pt);
 \node[left] at (-.7,-.6) {$\nu$};\node[right] at (2.7,-.6) {$\bar\nu$};
 \node at (1,-1.1) {1PI: $q^2d_Z^2$};
\begin{scope}[xshift=5cm,yshift=.5cm]
 \coordinate (a) at (0,0);\coordinate (b) at (2.5,0);
 \draw[ferm] (-.6,0)--(a);\draw[ferm] (b)--(3.1,0);
 \draw[ferm] (a) to[bend left=45] node[above] {$N$}(b);
 \draw[boson] (a) to[bend right=45] node[below] {$\gamma,Z$}(b);
 \node[dip] at (a) {};\node[dip] at (b) {};
 \node[left] at (-.6,0) {$\nu$};\node[right] at (3.1,0) {$\nu$};
 \node[align=center] at (1.25,-1.6) {External-leg self-energy\\$\MN^2(d_\gamma^2+d_Z^2)$};
\end{scope}
\end{tikzpicture}
\caption{Neutrino 1PI triangle (left; the conjugate orientation is understood)
and the self-energy supplying WFR (right). Their UV mass dependence is
qualitatively different: only the external-leg contribution supplies the
$\MN^2$ term in the neutral-current residue.}
\label{fig:znu-topologies}
\end{figure}

\subsubsection{Dimension-eight interpretation of the residues}

The second and third lines of the charged-lepton residue scale as
$v^2q^2C^2$ and $v^2m_W^2C^2$; the corresponding neutrino terms scale as
$v^2q^2C^2$ and $v^2m_Z^2C^2$. Since $m_V^2\sim g^2v^2$, these are
higher-derivative or additional-Higgs dimension-eight structures at double
insertion order. They must not be absorbed into the $\MN^2C^2$ coefficients
of $Q_{Hl}^{(1,3)}$. General dimension-eight renormalization at order
$\Lambda^{-4}$ requires such double-insertion terms~\cite{Chala:2021pll}.
This framework organizes the double-insertion renormalization; the
expressions above give the HNL-specific coefficients.

The two neutral vertices determine the combinations of residues relevant
to the pole response. Resolving a complete nonredundant dimension-eight
fermionic basis would require additional Green functions and EOM
reduction. The independent finite current coefficients require their own
boundary conditions; the common prescription used here is specified in
appendix~\ref{app:local-pole-closure}.

For the observable calculation the renormalized couplings enter,
in the massless limit, through
\begin{equation}
 \frac{\Delta\Gamma_f}{\Gamma_f^{\rm SM}}
 =\frac{2\,\mathrm{Re}[g_{fL}^*\Delta g_{fL}^{\rm ren}
                         +g_{fR}^*\Delta g_{fR}^{\rm ren}]}
        {|g_{fL}|^2+|g_{fR}|^2},
 \label{eq:width-template}
\end{equation}
with the required input and normalization effects included in
$\Delta g^{\rm ren}$. A zero right-handed \emph{1PI} dipole vertex does not
imply vanishing universal input shifts for right-handed charged leptons.
The finite response entering the forecast is given in
appendix~\ref{app:local-pole-closure}.

\subsection{Ultraviolet poles and counterterms}
\label{sec:master-record}

For compactness define $K=3\MN^2/(64\pi^2\epsilon_{\rm UV})$,
$B=g'^2C_B^2$ and $G=g^2C_W^2$. Table~\ref{tab:master-poles} distinguishes
the loop pole from its cancelling local term in the \emph{same} convention.
The quantities $\Gamma_f^{\rm FC,UV}$ refer precisely to the partial assembly
specified in appendix~\ref{sec:zpole}. SM parameter counterterms already
included there must not be added a second time. For self-energies and vertex
rows the table specifies cancellation at the Green-function level; only rows
with a named $C_i$ are Wilson-coefficient counterterms.

\begin{table}[tbp]
\centering
\small
\renewcommand{\arraystretch}{1.45}
\begin{tabularx}{\textwidth}{@{}>{\raggedright\arraybackslash}p{.19\textwidth} >{\raggedright\arraybackslash}X >{\raggedright\arraybackslash}p{.19\textwidth} >{\raggedright\arraybackslash}p{.18\textwidth}@{}}
\toprule
Quantity & Loop pole & Cancelling local term & EFT interpretation\\
\midrule
$F_{ij}(q^2)$ & $d_i d_j q^2/(6\pi^2\epsilon_{\rm UV})$
 & Negative of pole & Bosonic dim-8; $\Pi=q^2F$\\
$F_{WW}(q^2)$ & $d_W^2q^2/(12\pi^2\epsilon_{\rm UV})$
 & Negative of pole & Bosonic dim-8\\
\shortstack[l]{$C_{HB},C_{HW}$\\$C_{HWB}$} & $0$ in the displayed two-dipole sector
 & $0$ at this order & Finite dim-6 thresholds, eqs.~\eqref{eq:match-HB}--\eqref{eq:match-HWB}\\
$\Sigma_{eL},\Sigma_{\nu L}$ & Eqs.~\eqref{eq:Sigma-e}, \eqref{eq:Sigma-nu}
 & Negative of pole in the same $\Sigma$ convention
 & $\MN^2$: dim-6; $p^2,m_V^2$: dim-8\\
$C_{Hl}^{(1)}$ & $-K(4B+3G)$ & $+K(4B+3G)$ & Dim-6 current\\
$C_{Hl}^{(3)}$ & $+K(4B-G)$ & $-K(4B-G)$ & Dim-6 current\\
$\Gamma_{eL}^{\rm FC}$ & eq.~\eqref{eq:Zee-final}
 & $-\Gamma_{eL}^{\rm FC,UV}$ & Dim-6 and dim-8 projections\\
$\Gamma_{\nu L}^{\rm FC}$ & eq.~\eqref{eq:Znu-final}
 & $-\Gamma_{\nu L}^{\rm FC,UV}$ & Dim-6 and dim-8 projections\\
\bottomrule
\end{tabularx}
\caption{Ultraviolet poles and cancelling local terms for one flavour and real couplings.
For the neutral gauge-basis entry $i,j\in\{B,3\}$. A finite threshold is not
an absent effect, and a cancelling local term is not an independently fixed
renormalized Wilson coefficient.}
\label{tab:master-poles}
\end{table}

The bosonic dimension-eight Wilson counterterms, in precisely the Green
basis of eq.~\eqref{eq:dim8-green}, remain those of
 eq.~\eqref{eq:dim8-cts}. At coefficient level their loop-pole counterparts are
\begin{align}
 C_{2B,H}^{\rm pole}&=-\frac{g'^2C_B^2}{12\pi^2\epsilon_{\rm UV}},&
 C_{2W,H}^{\rm pole}&=-\frac{g^2C_W^2}{48\pi^2\epsilon_{\rm UV}},\nonumber\\
 C_{2WB,H}^{\rm pole}&=+\frac{gg'C_BC_W}{12\pi^2\epsilon_{\rm UV}}.
\end{align}
Electromagnetic gauge invariance fixes the transverse tensor structure
but still allows a UV pole multiplying $M_N^2$ in $F_{ij}$, or
$M_N^2q^2$ in the self-energy. Its absence here follows from cancellation
between the terms of the loop function. Thus the dimension-six bosonic
threshold is finite, whereas its momentum-dependent slope requires the
dimension-eight counterterms. These are distinct consequences of the
explicit amplitude and of gauge symmetry.

\subsection{\texorpdfstring{Finite current checks: $Zee$ and $Z\nu\nu$}{Finite current checks: Zee and neutrinos}}
\label{app:currentaudit}
The chiral projector is normalized by
${\cal P}_L[a\gamma^\mu P_L]=a$ and
${\cal P}_L[a\gamma^\mu P_R]=0$; the broken-phase amplitudes are
compared with the unbroken current matching in the same coupling-sign
convention.
For the pure $C_W^2$ sector define
\begin{equation}
 {\cal F}_e=\frac{e d_W^2\MN^2}{16\pi^2s_Wc_W}K_{\MSbar}.
\end{equation}
The local dimension-six terms are
$\Delta g_{eL}=-{\cal F}_e$ and
$\Delta g_{\nu L}=-{\cal F}_e/2$, as follows directly by substituting
 eqs.~\eqref{eq:finite-Hl1}--\eqref{eq:finite-Hl3} into
 eqs.~\eqref{eq:current-map-neutrino}--\eqref{eq:current-map}. The electron result combines 1PI and
wave-function contributions. In the neutrino channel, the zero-momentum 1PI contribution vanishes:
\begin{equation}
 {F_{\nu L}^{\rm 1PI}(0)\big|_{C_W^2}=0}.
\end{equation}
The cancellation holds exactly in $M_N/m_Z$ at $q^2=0$. Its
compatibility with the generic linear $R_\xi$ gauge follows from
appendix~\ref{app:gauge-current-check}; it does not imply a vanishing
vertex at nonzero $q^2$.
The dimension-six current threshold in this channel comes from WFR.

The cancellation can be checked at the level of scalar integrals,
retaining the nontrivial $C_0$ term. Put $a=M_N^2$, $b=m_Z^2$,
$K=1+\Delta+\ln(\mu^2/b)$ and $L=\ln(a/b)$, with the common regulated pole
$\Delta$ left arbitrary. The scalar integrals are
\begin{align}
 A_a&=a(K-L),& A_b&=bK,& B_{0a}&=K-L,\\
 B_{0b}&=K,& B_{ab}&=K-\frac{aL}{a-b},& C_{0ab}&=-\frac{L}{a-b}.
\end{align}
The full scalar bracket multiplying the 1PI prefactor is
\begin{align}
 {\cal B}_\nu={}&2A_a-3A_b+(3a-2b)B_{0a}
 +2bB_{0b}+(-5a+3b)B_{ab}\nonumber\\
 &+(4ab-2b^2)C_{0ab}=0.
\end{align}
Both the pole and finite parts cancel, with a regular limit at $a=b>0$.

\subsection{Finite-current subtraction and consistency checks}
The physical-measure result, after stripping the common mass and Dirac
prefactor, contains
\begin{align}
 K_{\rm bare}&=-\frac6\epsilon+1+6\gamma_E-6\ln(4\pi)
                  -6\ln\frac{\mu^2}{\MN^2},\\
 K_{\rm bare}+6\overline\Delta&=1-6\ln\frac{\mu^2}{\MN^2},
 \qquad\overline\Delta=\frac1\epsilon-\gamma_E+\ln(4\pi).
\end{align}
The finite constant is obtained from the full integral after
$\MSbar$ subtraction, including the numerator dependence on $D$. The identities
\begin{align}
 C_1+C_3&=\frac{\MN^2g^2C_W^2}{32\pi^2}K_{\MSbar},\\
 -C_1+C_3&=-\frac{\MN^2(4g'^2C_B^2+g^2C_W^2)}{64\pi^2}K_{\MSbar}
\end{align}
reconstruct the direct charged-lepton and neutrino shifts. In particular,
\begin{equation}
 \delta g_{\nu L}^{\rm current}=-\frac{g_Z\MN^2(d_B^2+d_3^2)}{16\pi^2}K_{\MSbar}.
\end{equation}
The two neutrino vertex diagrams are combined before the zero-momentum
limit, so their sum has no spurious $1/q^2$ singularity.

For a threshold $\Delta C_i=A_i[1-6\ln(\mu^2/\MN^2)]$, the derivative
with respect to $\ln\mu$ is $-12A_i$ at this order. Running and local contributions must
compensate the scale dependence in a complete prediction; varying only this
logarithm is not such a prediction. For example, a vanishing independent coefficient at $\mu=M_N$
leaves the threshold $\Delta C_i(M_N)=A_i$ below the heavy-state mass.
The forecast instead specifies the common boundary at $\Lambda_B$,
as in eq.~\eqref{eq:common-finite-boundary}.
Running of dimension-six Higgs--neutrino operators, as studied in
ref.~\cite{Chala:2020pbn}, provides context but is not a direct calculation
of these quadratic dipole thresholds.

\subsection{Projector normalization and contact-diagram identities}
\label{app:validation}

For massless external states, the left-handed projector must separate
the two chiral currents. With
$p_1\cdot p_2=q^2/2$, its normalization is fixed by:
\begin{equation}
 {\cal P}_L[a\gamma^\mu P_L]=a,\qquad
 {\cal P}_L[a\gamma^\mu P_R]=0.
 \label{eq:projector-unit-tests}
\end{equation}
The trace projection uses the spinor ordering of the physical external
states. For a decay chain $\bar u(p_1)M^\mu v(p_2)$, the corresponding
spin-summed trace contains $\slashed p_1 M^\mu\slashed p_2$; the
Born normalization alone does not fix this ordering for a general loop
chain. The tree amplitude in eq.~\eqref{eq:FC-sign} fixes the overall sign
relating the projected coefficient to the physical coupling.

The normalized projectors reconstruct the unbroken current poles from
the broken-phase amplitudes. The neutrino 1PI pole has no $M_N^2$ or
$m_Z^2$ term, while its external residue supplies the leading mass term.
The corresponding finite contributions are discussed in
appendix~\ref{app:currentaudit}.

For the charged-lepton mass term, $WNW$ denotes the triangle with one
internal HNL and two charged $W$ lines, while contact 1 and contact 2
place the four-point dipole vertex at the two ends of the charged line
in figure~\ref{fig:zee-topologies}. The superscripts FC and UV denote
the FeynCalc sign convention and the ultraviolet pole, respectively;
$|_{M_N^2}$ selects the term proportional to the squared HNL mass.
Their contributions are
\begin{align}
 \Gamma_{WNW}^{\rm FC,UV}\big|_{\MN^2}
 &=+\frac{3ec_Wd_W^2\MN^2}{8\pi^2s_W\epsilon_{\rm UV}},\nonumber\\
 \Gamma_{\rm contact,1}^{\rm FC,UV}\big|_{\MN^2}
 =\Gamma_{\rm contact,2}^{\rm FC,UV}\big|_{\MN^2}
 &=-\frac{9ec_Wd_W^2\MN^2}{16\pi^2s_W\epsilon_{\rm UV}},\nonumber\\
 \Gamma_{WNW+\rm contacts}^{\rm FC,UV}\big|_{\MN^2}
 &=-\frac{3ec_Wd_W^2\MN^2}{4\pi^2s_W\epsilon_{\rm UV}}.
\end{align}
Their sum reproduces the $\MN^2$ piece of eq.~\eqref{eq:Zee-1PI}.
This diagram-level identity illustrates why the non-Abelian completion
cannot be dropped.

At order $\MN^2$ the two broken-phase residues invert to
\begin{equation}
 C_{Hl}^{(1),\rm pole}
 =-\frac{\Delta g_{\nu L}^{\rm pole}+\Delta g_{eL}^{\rm pole}}{g_Zv^2},
 \qquad
 C_{Hl}^{(3),\rm pole}
 =\frac{\Delta g_{\nu L}^{\rm pole}-\Delta g_{eL}^{\rm pole}}{g_Zv^2}.
\end{equation}
Only the $\MN^2$ terms are inserted in these equations. Retaining the
$q^2$ or electroweak-mass pieces would incorrectly fold dimension-eight
structures into a dimension-six matching relation.

\subsection{Electromagnetic Ward identity and scale dependence}
\label{app:photon-Ward-finite}
The unbroken electromagnetic symmetry provides a check on the relative
normalization of the charged triangle, the two non-Abelian contact diagrams
and the external-electron residue. We obtain their photon vertices from the
corresponding neutral-current scalar amplitudes by replacing
$g c_W\to e$ in the $VWW$ coupling and $2g c_Wd_W\to2ed_W$ in the contact
couplings. The neutral-dipole triangle vanishes in the zero-momentum limit.
In the FC convention of eq.~\eqref{eq:FC-sign}, the required identity is
\begin{equation}
 \Gamma^{\gamma,\mathrm{1PI},\mathrm{FC}}_{eL}(0)
 +e\Sigma_{eL}(0)=0.
 \label{eq:photon-Ward-finite}
\end{equation}
Both terms change sign on conversion to physical coupling shifts.
The UV pole cancels algebraically for arbitrary $M_N$ and $m_W$:
\begin{equation}
 \Gamma^{\gamma,\mathrm{1PI},\mathrm{FC},\mathrm{UV}}_{eL}(0)
 =-\frac{3ed_W^2}{8\pi^2\epsilon_{\rm UV}}(2M_N^2+m_W^2).
\end{equation}
For the finite parts, evaluated in the same $\MSbar$ convention, we test
the three scale pairs
\begin{equation}
 (M_N,\mu)\in\{(1000,1000),\ (200,500),\ (5000,2000)\}\ \mathrm{GeV},
\end{equation}
with
$m_W=80.379$ GeV. At 100-digit precision the cancellation residual,
normalized to $|e\Sigma_{eL}^{\rm fin}(0)|$, is below $1.1\times10^{-35}$
at $q^2=10^{-30}\,\mathrm{GeV}^2$ and below $1.1\times10^{-45}$ at
$q^2=10^{-40}\,\mathrm{GeV}^2$. The decrease with $q^2$ also checks that the
zero-momentum limit is approached smoothly. This calculation uses the
neutral-current scalar reductions with the photon coupling replacements;
it does not constitute a separate generation of photon diagrams.
It tests the finite contact--triangle--residue cancellation, including the
finite terms inherited from $D$-dimensional numerator algebra. It does not
constrain the transverse $q^2$ coefficient of the charged vertex.

We also differentiate the finite scalar amplitudes with respect to
$\ln\mu^2$. For each of the four electron vertex groups, the neutrino
vertex and the exact external-lepton residue, the derivative equals the
corresponding coefficient of $1/\epsilon_{\rm UV}$. These identities hold
before the heavy-mass expansion and therefore check the scale-dependent
terms retained at both local orders. For $C_B=0$, the dimension-six direct
current shifts satisfy $\Delta g_{\nu L}/\Delta g_{eL}=1/2$; this relation
does not include the universal input corrections. Substitution of the
physical dipoles in the charged-lepton pole also reproduces the numerator
$17=12+5$ in the $q^2C_W^2$ term. In the physical coupling convention,
the momentum-dependent poles in eqs.~\eqref{eq:Zee-1PI} and~\eqref{eq:Znu-1PI}
equal the $L_B$ derivatives of the corresponding finite terms in
eqs.~\eqref{eq:complete-finite-electron}--\eqref{eq:complete-finite-currents}. The same relation holds for the
mass terms after including the external residues; the common boundary
therefore preserves the logarithmic slopes fixed by the scalar poles.

\subsection{Gauge-parameter independence of the direct currents}
\label{app:gauge-current-check}
We compute the HNL contributions to $Z\to e^+e^-$ and
$Z\to\nu\bar\nu$, together with the charged-lepton and neutrino
self-energies, in a generic linear $R_\xi$ gauge and at $\xi=1$.
These amplitudes receive five, two, one and two diagrams, respectively,
including the two non-Abelian contact diagrams in the charged-lepton
vertex. We retain the $D$-dimensional fermion chains and generic real
dipole coefficients in the massless-light-lepton limit.
No Goldstone or ghost diagrams contribute to this inventory at one loop
with two dipole insertions, vanishing active--sterile mixing and massless
light leptons. Ghosts have no fermion vertex, while a Goldstone-containing
dipole vertex also contains a gauge field; it does not replace a
three-point $N\ell V$ vertex in these one-loop current and self-energy
graphs. Mixing-induced Goldstone couplings are absent in the stated limit.
This statement is restricted to these amplitudes and perturbative order.
The reference amplitudes are nonzero and quadratic in the dipoles;
setting all gauge parameters to one recovers the reference before
conversion and contraction.

For every diagram, the generic-gauge minus Feynman-gauge difference
vanishes algebraically before integration. The relevant gauge parameters are
$\xi_W$ for the charged-lepton vertex and self-energy, $\xi_Z$ for
the neutrino vertex, and $\xi_A,\xi_Z$ for its self-energy. The
cancellation follows from the transverse
single-gauge dipole vertex,
\begin{equation}
 k_\mu\sigma^{\mu\nu}k_\nu=0,
\end{equation}
which removes longitudinal propagator contributions. Each internal gauge
line in the listed diagrams has at least one endpoint on a single-gauge
dipole vertex. This also holds for the contact diagrams: the other end
of the internal gauge line is a pure dipole, so transversality does not
require the two-gauge contact vertex itself to be separately transverse.
Thus the equality
holds for both the poles and finite parts, and differentiating the equal
self-energies also gives equal external residues. The retained bosonic
vacuum-polarization bubbles contain no internal gauge propagator; their
input and external-gauge-boson corrections are constructed from those
same two-point functions. Together these checks test gauge-parameter
independence of the retained pole ingredients.

\subsection{Assembly and independent local directions}
\label{app:local-pole-closure}

We assemble the local pole response at one loop with two real dipole
insertions, retaining terms of orders $M_N^2$ and $M_N^0$. We assume
aligned flavours and neglect light-lepton Yukawa couplings and
active--sterile mixing. The input parameters are $e,M_W,M_Z$. We describe the bosonic derivative operators through their corrections
to gauge propagators. Using the gauge-field equations of motion would
express the same effects as current interactions; including both
representations simultaneously would count them twice. The resulting physical response
can be written as
\begin{equation}
 \delta a_{f\chi}=A_T T^3_{f\chi}+A_Q Q_f
 +\delta_{f\chi,\nu_L}\,\delta a_\nu^{\rm dir}
 +\delta_{f\chi,e_L}\,\delta a_e^{\rm dir}.
 \label{eq:pole-four-combinations}
\end{equation}
Here the last two terms denote the corresponding universal flavour
classes, not an additional sum over three heavy states. Bosonic closed
loops carry the heavy-state multiplicity of the benchmark; an aligned
external-flavour current does not.

\paragraph{Independent current checks.}
The two-lepton--two-doublet amplitudes reduce algebraically to the dipole
self-energy master integral after aligning loop momenta and including
the appropriate coupling factors. A charged component, not used to determine the two
neutral leading coefficients, verifies the triplet relation. The tree-level fermionic $Z$ vertex fixes the coupling sign, and the
$ZhG^0$ vertex fixes the momentum convention of the derivative scalar
current. Here $G^0$ is the neutral would-be Goldstone boson in the Higgs
doublet. Comparing these two vertices ensures that the fermion and
scalar currents use the same action and Fourier conventions. With the resulting Wilson dictionary,
$\delta a_\nu^{(6)}=-v^2(C_{Hl}^{(1)}-C_{Hl}^{(3)})/2$ and
$\delta a_e^{(6)}=-v^2(C_{Hl}^{(1)}+C_{Hl}^{(3)})/2$ reproduce the
Born-normalized loop shifts. The Wilson dictionary and action-to-amplitude map are fixed together by
these Born calibrations.

The doublet fields used in this check are auxiliary matching sources,
not Higgs observables. In particular, putting $v=0$ in a two-doublet
source cannot determine operators containing four doublet fields.
Keeping instead the terms of order $v^2p$ in the neutral sources
provides two independent checks. Let
$S=H^\dagger H$, $K_\mu=iH^\dagger\overleftrightarrow D_\mu H$,
$J_\mu=\bar L\gamma_\mu L$, and
\begin{equation}
 Q_t=S J_\mu K^\mu,\qquad
 Q_{\nu t}=J_{\nu H,\mu}K^\mu,\qquad
 J_{\nu H,\mu}=\frac{S J_\mu-(H^\dagger\sigma^I H)J^I_\mu}{2}.
\end{equation}
In the neutral background $J_{\nu H}=S\bar\nu_L\gamma\nu_L$.
For $L_N=\ln(\mu^2/M_N^2)$ the independently extracted finite combinations
in this representative are
\begin{align}
 C_t&=\frac{e^4 C_W^2}{128\pi^2s_W^4}(7-6L_N),\\
 C_t+C_{\nu t}&=\frac{e^4(c_W^2 C_W-2s_W^2C_B)^2}
 {256\pi^2s_W^4c_W^4}(7-6L_N).
 \label{eq:aux-background-finite}
\end{align}
They reproduce the constant dimension-eight direct shifts
$-v^4 C_t/4$ and $-v^4(C_t+C_{\nu t})/4$, respectively. The coefficients
of $q^2$ are fixed by the independently normalized neutral vertices.
The neutral action reconstructs both constants and both $q^2$ coefficients;
it is not claimed to determine every off-shell dimension-eight Wilson
coefficient. In obtaining eq.~\eqref{eq:aux-background-finite}, the
$D$ dependence multiplies the ultraviolet poles before the finite part
is taken. The summed vacuum integral is proportional to
$\int d^D\ell/[\ell^2(\ell^2-M_N^2)]$ and is infrared convergent at the origin.

\paragraph{Soft regions and normalization.}
A loop containing heavy and light particles can receive contributions
from momenta of order $M_N$ and from momenta of order the electroweak
scales. The first region determines the local threshold; the second is
reproduced by light-field diagrams in the low-energy theory. For the
orders considered here, the chiral structure makes this separation
particularly simple. For the lepton-number-conserving open chains, the chiral
projectors remove the heavy-mass numerator. The soft expansion therefore
starts as
\begin{equation}
 \frac{\slashed\ell}{\ell^2-M_N^2}
 =-\frac{\slashed\ell}{M_N^2}+O(M_N^{-4}).
\end{equation}
The associated tree-level light-field interaction has the field content
$\bar L L H^2 X^2 D$ and canonical dimension ten. With no vertex growing
with $M_N$, a light-region loop in dimensional regularization cannot turn
its $C^2/M_N^2$ coefficient into a term of order $M_N^0$ or $M_N^2$.
This power counting has been checked against the heavy propagators and
chiral chains of the current, residue and auxiliary amplitudes.
In the two-point bosonic loop the heavy-mass numerator likewise vanishes
by the odd gamma trace. The argument is restricted to the stated
lepton-number-conserving approximation; it does not remove the soft terms
needed for an exact low-mass prediction or apply to Majorana
lepton-number-violating amplitudes.

For each direct current the assembly is
$\delta a_f^{\rm dir}=\delta a_f^{\rm vertex}+a_f^{\rm SM}\sigma_f$.
Here $a_f^{\rm SM}=T_f^3-s_W^2Q_f$ is the dimensionless SM chiral
coupling and $\sigma_f$ is the net fractional external-fermion residue
in our self-energy convention. Each external leg contributes half of
this residue, so the pair produces $a_f^{\rm SM}\sigma_f$.
The Wilson-derived direct shifts already contain this contribution.
The universal pole residue and input shifts enter only through $A_T,A_Q$;
adding EOM-induced bosonic currents on top of the unreduced propagator
representation would double count them. These statements also account
for the light-reducible external-leg and propagator attachments relevant
to eq.~\eqref{eq:pole-four-combinations}.

\paragraph{Common finite prescription.}
We use the NDR prescription of the open-chain calculation, with
$D$ retained through the Laurent separation and the specified physical
external-state projections. Changing the prescription for $\gamma_5$ requires a consistent finite
conversion of the renormalized action and its independent coefficients;
differences between bare diagrams alone do not define that conversion. For a threshold $T_i(L_N)$ linear in $L_N$, a
common independent boundary at $L_B=\ln(\Lambda_B^2/M_N^2)$ gives
\begin{equation}
 C_i^{\rm total}=T_i(L_N)+
 \frac{\partial T_i}{\partial L_N}(L_B-L_N)
 +C_i^{\rm independent}(\Lambda_B)
 =T_i(L_B)+C_i^{\rm independent}(\Lambda_B).
 \label{eq:common-finite-boundary}
\end{equation}
This rule applies to both orders in the heavy-mass expansion and to both current and bosonic
coefficients. The logarithmic slopes agree with the poles, and the
subtraction-scale dependence cancels at the retained order. Independent
finite ultraviolet coefficients are not predicted by the dipole threshold.

The finite functions entering eq.~\eqref{eq:pole-four-combinations}
can be written compactly in terms of $s=s_W$, $c=c_W$,
$L=L_B$, $q=M_Z^2$, and $D_Z=c^2C_W-2s^2C_B$.
For each external lepton flavour, the direct terms are
\begin{align}
 \delta a_e^{\rm dir}={}&
 \frac{e^2v^2M_N^2 C_W^2}{64\pi^2s^2}(6L-1)
 +\frac{e^4v^4C_W^2}{512\pi^2s^4}(6L-7)\nonumber\\
 &+\frac{e^2v^2q}{1152\pi^2s^2}
 \left[72s^2 C_BC_W(2L+3)-c^2C_W^2(102L+241)\right],\label{eq:complete-finite-electron}\\
 \delta a_\nu^{\rm dir}={}&
 \frac{e^2v^2M_N^2(4s^2C_B^2+c^2C_W^2)}{128\pi^2s^2c^2}(6L-1)
 \nonumber\\
 &+\frac{e^2v^2D_Z^2}{1024\pi^2s^4c^4}
 \left[e^2v^2(6L-7)-16s^2c^2q(2L+3)\right].
 \label{eq:complete-finite-currents}
\end{align}
No flavour multiplicity multiplies these direct terms. The universal
terms instead contain the sum over the three aligned HNLs. Define
\begin{gather}
 (R_{AA},R_{AZ},R_{ZZ},R_{WW})
 =3(d_\gamma^2,d_\gamma d_Z,d_Z^2,d_3^2),\nonumber\\
 f_0=-\frac{M_N^2}{2\pi^2},\qquad
 f'_0=\frac{L+11/6}{6\pi^2},\qquad
 f_Z=f_0+qf'_0,\qquad f_W=f_0+M_W^2f'_0,\nonumber\\
 \delta=\frac{R_{ZZ}f_Z-R_{WW}f_W}{2},\qquad
 x=\frac{R_{AA}f_0}{2}-\frac{c^2}{s^2}\delta,\qquad
 y=\frac{R_{AA}f_0}{2}+\delta,\nonumber\\
 u=sc(y-x)+R_{AZ}f_Z,\qquad
 z=\frac{R_{ZZ}}{2}(f_Z+qf'_0).
\end{gather}
Then the two universal coefficients are
\begin{equation}
 A_T=c^2x+s^2y-z,\qquad
 A_Q=-s^2y-sc\,u+s^2z.
 \label{eq:complete-finite-universal}
\end{equation}
Here $x,y$ encode the input shifts, $u$ the neutral mixing, and $z$ the
$Z$-pole residue. These expressions use the SM-evolved electromagnetic
input defined in section~\ref{sec:ewinputs}. Together with
$\delta a_f=A_TT_f^3+A_QQ_f+\delta a_f^{\rm dir}$, they give the full
local response through $M_N^0$; right-handed direct lepton terms vanish
in the massless-lepton limit.
The separate radiative-decay analysis is described in
section~\ref{sec:decay-contrast}.

\paragraph{Gauge-invariant ultraviolet profiling at the pole.}
The four-dimensional response above can be spanned by independent,
gauge-invariant local ultraviolet coefficients. This is more restrictive
than simply introducing four arbitrary shifts of measured couplings.
Besides $Q_{Hl}^{(1,3)}$, consider $Q_{HWB}$ and the dimension-eight
representative
\begin{equation}
 Q_{B\partial H}=(H^\dagger H)
 (\partial_\mu B^{\mu\nu})(\partial^\rho B_{\rho\nu}).
\end{equation}
The latter may be eliminated using gauge equations of motion, provided
all induced current and contact operators are retained. Here it is kept
in the propagator representation, without adding its EOM images again.
Define $b_{WB}$ as the constant off-diagonal $W^3$--$B$ transverse form
factor and $b_B$ as the coefficient of $q^2$ in the $BB$ form factor.
These coordinates absorb the nonzero vev and normalization factors of
the two Wilson coefficients; $b_{WB}$ is dimensionless and $b_B$ has
mass dimension $-2$. In the fixed $(e,M_W,M_Z)$ input prescription,
including the pole residue, their responses are
\begin{equation}
 \begin{pmatrix} A_T\\ A_Q\end{pmatrix}_{\rm UV}
 =\begin{pmatrix}
 c_W/s_W&-M_Z^2/2\\
 0&s_W^2M_Z^2/2
 \end{pmatrix}
 \begin{pmatrix}b_{WB}\\b_B\end{pmatrix}.
 \label{eq:pole-UV-bosonic-span}
\end{equation}
The determinant is $M_Z^2s_Wc_W/2\ne0$. Together with the two
independent lepton-current shifts this gives rank four. Writing $d_f\equiv\delta a_f^{\rm dir}$ for $f=\nu,e$, any
dipole-induced local response can consequently be cancelled by choosing
\begin{equation}
 b_B=-\frac{2A_Q}{s_W^2M_Z^2},\qquad
 b_{WB}=-\frac{s_W}{c_W}\left(A_T+\frac{A_Q}{s_W^2}\right),\qquad
 (d_\nu,d_e)_{\rm UV}=-(d_\nu,d_e).
\end{equation}
This cancellation precedes the observable map and covariance weighting.
With unrestricted coefficients in these directions, profiling therefore
removes all dipole sensitivity of the retained local pole response,
including in the presence of a nonzero experimental residual. Two free
lepton currents alone span rank two; adding only $Q_{HWB}$ gives rank
three and does not generally remove the subleading charge response.

The same four directions are already spanned by dimension-six Warsaw
operators. For $Q_{HD}=|H^\dagger D_\mu H|^2$, the tree-level mass
relations are
\begin{equation}
 M_W^2=\frac{g^2v^2}{4},\qquad
 M_Z^2=\frac{(g^2+g'^2)v^2}{4}
 \left(1+\frac{v^2}{2}C_{HD}\right).
\end{equation}
Holding $e,M_W,M_Z$ fixed gives
$\delta s^2=-c_W^2v^2C_{HD}/2$ for the Lagrangian mixing angle.
Consequently, in the dimensionless coupling convention of
 eq.~\eqref{eq:pole-four-combinations},
\begin{equation}
 \begin{pmatrix}A_T\\ A_Q\end{pmatrix}_{\rm UV}^{(6)}
 =v^2\begin{pmatrix}
 c_W/s_W &(1-2s_W^2)/(4s_W^2)\\
 0&1/4
 \end{pmatrix}
 \begin{pmatrix}C_{HWB}\\ C_{HD}\end{pmatrix}.
 \label{eq:pole-dim6-span}
\end{equation}
Its determinant is $v^4c_W/(4s_W)\ne0$. The two independent
$C_{Hl}^{(1,3)}$ directions then complete the rank-four response.
Thus the identifiability result at the fixed $Z$ pole does not rely on
an independent dimension-eight coefficient. This statement concerns
linearized pole couplings; the momentum dependence and correlations
with other observables distinguish these operator choices.

The degeneracy applies to the retained local pole response. Off-pole,
charged-current and direct-production observables probe additional
structures, while UV correlations or finite priors restrict the allowed
compensation. The boundary variations in eq.~\eqref{eq:common-finite-boundary}
therefore represent different hypotheses, not a probability distribution.

\section{Finite-mass dependence and the local expansion}
\label{app:finite-mass-remarks}

Section~\ref{sec:lowmass-validity} distinguishes the heavy-mass expansion
from the validity of the dipole EFT. Here we give the exact two-point
remainder, specify the ingredients of a finite-mass pole prediction and
compare the assembled amplitudes with their local limit. These results
explain the mass qualifications on the sensitivity curves.

\subsection{Nonlocal two-point remainder}
\label{sec:nonlocal-check}
The nonlocal part of the two-point function provides a quantitative
diagnostic of the heavy-mass expansion. Define the twice-subtracted remainder of eq.~\eqref{eq:F0-exact},
\begin{equation}
 R(t)=F_0(t)-F_0(0)-tF_0'(0),\qquad z=t/\MN^2.
\end{equation}
The constant and linear local terms, including their UV and finite-scheme
dependence, cancel. For $0<z<1$ the expression gives
\begin{align}
 R(t)&=\frac{\MN^2}{6\pi^2}r(z),\\
 r(z)&=\frac{2-z-(1-z)(2+z)b(z)}{z}+1-\frac76z,\label{eq:nonlocal-exact}\\
 b(z)&=2+\frac{1-z}{z}\ln(1-z),\label{eq:nonlocal-b}\\
 r(z)&=\frac{z^2}{2}+\frac{3z^3}{20}+\frac{z^4}{15}
       +\frac{z^5}{28}+\mathcal O(z^6).
 \label{eq:nonlocal-shape}
\end{align}
Since $F_0(0)=-\MN^2/(2\pi^2)$, the ratio is $R/F_0(0)=-r(z)/3$.
The expression agrees algebraically with the Package-X result. In particular, its constant and linear terms before subtraction
are proportional to
$-3+z[1/\epsilon-\gamma_E-\ln\pi+\ln(\mu_{\rm PX}^2/\MN^2)+11/6]$
in the scalar-integral convention of appendix~\ref{app:renormalization};
subtracting them reproduces eqs.~\eqref{eq:nonlocal-exact}--\eqref{eq:nonlocal-b}. This also avoids
discarding the finite effects of evanescent factors in the local terms.

The numerical values at $t=M_Z^2$ are given in
section~\ref{sec:finite-mass-summary}.

The summed neutrino vertex contains a UV pole proportional to
$q^2(c_W^2C_W-2s_W^2C_B)^2$. Its cancellation of spurious $1/q^2$ terms and
its checked zero-momentum limit do not remove the need for momentum-dependent
local counterterms at the pole. A finite-mass prediction must recombine the
vertex, WFR and input terms with consistently renormalized bosonic and
dimension-eight contributions. The same boundary prescription must cover the dimension-six currents and
the higher-derivative local terms. A boundary imposed at $\Lambda$ is
transported to $\MN$ by the corresponding running.

\subsection{Finite-mass pole amplitudes}
\label{sec:exact-mass-prescription}
The scalar functions $A_0,B_0,C_0$ can be evaluated with their full mass
and momentum dependence while keeping the perturbative order fixed.
This removes the expansion in electroweak scales over $M_N$, but leaves
the independent UV coefficients of the dipole EFT to be specified.
The relevant truncation parameters are the ratios appearing in each
amplitude, rather than a universal estimate based on $v^2/M_N^2$ alone.

For example, the universal two-point function is already known exactly in
eq.~\eqref{eq:F0-exact}. A useful organization of its renormalized loop part is
\begin{equation}
 F_{0,\rm loop}^{\rm ren}(t;\mu)
 =-\frac{\MN^2}{2\pi^2}+t\,b_{\rm loop}(\mu)
   +\frac{\MN^2}{6\pi^2}r(t/\MN^2),
 \label{eq:exact-mass-decomposition}
\end{equation}
where $r$ is the unexpanded function in eqs.~\eqref{eq:nonlocal-exact}--\eqref{eq:nonlocal-b} and
$b_{\rm loop}$ is the renormalized slope in the declared subtraction convention.
This decomposition is exact for this loop below its $N$--massless-particle
threshold, not just a truncation at $t^2$. The full two-point matrix also
contains independent local terms,
\begin{equation}
 F_{ij}^{\rm ren}(t)=d_i d_j F_{0,\rm loop}^{\rm ren}(t)
                    +F_{ij}^{\rm local}(t;\mu).
\end{equation}
The local matrix contains the independent UV contributions and need
not share the gauge-basis alignment of the dipole term. Its finite boundary
and the subtraction convention must therefore be specified even when
the loop function is evaluated without a mass expansion.

The analogous physical neutral-current correction must combine all retained
pieces at the pole,
\begin{equation}
 \delta g_f^{\rm ren}(M_Z^2)=
 \delta g_f^{\rm 1PI}(M_Z^2)+\delta g_f^{\rm external\ fields}
 +\delta g_f^{\rm inputs}+\delta g_f^{\rm local},
 \label{eq:finite-mass-coupling-assembly}
\end{equation}
with the same sign, coupling normalization and subtraction convention.
The schematic external-field term includes both fermion and gauge-field
normalization and neutral mixing where applicable. The input term uses the
chosen $\{\alpha,M_W,M_Z\}$ scheme with self-energies evaluated at the
appropriate momenta. The full mass dependence of a pole observable requires these
contributions together; the exact neutrino vertex or two-point function
alone is insufficient.

At double dipole insertion the dimension-eight local terms enter at the
same order in the portal expansion as the loop. Their finite boundary
conditions are part of a physical finite-mass prediction. Below we test
both the asymptotic expansion and its effect on the projected reach at
the masses represented in the figures.

The heavy-mass expansion of the pole amplitudes is controlled by ratios
such as $M_Z^2/M_N^2$ and $M_W^2/M_N^2$.
For example, $M_Z^2/M_N^2$ decreases smoothly from about $0.21$ at
200 GeV to $0.033$ at 500 GeV. There is no physical transition at 500 GeV:
the line-style change in figure~\ref{fig:direct-indirect} flags the more
cautious interpretation of the low-mass extrapolation. Neither that
reference mass nor a small two-point remainder alone bounds the error of
all current and input contributions; table~\ref{tab:finite-mass-reach}
quantifies their combined effect in representative directions. This heavy-mass expansion issue is
distinct from the ultraviolet coefficient-scale guide in the same figure.

The low-mass curves in the direct-search comparison illustrate the local
prediction under these qualifications. A fit below $M_Z$ would also
require the causal continuation and possible real $Z\to N\nu$ final
states, beyond shifts of light-fermion widths.

\subsection{Heavy-mass expansion of the assembled pole response}
\label{app:assembled-heavy-check}
We compare the local response with the unexpanded scalar amplitudes,
combining the Born-normalized electron and neutrino vertices, one net
external-fermion residue for each pair of external legs, and the universal
vacuum-polarization and input corrections. The exact and expanded
expressions use the same $\MSbar$ convention and the same input scheme.
No independent local UV coefficients are added to either side of this
comparison. Writing
\begin{equation}
 \delta a_f^{\rm ex}
 =\delta a_f^{(6)}+\delta a_f^{(8)}+\mathcal R_f,
\end{equation}
where the two local terms scale as $M_N^2$ and $M_N^0$ at fixed dipoles,
we evaluate $M_N=1,2,5,10,20,50$ TeV and
$\mu/M_N=1/2,1,2$, with $C_B=0.3\,\mathrm{TeV}^{-2}$ and
$C_W=0.2\,\mathrm{TeV}^{-2}$. The calculation uses 65-digit arithmetic.
For the nonvanishing leading responses, the endpoint logarithmic slopes
of $|\delta a_f^{\rm ex}/\delta a_f^{(6)}-1|$ lie between $-2.003$ and
$-1.997$. The labels $f=\nu_L,e_L,e_R$ refer to the left-handed neutrino and
the left- and right-handed charged-lepton couplings. For each of these
couplings and each of the three choices of $\mu/M_N$, the omitted
remainder $\mathcal R_f$ is smaller than $8\times10^{-4}$ of the
subleading local term $\delta a_f^{(8)}$ at the sampled masses from
5 to 50 TeV. The leading $e_R$ response vanishes identically, so
only the subleading comparison is used in that channel.
The exact self-energy expansion also reproduces the finite combination
$7-6\ln(\mu^2/M_N^2)$, while scalar quadrature of the bosonic kernel
recovers its $11/6$ slope constant.

The asymptotic agreement tests the expansion of the same underlying
amplitudes, rather than an independent calculation of the diagrams.
For the masses shown in the reach plots, a more direct measure is
\begin{equation}
 \epsilon_O=\frac{\|\delta O_{\rm ex}-\delta O_{\rm loc}\|_V}
 {\|\delta O_{\rm loc}\|_V},\qquad
 \|x\|_V^2=x^TV^{-1}x.
\end{equation}
Table~\ref{tab:finite-mass-reach} compares the assembled responses with
the same inputs and common UV boundaries. It includes the full scalar
mass dependence of the lepton currents and external residues, as well
as the bosonic functions. The small differences at 500 GeV and 1 TeV
support the local approximation in the displayed benchmark directions.
The larger effect at 200 GeV motivates retaining the distinct line style
for the low-mass extrapolation; none of these reference masses is a
physical threshold.\footnote{The reach ratio is
$C_{95}^{\rm ex}/C_{95}^{\rm loc}
=(\|\delta O_{\rm loc}\|_V/\|\delta O_{\rm ex}\|_V)^{1/2}$
at fixed coefficient direction and confidence threshold. These are
sampled comparisons, not a uniform bound near every possible cancellation.}
\begin{table}[htbp]\centering\small
\begin{tabular}{rrr}\toprule
$M_N$ [GeV] & $\max\epsilon_O$ & $\max|C_{95}^{\rm ex}/C_{95}^{\rm loc}-1|$\\\midrule
200 & $10.97\%$ & $4.29\%$\\
300 & $2.18\%$ & $0.89\%$\\
500 & $0.337\%$ & $0.143\%$\\
1000 & $0.0278\%$ & $0.0121\%$\\\bottomrule
\end{tabular}
\caption{Finite-mass corrections on the two axes and two diagonals.
Each entry is the largest value among these directions, $r_B=2,5$, and
the conservative and aggressive SM covariance scenarios, with the
SM-evolved electromagnetic input used throughout.}
\label{tab:finite-mass-reach}\end{table}

\section{Higgs-background relations and radiative-decay kernels}
\label{app:higgs-details}
The radiative-decay analysis in section~\ref{sec:higgs} uses the Higgs
partners of the dipole vertices. We give the auxiliary two-Higgs
structures that distinguish derivative terms, the tensor reduction
relating the decay bubble to the two-point kernel, and its soft-Higgs
limit. These results concern the neutral two-dipole contribution, rather
than a matching of the complete Higgs sector.

\subsection{Independent Higgs-background structures}
\label{sec:higgs-matching-update}
\paragraph{Two independent derivative structures.}
Independent Higgs momenta distinguish derivative interactions that
reduce to the same structure in a constant Higgs background. The CP-even
$hhVV$ amplitude can be resolved onto five off-shell tensors in
$D=4-2\epsilon$, keeping the external virtualities independent. The
separately calculated $\gamma Z$ and $ZZ$ amplitudes obey the neutral
coupling relations. For $X_{\mu\nu}^{\gamma}=hF_{\mu\nu}$, define
\begin{equation}
 {\cal C}_{\gamma}=\frac{n_F e^2(2C_B+C_W)^2}{16\pi^2},
 \qquad L_N=\ln\frac{\mu^2}{M_N^2}.
\end{equation}
The local finite result, through four powers of external momenta, is
represented in these projections by
\begin{align}
 \Delta{\cal L}_{hh\gamma\gamma}={\cal C}_{\gamma}\bigg[
 &\frac{M_N^2}{4}X^{\gamma}_{\mu\nu}X_{\gamma}^{\mu\nu}
 +\frac12\left(\frac{L_N}{6}-\frac1{36}\right)
 (\partial_\rho X^{\gamma}_{\mu\nu})(\partial^\rho X_{\gamma}^{\mu\nu})
 \nonumber\\[-1mm]
 &-\frac12\left(\frac{2L_N}{3}+\frac59\right)
 (\partial_\mu X_{\gamma}^{\mu\nu})
 (\partial_\rho X_{\gamma}^{\rho}{}_{\nu})\bigg].
 \label{eq:hh-finite-local-update}
\end{align}
The coefficient includes the three flavour copies of the aligned
benchmark through $n_F=3$. The charged $W^+W^-hh$ kernel agrees in all five
CP-even projections, in the massless charged-lepton and vanishing Yukawa
limit. Its normalization is obtained by replacing
${\cal C}_{\gamma}$ with $n_F e^2C_W^2/(16\pi^2s_W^2)$.
This equality holds after reduction, not term by term in the integrands.

For the electroweak composite field
\begin{equation}
 {\cal X}_{\mu\nu}=
 \left(g'C_B B_{\mu\nu}{\bf1}
       +\frac{gC_W}{2}\sigma^I W^I_{\mu\nu}\right)\widetilde H,
\end{equation}
a compatible local representative is
\begin{align}
 \Delta{\cal L}_{\rm rep}={}&
 \alpha_0{\cal X}_{\mu\nu}^{\dagger}{\cal X}^{\mu\nu}
 +\alpha_1(D_\rho{\cal X}_{\mu\nu})^{\dagger}D^\rho{\cal X}^{\mu\nu}
 +\alpha_2(D_\mu{\cal X}^{\mu\nu})^{\dagger}D_\rho{\cal X}^{\rho}{}_{\nu},\\
 \alpha_0={}&\frac{n_F M_N^2}{8\pi^2},\qquad
 \alpha_1=\frac{n_F(6L_N-1)}{144\pi^2},\qquad
 \alpha_2=-\frac{n_F(6L_N+5)}{36\pi^2}.
 \label{eq:hh-alpha-update}
\end{align}
The two-point combination is recovered:
$\alpha_1+\alpha_2/2=-n_F(6L_N+11)/(144\pi^2)$.
The independent Higgs momenta separate the two derivative coefficients,
which a two-point function alone cannot distinguish. Equations
\eqref{eq:hh-finite-local-update}--\eqref{eq:hh-alpha-update} describe
these two-field-strength projections. Fixing additional non-Abelian
structures would require higher-point amplitudes beyond this sector.

Finite rational terms are essential even in this CP-even calculation.
If the projected vector is $P/\epsilon+F$ and its Gram matrix is
$G(D)=G_4-2\epsilon G'_4+\cdots$, the finite coefficients obey
\begin{equation}
 G_4 c_{\rm fin}=F+2G'_4 c_{\rm pole}.
\end{equation}
Keeping only the four-dimensional Gram matrix before subtracting the
pole would change the finite constants.

The finite $hZee$ projections with generic couplings provide an
additional check of the current convention. They do not fix coefficients
invisible to the pole projections. A complete non-Abelian higher-point
matching and a $ZH$ production likelihood lie outside this analysis.

\subsection{Tensor reduction of the neutral Higgs bubble}
\label{app:higgs-tensor-proof}
Let $S^\mu(k)=[\gamma^\mu,\slashed k]$, with irrelevant common dipole
normalization suppressed. The tensor integral multiplying the CP-even
trace has the form $I_{\alpha\beta}=A g_{\alpha\beta}+B p_\alpha p_\beta$.
The Clifford identities
\begin{align}
 \gamma_\alpha S^\nu(k)\gamma^\alpha&=(D-4)S^\nu(k),\\
 \slashed p S^\mu(p)\slashed p&=-p^2 S^\mu(p)
\end{align}
imply
\begin{equation}
 I_{\alpha\beta}\mathrm{tr}
 [S^\mu(p)\gamma^\alpha S^\nu(k)\gamma^\beta]
 =[(D-4)A-p^2B]\,\mathrm{tr}[S^\mu(p)S^\nu(k)].
\end{equation}
Thus the same scalar kernel that appears when the gauge momenta coincide
also multiplies the physical Higgs tensor. The factor $(D-4)A$ must not be
dropped before integration: its UV pole supplies a finite contribution,
already retained in $F_0$. The CP-odd trace cancels in the real,
CP-even coupling sum considered here. Each gauge Ward identity follows
from contracting its commutator with its own momentum.

\subsection{Soft-Higgs relation}
\label{app:additional-consistency}
At fixed $M_N$ and Wilson coefficients, $d_\gamma$ is proportional to $v$
and the neutral fermion-bubble kernel is independent of $v$. Consequently
\begin{equation}
 -\frac14\frac{\partial}{\partial v}
       [n_Fd_\gamma^2F_0(0)]
 =-\frac{n_Fd_\gamma^2F_0(0)}{2v}.
\end{equation}
This reproduces the zero-momentum two-dipole Higgs-bubble coupling. The
relation applies to this bubble contribution, with the heavy mass held
fixed, and is not a low-energy theorem for the entire Higgs amplitude.

\section{Statistical implementation of the FCC-ee forecast}
\label{app:forecast-methods}
The forecast in our analysis compares the dipole-induced shifts with the correlated experimental and SM uncertainties. This appendix specifies the input-coordinate conversion, the confidence construction and their numerical checks, supporting the results in section~\ref{sec:fcc}.

\subsection{Observable set and correlated uncertainties}

We use the twelve nonredundant pole pseudo-observables defined in
section~\ref{sec:fcc-channel-extension}.
The projections and correlated SM covariance scenarios are taken from
the public dataset accompanying ref.~\cite{Armadillo:2026mvp}.
This set includes all three charged-lepton flavours. The corresponding
$A_{\rm FB}$ and epsilon combinations contain overlapping information
and are not added as independent measurements. The BSM
correction is the linear derivative of the tree pole observables with
respect to the matched chiral couplings; the precision SM uncertainty
model is additive. Missing higher-order BSM electroweak corrections are
not computed by this procedure.

The public covariance is supplied jointly with an electromagnetic input
coordinate in the $\{G_F,M_W,M_Z\}$ scheme. We replace $G_F$ by that
measured coordinate rather than adding a second parametric-error matrix.
For the joint residual vector $(\delta O,\delta\alpha_{\rm src})$, define
\begin{equation}
 K=(I,-r),\qquad r_i=
 \frac{\partial O_i/\partial G_F}
 {k_{\rm src}\,\partial\Delta\alpha/\partial G_F},\qquad
 V_{\alpha M_W}=K(V_{\rm exp}+V_{\rm SM})K^T,
 \label{eq:theory-input-conversion}
\end{equation}
where the forecast uses the local affine convention
$\delta\alpha_{\rm src}=k_{\rm src}\delta\Delta\alpha$ and
$k_{\rm src}=\alpha_{\rm ref,src}^2/\alpha(0)$ with
$\alpha_{\rm ref,src}=0.00780832$. This convention is an explicit
covariance-transfer assumption. The central SM-evolved coupling is
$\bar\alpha_Z^{-1}=128.9468$, using $\Delta\alpha_{\rm SM}=0.05903$.
The source coordinate has experimental uncertainty $4.81098\times10^{-8}$;
including the source SM covariance gives marginal uncertainties
$(7.429,3.905,2.016)\times10^{-7}$ in the current, conservative and
aggressive scenarios. These are uncertainties of $\alpha_{\rm src}$,
whose full correlations are propagated by $K$, rather than additional
independent errors on $\bar\alpha_Z$. We do not identify the source central
value with our running-$\alpha$ central value: the expected residuals
are recentered at the SM. The underlying SM Jacobians are obtained from
ref.~\cite{Mildner:2024wbl}. The $G_F$ error direction cancels and the
shared-input derivatives obey the chain rule in the new scheme. The
uncertainty of the electromagnetic measurement is included once; the
$G_F$ measurement is not retained as another observable.

The normalization of the input coordinate is fixed by separating the
nonparametric and shared-input contributions to the public covariance.
The source Jacobian reconstructs the remaining joint parametric part. The
maximum residual, normalized to the product of marginal uncertainties,
is below $2\times10^{-6}$. A consistent rescaling of the source coordinate, its covariance and its
Jacobian leaves the forecast unchanged. As a diagnostic, changing only
the normalization to our running-$\alpha$ reference changes the four
sections at 1 TeV in table~\ref{tab:final-FCC} by less than $0.36\%$ across the three theory
scenarios. We use the source normalization throughout; this comparison tests the
coordinate convention and is not included as an additional uncertainty.

The finite dipole charge condition is also specified by this choice of
input. With photon inverse propagator $q^2[1+F_{AA}(q^2)]$ and
$\Pi_{AZ}(0)=0$, the parameter shift that holds the Thomson charge fixed is
\begin{equation}
 \left.\frac{\delta e}{e}\right|_{\rm dip}=\frac{F_{AA}(0)}2,\qquad
 \Delta F=\mathrm{Re}[F_{AA}(M_Z^2)-F_{AA}(0)]
 =\frac{3d_\gamma^2M_Z^2}{6\pi^2}\left(L_B+\frac{11}{6}\right)
\end{equation}
through the local order retained. Thus $\Delta F$ remains in the
renormalized photon propagation. An ideal effective charge defined from
photon exchange at $M_Z$ would instead absorb this difference, shifting
the fixed-input response by
$(T_f^3-s_W^2Q_f)\Delta F/2$. Comparing these two input prescriptions with
the same covariance changes the coefficient reach by less than $0.96\%$
at 500 GeV and $0.26\%$ at 1 TeV on the axes and diagonals, for both
boundaries and the conservative/aggressive scenarios. This comparison
isolates photon propagation, not the additional process-dependent
corrections to an experimental extraction.\footnote{The SM conversion
$\alpha_{\MSbar}(M_Z)=\alpha(0)/[1-\Delta\alpha-0.007127]$ used in
ref.~\cite{Celada:2024mcf} gives the same Jacobian form
$\alpha_{\rm ref}^2/\alpha(0)$. It is distinct from the BSM term $\Delta F$;
a consistent coordinate conversion cannot replace its treatment.}

The current, conservative and aggressive SM scenarios describe different
levels of theoretical calculation accuracy, with the correlations of the
public forecast. We represent these projected uncertainties by Gaussian
nuisance parameters. The coverage statement below is conditional on
this statistical model; a fixed unknown theoretical bias would require
a different treatment.

\paragraph{LEP precision as a reference.}
The LEP reference in table~\ref{tab:final-FCC} uses the
flavour-universal pole data distributed with ref.~\cite{Mildner:2024wbl},
including the luminosity correction of ref.~\cite{Janot:2019oyi}.
The experimental standard deviations of
$(\Gamma_Z,\sigma_h^0,R_\ell,A_{\rm FB}^{0,\ell})$ are
$0.0023\,\mathrm{GeV}$, $0.0325\,\mathrm{nb}$, $0.0247$ and $0.0010$,
respectively.
Their correlations with the inputs are retained. If $C$ denotes the
joint measurement covariance of observables $O$ and inputs $p$, the
covariance of the pole residuals is
\begin{equation}
 V_{\rm LEP}=C_{OO}+J C_{pp}J^T-JC_{pO}-C_{Op}J^T+V_{\rm SM},
 \qquad J=\frac{\partial O_{\rm SM}}{\partial p}.
\end{equation}
Here $p=(\Delta\alpha,M_W,M_Z,M_H,m_t,\alpha_s)$, with the same central
values as in the FCC-ee response; $V_{\rm SM}$ contains the current
SM theory uncertainties. The dipole response follows from the same
coupling shifts as the FCC-ee forecast, with
$\delta A_{\rm FB}^{0,\ell}=(3/2)A_\ell\delta A_\ell$ in the
flavour-universal limit. Both response matrices have rank three.
Centring the reference on the SM removes dependence on fluctuations
in the observed LEP central values, so that the comparison measures
expected precision rather than the relative strength of observed limits.

\paragraph{Perturbative accuracy and running.}
The forecast combines precision SM predictions with a leading one-loop
portal response. To specify its perturbative accuracy, we distinguish
higher-order SM improvements from corrections to the BSM amplitude.
One-loop electroweak SMEFT predictions and input-scheme dependence are
studied in refs.~\cite{Dawson:2019clf,Biekotter:2025nln}; loop sensitivity
at Tera-$Z$ is explored in refs.~\cite{Maura:2024zxz,Gargalionis:2024jaw}.
Recent two-loop and NLL studies~\cite{Born:2026tgm,Haisch:2026eyq}
address higher perturbative orders than the portal contribution considered here.
The gauge and Yukawa contributions to the dimension-six $\nu$SMEFT
running are available in refs.~\cite{Datta:2020ocb,Ardu:2024tzb}, with
an implementation in \texttt{wilson}~\cite{Aebischer:2024csk}. Inserting
one-loop running of $C_B,C_W$ into our one-loop quadratic threshold
changes that contribution at two-loop order. We therefore omit this effect at fixed one-loop order. Large logarithms
would instead require a consistent renormalization-group improvement of
the threshold calculation.

\subsection{Common boundary and confidence construction}
A common boundary prescription gives an unambiguous meaning to dipole
dominance: every independent local coefficient is set to zero at the
same scale $\Lambda_B$, in the same subtraction scheme. Equation~\eqref{eq:common-finite-boundary}
then combines its evolution with the HNL threshold. We compare
$r_B=\Lambda_B/M_N=2,5$. Running of the dipoles inside the one-loop
response is beyond the fixed order used here. The two choices are UV
hypotheses; shading between them is neither a confidence interval nor an
estimate of the omitted heavy-mass terms.

\begin{samepage}
For illustration, the same-sign section $|C_B|=|C_W|$ at $M_N=1$ TeV
varies with the common boundary as follows, in $\mathrm{TeV}^{-2}$:
\begin{center}
\begin{tabular}{lrrrrr}\toprule
$r_B$ & 1.2 & 1.5 & 2 & 5 & 10\\\midrule
Conservative & 0.2690 & 0.2484 & 0.2201 & 0.1614 & 0.1375\\
Aggressive & 0.1779 & 0.1615 & 0.1431 & 0.1069 & 0.0919\\\bottomrule
\end{tabular}
\end{center}
\end{samepage}
The reach strengthens with increasing $r_B$ because the longer evolution
interval enhances the induced local response. This dependence concerns
the UV boundary, while the experimental precision is unchanged.
Identifying the boundary scale with a mediator mass requires the
power counting of a specified completion, particularly when that mass
is close to $M_N$. Similarly, finite constants and independent Wilson
coefficients must be transformed together under a change of subtraction
scheme. Their combined prediction is scheme independent to the retained
order; varying only the constants at a fixed zero boundary defines a
different physical hypothesis.

At fixed $M_N,r_B$ the observable shift is
\begin{equation}
 \delta O=B t,\qquad t=(C_B^2,C_BC_W,C_W^2)^T.
\end{equation}
The $12\times3$ matrix $B$ contains the response of each observable
to the three quadratic monomials, at the specified mass and boundary.
After weighting by the inverse covariance and projecting out any
profiled ultraviolet directions, the linear signal space has rank $k$.
We construct a Gaussian-mean confidence ellipsoid in this space and
intersect it with the surface generated by the two dipole coefficients. For the expected SM dataset, the boundary is
\begin{equation}
 t^T B^TV^{-1}B t=\chi^2_k(0.95),\qquad
 \chi^2_3(0.95)\simeq7.81.
 \label{eq:final-confidence}
\end{equation}
The rank is that of the linear space of quadratic monomials, not the
number of dipole parameters. Their physical image satisfies
$t_2^2=t_1t_3$, $t_1,t_3\geq0$, but the three-component Gaussian-mean
region is constructed before imposing this relation. Intersecting that
region with the physical surface preserves inclusion of the true point
whenever it belongs to the Gaussian-mean region. This provides a direct
coverage argument without relying on a regular two-parameter tangent
plane at the SM origin. The value $5.99$ often used for two-parameter contours is instead the
95th percentile of a chi-square distribution with two degrees of
freedom. It applies to a regular two-parameter likelihood-ratio
construction, not automatically to any plot with two axes. Here the
observable shifts are quadratic in $C_B,C_W$, and their first derivatives
vanish at the SM point. The usual regularity argument for that
likelihood-ratio prescription therefore does not apply there. Our
threshold $7.81$ follows from the three-dimensional Gaussian-mean region
specified above; it is not an approximation to the two-degree-of-freedom
threshold. A different construction on the physical surface would need
its own coverage calibration. The numerical dependence on the chosen
threshold is weak: $C_{\rm reach}\propto(\chi^2_{\rm crit})^{1/4}$,
so thresholds of $5.99$ and $3.84$ would rescale the sections by $0.936$
and $0.837$, respectively, at fixed response and covariance. These
factors describe the threshold dependence only; they do not assign
coverage to another construction or make the direct-search likelihood
equivalent to the precision forecast.

The inverse covariance here denotes its action after any specified
nuisance projection. The statistic measures distance from the expected
SM mean, rather than from a best fit constrained to the quadratic
surface. This choice is useful at $C_B=C_W=0$, where the ordinary
coupling Jacobian vanishes and the regularity assumptions of a
two-parameter Wilks construction fail. With known Gaussian covariance the construction has nominal
pointwise coverage; $200000$ Gaussian trials for each of ranks $1,2,3$
give coverages $0.95006,0.94966,0.94964$. Axis and diagonal values are
sections of one joint region, not separately profiled 1D intervals.

The passage from coupling shifts to measured widths and asymmetries
introduces a further, separate approximation. For example, a partial
width contains $(a_f^{\rm SM}+\delta a_f)^2$: the linear observable
response retains $2a_f^{\rm SM}\delta a_f$ and omits $(\delta a_f)^2$.
Evaluating the full tree-level expressions with the same calculated
coupling shifts tests this approximation without calculating any new
amplitudes. For the four coefficient sections at 1 TeV, both boundaries
and the conservative and aggressive scenarios, the difference from the
linear response is below $7\times10^{-5}$ of the signal, measured with
the covariance-weighted norm. Linearizing the observable map is thus
numerically adequate for these benchmarks. This comparison does not
estimate corrections to $\delta a_f$ itself, such as higher-order portal
amplitudes or terms omitted from the heavy-mass expansion.

The entries in table~\ref{tab:final-FCC} follow from the information
matrix $B^TV^{-1}B$ with the threshold $\chi^2_3(0.95)$.
Independent reconstruction from the response and covariance matrices
agrees with all 32 FCC-ee entries to the displayed precision; the LEP
reference uses the covariance specified above.

\section{Ultraviolet interpretation: power counting and correlated responses}
\label{app:UV-interpretation-details}
A UV interpretation of the FCC-ee reach must address two issues:
the conditions under which dipoles dominate the pole observables, and
the effect of additional interactions. To develop the interpretation
of section~\ref{sec:fcc}, we first
discuss the size of the relevant coefficients and the stability of
negligible mixing. We then express reinforcement and compensation in
observable space and illustrate them with heavy-lepton representations.

\subsection{Radiative stability and ultraviolet power counting}
The tree/loop classification in weakly coupled UV completions is discussed
in refs.~\cite{Arzt:1994gp,Einhorn:2013kja,Craig:2019wmo} and implemented
for selected one-loop SMEFT operators in the SOLD
dictionary~\cite{Guedes:2023azv}. Strong-sector power counting, such as
SILH~\cite{Giudice:2007fh}, rests on different dynamical
assumptions. These classifications organize coefficient sizes according to the
underlying dynamics; probabilities require additional assumptions on
the model parameters.

Neutrino-mass naturalness constrains large Dirac magnetic moments in
ref.~\cite{Bell:2005kz}. Explicit constructions include the
leptoquark model in ref.~\cite{Brdar:2020quo}, the model accompanying
Higgs production in ref.~\cite{Butterworth:2019iff}, and the one-loop matching
example of ref.~\cite{Chala:2020vqp}. Symmetry-based separation of
mass and magnetic moments is discussed in ref.~\cite{Huang:2022pce};
ref.~\cite{Pages:2025odc} instead uses electroweak representations to
select the weak dipole contribution while suppressing the associated
mass diagram. These mechanisms motivate suppressed mixing or selected dipole
directions. Establishing dipole dominance in the pole observables also
requires the currents and bosonic operators generated by the same
completion. The interplay with mixing at GeV masses and the wider UV
phenomenology are discussed in refs.~\cite{Bertuzzo:2024eds,Kang:2025yep}.

\paragraph{Radiative stability of negligible mixing.}
The negligible-mixing benchmark requires a radiatively consistent
boundary condition or a protecting symmetry. The operator
$C_{LNH}(\bar L\widetilde H N)(H^\dagger H)$ is generated by dipole
running~\cite{Chala:2020pbn}. In the convention
\[
\mathcal L\supset-Y_N\bar L\widetilde HN+C_{LNH}(\bar L\widetilde HN)
(H^\dagger H)+\mathrm{h.c.},
\]
the Dirac mass is
\begin{equation}
 m_D=\frac{v}{\sqrt2}\left(Y_N-\frac{v^2}{2}C_{LNH}\right),
 \qquad |\theta|\simeq\frac{|m_D|}{M_N}.
\end{equation}
For a single Majorana state, or independent real aligned copies without
cancellations, the seesaw contribution is $|m_\nu|\simeq|m_D|^2/M_N$.
Requiring this contribution below an illustrative $0.1$ eV gives
\begin{equation}
 |m_D|\lesssim0.32\,\mathrm{MeV}\sqrt{M_N/(1\,\mathrm{TeV})},
 \qquad
 |\theta|\lesssim3.2\times10^{-7}\sqrt{(1\,\mathrm{TeV})/M_N}.
\end{equation}
Two transition dipoles can also induce a lepton-number-violating mass. The cutoff power counting discussed in
ref.~\cite{Brdar:2020quo} gives
\begin{equation}
 |\delta m_\nu|_{\rm NDA}\sim
 \frac{|\mu_\nu|^2\Lambda_{\rm UV}^2M_N}{16\pi^2}
 =\frac{|d_\gamma|^2\Lambda_{\rm UV}^2M_N}{4\pi^2},
 \qquad \mu_\nu=2d_\gamma.
\end{equation}
For $|d_\gamma|=10^{-5}\,\mathrm{GeV}^{-1}$, $M_N=1$ TeV and
$\Lambda_{\rm UV}=2$ TeV, this estimate is of order $10$ MeV, far above
sub-eV masses. The estimate is sensitive to the cutoff, order-one factors and local
mass counterterms. Its role is to quantify the mass-protection problem:
a small renormalized mixing does not suppress this separate contribution. A viable realization near the conditional reach
needs mass protection or cancellations as well as control of the EWPO
operators. Approximate lepton number with pseudo-Dirac partners changes
the Majorana/flavour benchmark and requires the coupling decomposition
of section~\ref{sec:majorana-dirac} and a separate treatment of
lepton-number-violating observables;
it cannot be assumed to protect the three independent real Majorana
states used here.

\subsection{Dipole dominance in ultraviolet completions}
\label{sec:dipole-dominance}
The relative size of dipole loops and independent local interactions
can be expressed in terms of the physical mediator scale
$\Lambda_{\rm UV}$, which need not equal the boundary scale $\Lambda_B$:

\begin{equation}
 C_X=\frac{c_X}{\Lambda_{\rm UV}^2},\qquad
 D_6=\frac{u_6}{\Lambda_{\rm UV}^2},\qquad
 D_8=\frac{u_8}{\Lambda_{\rm UV}^4}.
\end{equation}
For a representative quadratic combination $c^2$, the threshold-to-UV
ratios scale as
\begin{equation}
 \frac{|\Delta D_6|}{|D_6|}\sim
 \frac{|\kappa_6|c^2}{16\pi^2 r_{\rm UV}^2|u_6|},\qquad
 \frac{|\Delta D_8|}{|D_8|}\sim
 \frac{|\kappa_8|c^2}{16\pi^2|u_8|},\qquad
 r_{\rm UV}=\frac{\Lambda_{\rm UV}}{M_N}.
 \label{eq:dipole-dominance-counting}
\end{equation}
The $\kappa_i$ include gauge and flavour factors, finite terms and
logarithms; mixed dipoles replace $c^2$ by the relevant product. These
ratios compare coefficients in the same operator convention and do not
replace observable-level cancellation tests.

For illustration, with $|\kappa_6|=|c|=1$, a tree-sized
$|u_6|=1$ gives ratios $1.58\times10^{-3}$ and
$2.53\times10^{-4}$ at $r_{\rm UV}=2,5$. For
$|u_6|=1/(16\pi^2)$ the ratios become $1/4$ and $1/25$.
Dominance therefore requires, parametrically,
$|c|\gtrsim4\pi r_{\rm UV}/\sqrt{|\kappa_6|}$ or
$|c|\gtrsim r_{\rm UV}/\sqrt{|\kappa_6|}$, respectively.
A loop-generated dipole can introduce an additional suppression.
Weakly coupled power counting therefore generally favours independent
UV terms over a double insertion of loop-generated dipoles. Dipole
dominance would require a mechanism suppressing those terms, such as a
selection rule or correlated matching. Assessing such a mechanism requires
the full set of operators contributing to the pole response.

If one additionally identifies $\Lambda_{\rm UV}=\Lambda_B$, the
aggressive sections at 1 TeV correspond to dimensionless dipoles
$(|c_B|,|c_W|,|c|_{+},|c|_{-})\simeq
(1.74,0.79,0.57,0.73)$ for $r_B=2$ and
$(6.85,3.11,2.67,3.17)$ for $r_B=5$.
The $r_B=5$ scenario thus probes coefficients larger than unity at this
mass. Their viability depends on the matching couplings, multiplicities
and loop expansion of a specified UV completion; the coefficient size
alone supplies no universal perturbativity boundary.

An explicit scale estimate follows from the approximately degenerate
mediators $X_E,X_N,\phi$ in appendix A of
ref.~\cite{Butterworth:2019iff}. Its leading matching gives
\[
 \frac{\alpha_{NA}}{\Lambda^2}\simeq
 \frac{e g_Lg_Xg_N}{192\pi^2M_{\rm med}^2}.
\]
In our normalization this corresponds to
\begin{equation}
 |g_Lg_Xg_N|\simeq192\pi^2M_{\rm med}^2|C_B+C_W/2|.
 \label{eq:Butterworth-size}
\end{equation}
For the illustrative coefficient $|C_B+C_W/2|=0.1\,\mathrm{TeV}^{-2}$
and $M_{\rm med}=2$ TeV, the required product is about $7.6\times10^2$,
or a geometric-mean coupling $9.1$. This is already outside a comfortably
weakly coupled regime; increasing $M_{\rm med}$ strengthens that conclusion.
The geometric mean summarizes the required coupling product; it does not
impose equal couplings, since the model assumes a hierarchy among them. Its matching also
produces $O_{LNH}$. The estimate exposes the difficulty of obtaining a
large dipole with perturbative mediators, while an exclusion of the UV
model would require its complete correlated EWPO response.

The apparent mass enhancement is also conditional. At fixed dimensionful
$C_X$, the threshold grows as $M_N^2C_XC_Y$ and the leading coefficient
reach improves approximately as $1/M_N$. In a family of UV theories with
fixed dimensionless $c_X$ and fixed $r_{\rm UV}$, instead,
$M_N^2C_XC_Y=c_Xc_Y/(r_{\rm UV}^4M_N^2)$ decreases. There is no
contradiction with decoupling in that limit.

Blind directions in explicit multi-field UV completions, including
finite matching and running, are exhibited in
ref.~\cite{Chala:2025asj}. Positivity of $W,Y$ in universal theories
follows under the assumptions of ref.~\cite{McCullough:2023szd}; it cannot
be applied as an unrestricted prior to our full response, which includes
direct lepton currents. The four directions constrained in the universal
analysis of ref.~\cite{Corbett:2025oqk} concern its own enlarged
bosonic parameter space, not a derivation of our four-combination map.

\subsection{Geometry of compensating contributions}

An additional interaction weakens the forecast only if it changes the
measured quantities in a sufficiently opposing direction. The covariance
provides the appropriate metric for comparing these changes, giving a
simple description of the examples in section~\ref{sec:UV-robustness}.

\paragraph{Magnitude and orientation of the additional response.}
Let $\mathbf d$ denote the dipole-induced shifts of the twelve
pseudo-observables and $\mathbf u$ the sum of all additional contributions,
in the same input scheme. For the SM-centred forecast define
\begin{equation}
 \|\mathbf d\|_V^2=\mathbf d^TV^{-1}\mathbf d,\qquad
 r=\frac{\|\mathbf u\|_V}{\|\mathbf d\|_V},\qquad
 \eta=\frac{\mathbf d^TV^{-1}\mathbf u}
 {\|\mathbf d\|_V\|\mathbf u\|_V}.
\end{equation}
Here $r$ compares the sizes of the two observable patterns, while
$\eta$ measures their alignment after accounting for measurement
uncertainties and correlations. Parallel patterns have $\eta=1$ and
opposite patterns have $\eta=-1$. For nonzero responses the change in
the quadratic statistic is exactly
\begin{equation}
 R\equiv\frac{\chi^2_{\rm dip+extra}}{\chi^2_{\rm dip}}
 =1+r^2+2r\eta .
 \label{eq:UV-response-angle}
\end{equation}
A negative overlap is necessary, but not sufficient, to weaken the
forecast: weakening requires $\eta<-r/2$. In particular, $\eta\geq0$
or $r\geq2$ guarantees $R\geq1$. The orientation is defined by the complete extra response in the
SM-centred forecast. It incorporates the relative precision and
correlations of all measured quantities.

\paragraph{Relative and fixed budgets.}
First consider a \emph{relative} budget that scales with the dipole
response at every coefficient value. If
$\|\mathbf u\|_V\leq\epsilon\|\mathbf d\|_V$ is assumed,
minimization over its direction gives
\begin{equation}
 \chi^2_{\min}=\max(1-\epsilon,0)^2\chi^2_{\rm dip}.
\end{equation}
Along a fixed dipole ray the response scales quadratically with the
coefficient magnitude. At an unchanged reference threshold the maximum
increase of the upper limit is therefore $(1-\epsilon)^{-1/2}$ for
$\epsilon<1$. For example, budgets of $20\%$ and $50\%$ allow increases
of approximately $12\%$ and $41\%$, respectively. These budgets are
assumptions on observable shifts, not probabilities or estimates of
natural UV Wilson coefficients. For $\epsilon\geq1$ no nonzero residual
is guaranteed in the unrestricted response space.

A different assumption fixes the extra response independently of the
dipole magnitude. Let $C_0$ be the nominal limit on a chosen ray,
$k=|C|/C_0$, and $\mathbf d_0=\mathbf d(C_0)$, so that
$\|\mathbf d_0\|_V=\sqrt{\chi^2_{\rm crit}}$ and
$\mathbf d(C)=k^2\mathbf d_0$. For the fixed antiparallel contribution
$\mathbf u=-\epsilon_0\mathbf d_0$, the acceptance condition is
\begin{equation}
 |k^2-\epsilon_0|\leq1,\qquad
 \sqrt{\max(0,\epsilon_0-1)}\leq k\leq\sqrt{1+\epsilon_0}.
 \label{eq:fixed-UV-budget}
\end{equation}
Thus the upper endpoint increases by $\sqrt{1+\epsilon_0}$:
$1.095$ and $1.225$ for $\epsilon_0=0.2,0.5$.
For a fixed contribution with $\epsilon_0>1$ the origin is excluded and
the allowed interval is detached from it. In contrast, profiling over a
fixed \emph{ball} $\|\mathbf u\|_V\leq\epsilon_0\|\mathbf d_0\|_V$
gives the same upper endpoint but retains the origin, since $\mathbf u=0$
is allowed. These fixed budgets differ from the relative budget above;
none of these geometric rescalings assigns a probability to a UV model.

To recover a point $k>1$ times the nominal coefficient limit requires
$R\leq k^{-4}$. Necessary conditions are
\begin{equation}
 |r-1|\leq k^{-2},\qquad
 \eta\leq-\sqrt{1-k^{-4}}.
 \label{eq:UV-compensation-required}
\end{equation}
The second condition follows by optimizing $r$ as well. A factor-two
weakening therefore requires contributions of comparable size with a
strongly opposing orientation. A symmetry or mediator representation
can enforce that orientation. Only four local combinations enter the
universal pole response, even though twelve pseudo-observables are
measured.
These diagnostic rescalings use the original threshold; recalibrated
$95\%$ confidence regions would require including the extra nuisance
parameters in the statistical construction.

\paragraph{Finite terms, logarithms and UV boundary conditions.}
In the fixed prescription the local dipole response can be separated as
$\mathbf d=\mathbf F+\mathbf K\log(\Lambda_B^2/M_N^2)$.
The dipole contribution to $\mathbf K$ is fixed by the EFT, whereas
additional UV operators can also contribute to running. At a fixed pole momentum, the logarithm of scales contributes to the
same local coupling combinations as the finite term. Free boundary
coefficients can consequently compensate their sum. The split between
finite terms and logarithms also depends on the scheme and reference
scale. A prediction retaining only the logarithmic response therefore
corresponds to a particular boundary assumption, rather than to a
minimum signal shared by all completions.

\subsection{Correlated currents from heavy leptons}
Heavy vector-like leptons provide concrete examples in which gauge
quantum numbers fix the relative changes of neutral-current couplings.
Their tree-level currents can reinforce or compensate a dipole signal,
depending on the representation. The matching of
ref.~\cite{Crivellin:2020ebi} fixes their relative signs. Translating to
our convention $\mathcal L_Z=-(g/c_W)Z_\mu\bar f\gamma^\mu
(a_{fL}P_L+a_{fR}P_R)f$, write
$(\delta a_{\nu L},\delta a_{\ell L},\delta a_{\ell R})
=x\,\mathbf p$, with $x=v^2|\lambda|^2/M_{\rm med}^2\geq0$.
The patterns are
\begin{equation}
\begin{array}{c|c|c}
\text{mediator}&(SU(2)_L,Y)&\mathbf p\\ \hline
N_s&(1,0)&(-1/4,0,0)\\
E&(1,-1)&(0,1/4,0)\\
\Sigma_0&(3,0)&(-1/16,-1/8,0)\\
\Sigma_1&(3,-1)&(1/8,1/16,0)\\
\Delta_1&(2,-1/2)&(0,0,-1/4)\\
\Delta_3&(2,-3/2)&(0,0,1/4)
\end{array}
\label{eq:VLL-response-patterns}
\end{equation}
The $E$-singlet result also follows from tree-level matching with
\textsc{Matchete}~\cite{Fuentes-Martin:2022jrf}. Integrating out $E$ in
$\mathcal L_{\rm SM}+\bar E(i\slashed D-M_E)E
-(y_E\bar\ell_L H E_R+\mathrm{h.c.})$
gives, for one lepton flavour, the current interaction
\begin{equation}
 -\frac{|y_E|^2}{2M_E^2}
 i\bigl[H_i^*(D_\mu H)^j-(D_\mu H)_i^*H^j\bigr]
 (\bar\ell_{Lj}\gamma^\mu\ell_L^i).
\end{equation}
Using $\delta_{il}\delta_{jk}=
(\delta_{ij}\delta_{kl}+\sigma^I_{ij}\sigma^I_{kl})/2$ gives
\begin{equation}
 C_{Hl}^{(1)}=C_{Hl}^{(3)}=-\frac{|y_E|^2}{4M_E^2},
 \qquad
 (\delta a_{\nu L},\delta a_{\ell L},\delta a_{\ell R})
 =\frac{v^2|y_E|^2}{M_E^2}(0,1/4,0).
\label{eq:matchete-E-check}
\end{equation}
Field redefinitions also generate a charged-lepton Yukawa operator.
It belongs to the effective action but leaves the current projection
unchanged in the light-lepton limit used here. This establishes the current normalization used in the response comparison.

Here $N_s$ denotes an additional neutral mediator, not an identification
with the dipolar HNL already included. Separate aligned copies for each
lepton flavour realize the universal diagonal patterns. We minimize the
quadratic statistic over the nonnegative strength of one pattern at a
time, independently of the dipole coefficients. The resulting displacement isolates the effect of that correlated
current sector on the dipole forecast.

At $M_N=1$ TeV with the aggressive covariance and both $r_B=2,5$,
the $E$ and $\Sigma_1$ patterns do not reduce the statistic on either
axis or either diagonal. A fixed nonzero contribution therefore tightens
the expected constraint on these sections; if its magnitude is profiled
freely down to zero, the minimum remains at zero additional contribution. The $\Sigma_0$ pattern instead
allows a same-sign diagonal weakening of approximately a factor
$1.5$--$2.2$ at the unchanged threshold.\footnote{For $r_B=2$ and unit coefficient
amplitude in $\mathrm{TeV}^{-2}$, the normalized covariance-weighted overlap
and optimal nonnegative mediator strength are $(\rho,x_\star)=(-0.9010,0.00405)$
for $\Sigma_0$ on $C_B=C_W$, and $(-0.7933,0.00302)$ for $N_s$ on $C_B=-C_W$.
Here $x=v^2|\lambda|^2/M_{\rm med}^2$ and
$x_\star=\max[0,-\langle n,d\rangle_V/\langle n,n\rangle_V]$.
These give weakening factors 1.518 and 1.282 at fixed threshold;
$x_\star$ scales quadratically with the coefficient amplitude.} The $N_s$ pattern gives a more
moderate effect on the opposite-sign diagonal. More importantly,
$N_s$ on the $C_B$ axis and $\Sigma_0$ on the $C_W$ axis are nearly
degenerate with the dipole response in this restricted test. The resulting small residual is particularly sensitive to additional
contributions. Gauge-related charged-current measurements, mediator
loops and other constraints would therefore be essential for assigning
an exclusion to these nearly degenerate cases. In particular, the
optimal strengths illustrate the geometry of compensation in pole
observables. Their realization in a viable mediator model would also
require consistency with mixing and direct-production constraints. The pole comparison
already establishes that the representation can select either a
reinforcing or a compensating direction.

In the dipole-generating $E+\phi$ model of
ref.~\cite{Beltran:2024twr}, the charged-lepton current is correlated
with the same Yukawa vector that enters one contribution to the dipole;
the latter also involves an antisymmetric flavour coupling and an HNL
coupling. For the dipole source sharing this Yukawa coupling, the $E$ tree current
is much larger in the heavy-mediator regime considered. The unequal
magnitudes prevent compensation between these two contributions. The
full response additionally includes scalar contacts, other thresholds
and mixing effects. A loop involving a different,
larger coupling can compete with a small tree contribution without
invalidating perturbation theory.

The examples identify reinforcing and compensating responses fixed by
mediator representations. Assigning probabilities to these possibilities
would require a specified UV model and a measure on its parameters.

\begin{table}[tbp]
\centering\small
\begin{tabularx}{\textwidth}{@{}XX@{}}
\toprule
Results established & Interpretation not implied\\
\midrule
Finite local pole response and a four-direction UV degeneracy
& Complete off-shell dimension-eight matching\\
Conditional FCC-ee sensitivity with specified UV boundaries
& A dipole bound valid for every UV completion\\
Benchmark-level direct--indirect comparison
& A combined likelihood or a complete flavour exclusion\\
Leading radiative Higgs-decay contrast
& A complete Higgs-production analysis\\
\bottomrule
\end{tabularx}
\caption{Scope of the results. The assumptions defining the conditional
forecasts are specified in sections~\ref{sec:ewinputs} and~\ref{sec:fcc}.}
\label{tab:claims-scope}
\end{table}

\section{Direct-search conventions and extensions beyond the pole}
\label{app:comparison-extensions}
This appendix follows the discussion in sections~\ref{sec:direct-comparison-plan}
and~\ref{sec:offpole-identifiability}. We first give the coupling and
flavour conversion used for the HL-LHC comparison, then distinguish
related searches and summarize what off-pole and charged-current
measurements could add. No additional production likelihood is included
in the present forecast.

\subsection{Coupling and flavour conversion}
\label{app:direct-comparison-details}
The conversion below fixes the normalization and flavour convention of
figure~\ref{fig:direct-indirect}.

The direct curves use the same-sign dilepton projections of
ref.~\cite{Vignaroli:2025pwn}. The observable comparison requires the
same operator normalization and diagonal flavour response on both sides.
The coordinate dictionary follows directly from the gauge-invariant
operators. For $C_W=kC_B$,
\begin{equation}
 C_B=\frac{\sqrt2 d_\gamma}{ev(1+k/2)},\qquad C_W=kC_B.
\end{equation}
The benchmark directions of ref.~\cite{Vignaroli:2025pwn} are
\begin{equation}
 \begin{array}{c|c}
 A&C_B=0\\
 B&k=2s_W^2/[c_W(c_W-\sqrt2)]\quad(d_W=d_Z)\\
 C&k=2s_W^2/c_W^2\quad(d_Z=0)\\
 D&C_W=0
 \end{array}
\end{equation}
The photon-blind direction $k=-2$ cannot be parameterized by $d_\gamma$.
A change from $\sigma^I/2$ to $\sigma^I$ in the portal operator also
requires a factor-of-two coefficient conversion.

For the comparison in figure~\ref{fig:direct-indirect}, both curves can
be interpreted in the single-HNL, flavour-universal benchmark of
ref.~\cite{Vignaroli:2025pwn}. At quadratic dipole order the diagonal
pole response depends on $(C_XC_Y^\dagger)_{ii}$ and its flavour trace.
For real universal coefficients, $C_X=C_X^{(0)}(1,1,1)^T$ and the
three-HNL aligned matrix $C_X=C_X^{(0)}I_3$ give the same diagonal
entries and trace. Thus the bosonic factor remains three in both cases,
and each diagonal current is unchanged. The relevant multiplicity is therefore the flavour trace, which includes
both heavy-state and active-flavour couplings. Off-diagonal flavour amplitudes
differ and their constraints are not included in this comparison.

The HL-LHC sensitivities are taken from figure~6 of
ref.~\cite{Vignaroli:2025pwn}, over the mass interval shown there. The HL-LHC criterion $S/\sqrt{S+B}=2$ and our
joint 95\% C.L. region are different statistical constructions.
The separate UV-scale guide assumes unit dimensionless coefficients:
$\Lambda_{\rm eff}=[\max(|C_B|,|C_W|)]^{-1/2}$. Above the guide, keeping
the mediator scale above $M_N$ requires larger dimensionless coefficients;
this is not a model-independent exclusion of EFT validity. For direct
production the partonic energies must also remain within the EFT domain.
The low-mass line-style change has a different purpose, described in
section~\ref{sec:finite-mass-summary}.

The comparison is restricted to the active--sterile dipole and the
stated flavour benchmark. The relation to other direct searches is
summarized next.

\subsection{Related operators and matching tools}
\label{app:literature-scope}
Several useful comparisons involve different operators or production
regimes. The dimension-five sterile--sterile moments of
refs.~\cite{Aparici:2010zz,Barducci:2022gdv} and the see-saw
portal studied at Higgs factories~\cite{Barducci:2020icf} cannot be mapped
directly onto our dimension-six active--sterile coefficients. In
particular, their quoted large cutoff reach is not a bound on $C_B,C_W$.
The active--sterile searches through meson decays, including NA62,
probe the low-mass regime~\cite{Barducci:2023hzo,Barducci:2024nvd}.
The gauge-invariant dimension-six portal at 3- and 10-TeV muon
colliders is studied in ref.~\cite{Barducci:2024kig}, including
single-flavour benchmarks. Comparison with our curves requires matching
the flavour assumptions and collider setting. Other active--sterile
searches are presented in
refs.~\cite{Magill:2018jla,Zhang:2023nxy,Brdar:2025iua,Beltran:2024twr}.
The Majorana transition moments between active neutrino flavours studied
in ref.~\cite{Frigerio:2024jlh} probe lepton-number and flavour violation
and are a distinct operator sector from the active--sterile portal here.
Broader sterile-neutrino EFT phenomenology is covered by
refs.~\cite{Alcaide:2019pnf,Fernandez-Martinez:2023phj,Duarte:2025zrg,Beltran:2025ilg}.
Mixing-induced EW fits and boson-decay corrections in seesaw scenarios,
refs.~\cite{Akhmedov:2013hec,Abada:2023raf}, are distinct from our
negligible-mixing benchmark.

Earlier one-loop electroweak analyses of top and bosonic effective
operators~\cite{Zhang:2012cd,Mebane:2013zga,Liu:2022vgo} provide
context for renormalization and indirect sensitivity, rather than the
portal-specific matching coefficients. The parameter-sensitivity measure of ref.~\cite{Barbieri:1987fn}
addresses a different question from our covariance-weighted response comparison.

Automated off-shell matching with
\texttt{matchmakereft}~\cite{Carmona:2021xtq} and the
\texttt{SOLD} dictionary~\cite{Guedes:2023azv} can assist UV-model
matching. Applying them to the present nonrenormalizable portal would
require its explicit implementation, the retained operator order and a
compatible finite prescription; these tools have not been used as an
independent validation of our two-insertion result.
\subsection{Energy dependence and ultraviolet degeneracies}
\label{app:beyond-pole-details}
Away from the $Z$ pole, energy dependence and new scattering structures
can distinguish interactions that have the same pole response. Their
usefulness depends on the local UV coefficients allowed to vary.

Measurements at distinct energies can separate the constant and slope
of a local amplitude. If both polynomial coefficients also have free UV
contributions, those terms can compensate the dipole polynomial at every
energy. The additional leverage therefore comes from restrictions on the
UV coefficients, new structures in the scattering amplitude, or mass
dependence that is not captured by that local polynomial.

At $M_N=1$ TeV the expansion
parameters $s/M_N^2$ are $0.0576$ and $0.133$ at 240 and 365 GeV;
$(s-M_Z^2)/M_N^2$ is about $0.0493$ and $0.1249$. These ratios characterize the hierarchy underlying the local expansion. At
$M_N=200$ GeV, $s/M_N^2=1.44,3.33$, so this local argument cannot be
used to forecast those runs. A numerical off-pole analysis requires the
complete $e^+e^-\to f\bar f$ amplitudes, including boxes, photon
interference and contact terms, together with the enlarged coefficient
space and an experimental covariance.

\subsection{\texorpdfstring{Prospects for a charged-current test with $W$ pairs}{Prospects for a charged-current test with W pairs}}

The process $e^+e^-\to W^+W^-$ probes charged currents and triple-gauge
interactions in addition to the neutral pole couplings. Doubly resonant
$W$ data can lift flat directions in restricted SMEFT fits
\cite{Berthier:2016tkq}; this does not by itself establish that the
four-direction degeneracy found here is removed.

The charged dipole also permits tree-level $t$-channel HNL exchange with
two insertions and an exact propagator proportional to $(t-M_N^2)^{-1}$.
With negligible active--sterile mixing it probes $C_W$, with no analogous
tree contribution on the pure $C_B$ axis. At quadratic dipole order the
rate depends on its interference with the SM amplitude, whereas the
squared HNL amplitude is quartic. The electron-flavour coupling enters
without a closed-loop flavour multiplier.

Angular distributions at 240 and 365 GeV could separate this exchange
from shifts of the $We\nu$, $Zee$, $WWZ$ and $WW\gamma$ interactions.
The relevant information is the part of that distribution which cannot
be reproduced by the allowed UV coefficients. Establishing its reach
requires the full gauge-related interference, $W$ decays and widths,
and correlated experimental uncertainties. This explains the potential
complementarity of $W$ pairs while delimiting the pole and decay
forecast presented here.

\bibliographystyle{JHEP}
\bibliography{hnl_dipole_ewpo_2026-10-30}

\end{document}